\documentclass[sn-mathphys-num,iicol]{sn-jnl}
\usepackage[utf8]{inputenc}
\usepackage{amssymb,amsmath,latexsym,amsthm,amsfonts,mathtools}

\usepackage{mathrsfs}
\usepackage{multirow}
\usepackage{graphicx}
\usepackage{textcomp,siunitx}
\usepackage{float,subcaption}
\usepackage{booktabs}
\usepackage{enumitem}

\usepackage{bm}
\usepackage{romannum}
\usepackage{anyfontsize}
\usepackage[hyperref]{xcolor}
\definecolor{Skgreen}{rgb}{0, 0.5, 0}
\usepackage{hyperref}
\hypersetup{colorlinks, breaklinks, citecolor=blue, linkcolor=Skgreen, urlcolor=blue}
\usepackage[nameinlink,noabbrev,sort&compress]{cleveref}

\crefname{assumption}{Assumption}{Assumptions}
\crefname{algocfline}{Algorithm}{Algorithms}
\theoremstyle{plain}
\newtheorem{theorem}{Theorem}
\newtheorem{lemma}{Lemma}
\newtheorem{corollary}{Corollary}
\newtheorem{proposition}{Proposition}
\newtheorem{assumption}{Assumption}
\theoremstyle{remark}
\newtheorem{remark}{Remark}

\allowdisplaybreaks
\theoremstyle{definition}
\newtheorem{definition}{Definition}

\newtheorem{problem}{Problem}

\newcommand{\sign}{\mathrm{sign}}
\newcommand{\sgmf}{\mathrm{sgmf}}
\usepackage{accents}

\allowdisplaybreaks
\begin{document}

\title[]{Three-Dimensional Nonlinear Guidance with Impact Time and Field-of-view Constraints}

\author[]{\fnm{Ashok R} \sur{Samrat}}\email{22m0019@iitb.ac.in}
\author*[]{\fnm{Swati} \sur{ Singh*}}\email{swatisingh@aero.iitb.ac.in}
\author[]{\fnm{Shashi Ranjan } \sur{Kumar}}\email{srk@aero.iitb.ac.in}

\affil[]{\orgdiv{Department of Aerospace Engineering}, \orgname{Indian Institute of Technology Bombay}, \orgaddress{\city{Mumbai}, \postcode{400076}, \state{Maharashtra}, \country{India}}}



\abstract{This paper addresses the time-constrained interception of targets at a predetermined time with bounded field-of-view capability of the seeker-equipped interceptors. We propose guidance laws using the effective lead angle and velocity lead angles of the interceptor to achieve a successful interception of the target. The former scheme extends the existing two-dimensional guidance strategy to a three-dimensional setting. We have shown that such an extension may result in high-frequency switching in the input demand, which may degrade the interceptor's performance. To overcome the potential limitations of such a guidance strategy, we propose an elegant solution using the velocity lead angles and the range error with a backstepping technique. Using the velocity lead angles as virtual inputs, the effective lead angle profile is subsequently regulated to satisfy the seeker's field-of-view bound. Unlike the existing strategies, the proposed guidance strategy does not rely on the time-to-go estimate, which is an appealing feature of the design, as the time-to-go estimate may not always be available with high precision. We provide a theoretical analysis of the error variable and subsequently analytically derive the bounds on achievable impact times. Numerical simulations are performed to support the theoretical findings. The performance of the proposed guidance strategy is compared with that of an existing one, and it has been shown to yield better performance. Finally, a study on different choices of virtual inputs is also provided.}

\keywords{Impact time, Three-dimensional engagement, Field-of-view constraints, Nonlinear guidance.}

\maketitle

\section{Introduction}
The primary goal of most terminal-constrained guidance laws is to intercept a target precisely, ensuring no miss distance for effective neutralization. A single interceptor may be countered by the advanced defense systems protecting high-value targets. However, when multiple interceptors are launched simultaneously in a salvo attack \cite{kumar2021deviated,zhao2017distributed}, the likelihood of success and the damage inflicted on the target is significantly increased. For a successful salvo attack, it is crucial to strike the target at a specified impact time \cite{MKim2015Lyap, SRK2015largeHE,Jeon2006ITCAntiShip}.
During the practical execution of any guidance strategy, it is essential that the designer incorporates the constraints due to the restricted field of view (FOV) of its seeker. The onboard seekers on interceptors pose a significant challenge in terms of their bounded field-of-view (FOV), which can cause interceptors to enter dead zones where the interceptor might lose track of the target. This is because imposing a desired impact time may lead the interceptor to move along a curve that occludes the target from its seeker. 
To address this, it is important to account for the seeker's FOV in the design of the guidance scheme.

Guidance strategies aimed at intercepting a target at a specified angle have been a long-standing research topic \cite{zhao2016impact,chen2019two,lyu2017prescribed,majumder2024three,he2015robust}. However, the concept of intercepting a target at a prespecified time gained significant attention starting in the early 2000s. The impact time control problem was first addressed in \cite{Jeon2006ITCAntiShip}, where the authors introduced a feedback term to the conventional proportional navigation guidance (PNG) law to minimize the impact time error. Later, in \cite{Lee2007ITAC}, the derivative of acceleration was included as an additional input to extend the optimal guidance law (OGL), thus accounting for the impact time constraint. However, this approach required integration to obtain the lateral acceleration command (control input), which led to further complications

 Noting that the above guidance laws were based on linear engagement kinematics, in \cite{Jeon2016NonLinear}, a guidance law was designed within a nonlinear framework. On the other hand,  the OGL failed at considerably large initial heading angles due to the inherent assumption of a small lead angle. A composite guidance law with a two-phase variation for the lead angle was presented in \cite{Cheng2018comp_AST}. In the first phase, the lead angle at the launch is maintained constant, and in the second phase, the lead angle linearly decreases to zero at target interception. Under this strategy, the interceptor does not take the collision course. However, pursuing a collision course would be more advantageous, as it maximizes the damage inflicted. 
A virtual target approach with two stages was proposed in \cite{Hu2018VirTar} where the impact time constraint was satisfied in the first stage by controlling the point of entry into the collision course, and then the collision course corresponding to the desired impact angle was tracked using proportional navigation guidance (PNG) law in the second phase. The issue here was the abrupt increase in acceleration at the point of switching from one stage to the next. A feedback controller over a biased PNG law was introduced in \cite{Chen2019OGL_AST} to drive the impact time of the nominal trajectory to the desired value. In \cite{Wang2019SvtDvt}, a differential geometric approach was adapted to design the guidance law by driving the angle between the desired and actual tangent vectors to zero. Although a time-to-go estimate was not used, the position of the virtual target had to be calculated at every instant. Moreover, the feasible range of impact time was not studied. 
In \cite{Harl2012ITACG}, a nonlinear control-based method was adopted, where the authors used a second-order sliding mode control (SMC) with LOS shaping for impact time-constrained guidance (ITCG). Authors of \cite{SRK2015largeHE} also employed SMC to suggest a guidance law for the large initial heading angle errors. Later, \cite{Saleem2016Lyap} proposed a strategy derived using the Lyapunov-based techniques. Both of these guidance strategies utilized time-to-go estimates. In the guidance law proposed in \cite{Saleem2016Lyap}, the lead angle of the interceptor was expected to monotonically decrease to zero, which in turn restricts the range of achievable impact time. Unlike the previous work, in addition to the guidance law, the autopilot was also designed in \cite{Sinha2021nested_AST} for a dual-controlled interceptor, having canard as well as tail control inputs. For a similar interceptor configuration, an integrated guidance and control (IGC) strategy was presented in \cite{Sinha2021integ_AST}, owing to the effectiveness of IGC over the separate guidance and control design. Realizing the importance of accounting for the seeker's FOV, some later works considered the FOV bound. The concept of adding a bias term to the PNG law was used in \cite{Zhang2014} to maintain the lead angle of the interceptor within its bounds. The guidance law posed an additional constraint on the interceptor's lead angle at launch being non-zero. Authors in \cite{Zhang2016FOV_Syslag} had also addressed the FOV bound, but their small-angle assumption restricts the overall performance. 
A guidance law with two terms was introduced in \cite{Shim2018FOV} to fulfill the impact time as well as the impact angle constraint. The first term was responsible for ensuring target interception as well as the impact angle, while the second term compensated for the possible impact time error. The second term contains a varying gain, which handles the FOV bounds.  
The look angle was shaped as a polynomial function of the relative range in \cite{Kim2023Trajec} to achieve terminal-constrained target interception. In \cite{H_J_Kim}, the time-to-go error was expressed as a range error, and the desired lead angle acted as a virtual control input for backstepping the range error. As a result, the FOV bound was easily imposed as a lead angle bound. The maximum amount of lateral acceleration that the control surfaces or thrusters can produce was bounded, introducing saturation of the control input. In addition to the FOV bound, the guidance law given by \cite{Swati2022FOVInputCons} considered the lateral acceleration saturation.

The above-mentioned strategies consider a planar engagement scenario to derive the guidance laws. Due to the strong interaction between the states in the lateral planes in 3D engagement, the guidance laws derived assuming a planar engagement perform poorly in actual 3D scenarios. Hence, the design of the guidance law for three-dimensional engagement requires special consideration. Similar to \cite{Jeon2006ITCAntiShip} for 2D, in \cite{He20193D}, the traditional 3D PNG law was modified with an additional feedback term representing the impact time constraint. However, the decoupling of the dynamics into two lateral planes causes the performance of the resulting guidance law to degrade at the operating conditions with significant coupling in pitch and yaw channels. As a solution, the authors of \cite{Sinha20213DITC} suggested guidance strategies based on the original dynamics without decoupling. In \cite{MKim2015Lyap}, a guidance law for three-dimensional engagement was derived using the Lyapunov stability theory. A guidance law that is designed for a stationary target can be extended for a constant-velocity target using the concept of predicted interception point (PIP) if the velocity of the target is known. The concept of PIP was used in \cite{Sinha2023PIPITCG3D} to present a guidance law for constant speed non-maneuvering targets. While the aforementioned approaches are suitable for three-dimensional engagement, they fail to consider the seeker's FOV constraint in 3D. Due to the fact that the heading of the interceptor in 3D engagements is determined by two heading angles (Euler angles), it is even more difficult to take into consideration the seeker's FOV bound. In \cite{10271537}, the authors provide a 3D ITAC guidance law that does not utilize the time-to-go estimate to achieve the impact constraints. Authors of \cite{Kumar2022BarLyap} introduced a barrier Lyapunov function-based strategy for an interceptor with a bounded FOV. To achieve large impact times, a two-stage strategy, where an additional stage based on deviated pursuit, was introduced at the beginning. The guidance law proposed in this work essentially resembles a deviated pursuit during the initial phase without the need to introduce additional stages. In \cite{He2021OptimalFOV}, the authors address the same problem, but the assumption of a small lead angle limits the performance of the guidance law. Moreover, these guidance laws rely on the time-to-go approximations.  To the best of the authors' knowledge, accounting for FOV constraint in 3D scenarios has still not been completely explored. This article presents a backstepping-based guidance law inspired by \cite{H_J_Kim}, which addressed this problem for a planar interceptor-target engagement. In the context of the above discussions, the main contributions of this paper are summarized below:
\begin{itemize}
     \item A nonlinear 3D guidance law is designed using a backstepping technique to achieve time-constrained interception of stationary targets while satisfying FOV constraints. Owing to the nonlinear framework, the proposed guidance law remains applicable even for the large initial heading errors of the interceptors. 
     
    \item The guidance laws are designed without decoupling the dynamics into two mutually perpendicular planes, thus preventing performance degradation due to the decoupling. 
    
    \item Lateral acceleration demand and look angle reduce to zero near target interception, thus ensuring high impact speed to maximize damage to the target. 

    \item Unlike \cite{H_J_Kim}, in this work, the heading angle has been used instead of the velocity lead angle to avoid oscillations in the lateral acceleration components. The guidance laws utilize a backstepping-based approach to regulate the range error and the heading angle error. The virtual control inputs have been designed in a manner such that the field-of-view constraints are satisfied.

    \item The proposed guidance law permits to control of the maximum lead angle reached through suitable selection of a tunable parameter. Furthermore, the influence of the tunable parameters in the guidance law is further examined using numerical simulation.

    \item In the proposed guidance law, the errors converge in the neighborhood of zero, and a suitable selection of Lyapunov function candidates ascertain the same. 
\end{itemize}
Additionally, as discussed in\cite{H_J_Kim}, in this work, we have obtained the minimum and maximum impact times that can be achieved sufficiently such that the FOV bounds are respected.

The remainder of this paper is organized as follows. \Cref{Sec:Problem_Formulation} provides the necessary background and formally states the problem statement for this paper. A detailed derivation of the guidance law is presented in \Cref{Sec:Main_Results}, followed by validation of proposed designs using simulation results in \Cref{Sec:Performance_Study}. Alternate choices of virtual inputs will be presented in \Cref{Subsec:Vir_Inp}. Finally, \Cref{Sec:Conclusion} provides some concluding remarks and directions for possible future research.

\section{Problem Formulation}
\label{Sec:Problem_Formulation}
\begin{figure}[!ht]
    \centering
    \includegraphics[width = 0.5\textwidth]{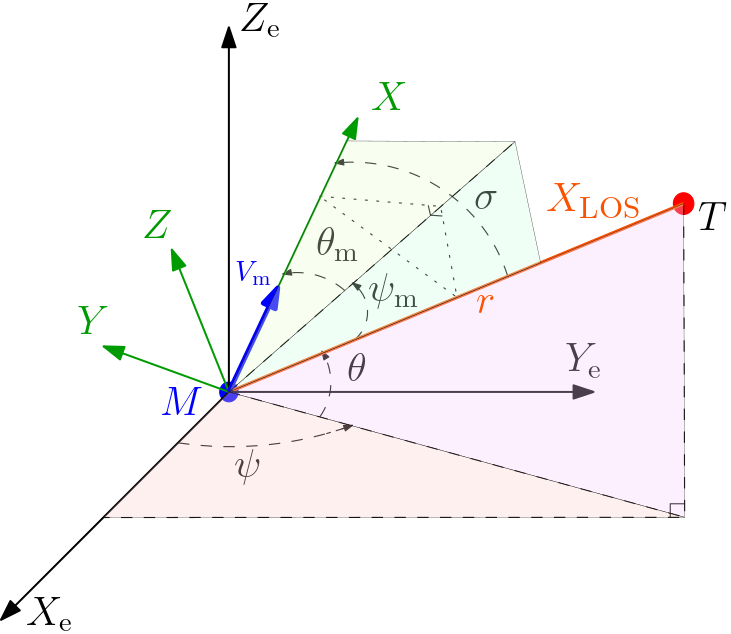}
    \caption{3D engagement Geometry}
    \label{fig:3D_engagement}
\end{figure}

Consider the engagement between an interceptor and a stationary target in the three-dimensional Euclidean space, as shown in \Cref{fig:3D_engagement}. 
We assume the interceptor to be a point-mass nonholonomic vehicle that can maintain a constant speed, $V_{\rm m}$, throughout the engagement. Let $\{X_{\rm e}, Y_{\rm e}, Z_{\rm e}\}$ represent a set of mutually orthogonal axes of the Earth-fixed frame. Without loss of generality, we assume that the origin of this frame is initially located at the interceptor's center of gravity. The line-of-sight frame is defined with respect to the Earth-fixed frame by a set of azimuth and elevation angles of $\psi$ and $\theta$ from $X_{\rm e}$--axis, whose $X$-- axis aligns with the line-of-sight between the interceptor and the target. Similarly, the axes system $\{X,Y,Z\}$ denotes the body frame of the interceptor with the axis $X$ aligned with the velocity vector of the interceptor. The interceptor-target relative separation is represented by $r$. The velocity lead angles or the heading angles of the interceptor, $\psi_{\rm m}$ and $\theta_{\rm m}$ provide the orientation of the velocity vector in the LOS frame along the azimuth and elevation directions, respectively. The effective lead angle of the interceptor, $\sigma$, is the angle between the velocity vector of the interceptor and the LOS between the interceptor and target.  

With the assumption that the interceptor autopilot is sufficiently fast, the dynamics of the interceptor can be neglected, and the guidance law can be designed considering the kinematic relations. The kinematic equations, \cite{Song1994Kine3D}, describing the relative engagement between the interceptor and the target in the spherical coordinate system are given by
\begin{subequations}
    \begin{align}
   \dot{r} &= -V_{\rm m} \cos \theta_{\rm m} \cos \psi_{\rm m}, \\
    \dot{\theta} &= -\frac{V_{\rm m} \sin \theta_{\rm m}}{r}, \\
    \dot{\psi} &= -\dfrac{V_{\rm m} \cos \theta_{\rm m} \sin \psi_{\rm m}}{r \cos \theta}, \\
    \dot{\theta}_{\rm m} &= \dfrac{a_{\rm z}}{V_{\rm m}}-\dot{\psi} \sin \theta \sin \psi_{\rm m}-\dot{\theta} \cos \psi_{\rm m},\label{eq:3D_Kine_eqns_theta_m_dot} \\
\nonumber    \dot{\psi}_{\rm m} &= \dfrac{a_{\rm y}}{V_{\rm m} \cos \theta_{\rm m}}+\dot{\psi} \tan \theta_{\rm m} \cos \psi_{\rm m} \sin \theta \\
    &~~-\dot{\psi} \cos \theta-\dot{\theta} \tan \theta_{\rm m} \sin \psi_{\rm m}.\label{eq:3D_Kine_eqns_psi_m_dot}
    \end{align}
    \label{eq:3D_Kine_eqns}
\end{subequations}
The interceptor lateral acceleration commands are denoted by $a_{\rm y}$ and $a_{\rm z}$, acting along the $Y$ and $Z$ axes of the body frame, respectively. Consequently, they directly influence the velocity lead angles along the azimuth ($\psi_{\rm m}$) and elevation ($\theta_{\rm m}$) directions, respectively. Throughout the rest of this article, $\theta_{\rm m}$ and $\psi_{\rm m}$ are also referred to as the heading angles for convenience. It is evident from \eqref{eq:3D_Kine_eqns} that the motions in the two lateral planes are highly coupled, and hence, decoupling the dynamics is to be avoided to prevent loss of performance.
The effective lead angle of the interceptor, $\sigma$, is related to the interceptor's heading angles through the following expression
\begin{equation}
    \cos{\sigma} = \cos{\psi_{\rm m}}~\cos{\theta_{\rm m}}.
    \label{eq:sigma_psi_m_theta_m}
\end{equation}
The dynamics of the effective lead angle can be obtained by differentiating \eqref{eq:sigma_psi_m_theta_m} with respect to time as 
\begin{equation}\label{eq:simplifyintersigma}
    \dot{\sigma} \sin{\sigma} = \sin{\theta_{\rm m}}\cos{\psi_{\rm m}}\dot{\theta}_{\rm m} + \sin{\psi_{\rm m}}\cos{\theta_{\rm m}}\dot{\psi}_{\rm m},
\end{equation}
which on substituting for $\dot{\theta}_{\rm m}$ and $\dot{\psi}_{\rm m}$ from
the kinematic relations given in \eqref{eq:3D_Kine_eqns} and performing further simplification, yields
\begin{equation}\label{sigma_dot}
\dot{\sigma} = \dfrac{V_{\rm m}\sin{\sigma}}{r} + \dfrac{\sin{\psi_{\rm m}}}{V_{\rm m}\sin{\sigma}}a_{\rm y} + \dfrac{\sin{\theta_{\rm m}}\cos{\psi_{\rm m}}}{V_{\rm m}\sin{\sigma}}a_{\rm z}.
\end{equation} 
It may be observed from \eqref{sigma_dot} that the interceptor's effective lead angle has a relative degree of one with respect to the lateral acceleration commands $a_{\rm y}$ and $a_{\rm z}$. Now, we present two essential temporal quantities for the impact time-constrained guidance in the following definition.  
\begin{definition}
    The time taken by the interceptor to intercept the target, starting from its launch instant or any pre-specified reference, is defined as the impact time. At any point during the engagement, the time-to-go refers to the remaining time before the target is intercepted. The desired value of time-to-go $t_{\rm go}^{\rm d}$ at any time instant $t_{\rm i}$, for a given desired impact time, $t_{\rm f}$, is given as follows:
\begin{equation}
    t_{\rm go}^{\rm d} = t_{\rm f} - t_{\rm i}.
\end{equation}
\end{definition}
\begin{assumption}\label{assum1}
    The angle of attack and the sideslip angles are assumed to be sufficiently small enough to be considered negligible. This, in turn, implies that the body frame coincides with the wind frame.
\end{assumption}
As a consequence of the above assumption, the seeker's FOV bound can be converted as a bound on the effective lead angle. Taking the FOV bound to be in the range $(-\pi/2, \pi/2)$, we get
\begin{equation}\label{sigma bound}
    |\sigma| \le \sigma_{\max} < \pi/2,
\end{equation}
where $\sigma_{\max}$ is the maximum permissible lead angle of the interceptor. Equation \eqref{sigma bound} intuitively implies that the velocity vector should remain within a cone with the LOS line as the axis and $\sigma_{\rm max}$ as the semi-vertical angle \cite{Kumar2022BarLyap}.

\begin{problem}
    The aim of this paper is to design nonlinear guidance schemes that enable the interceptor to achieve target interception with desired impact time requirements while complying with the interceptor's FOV bounds for three-dimensional engagements, given the interceptor's launch angle, speed, initial range-to-go, and value of maximum permitted FOV. 
\end{problem}
Owing to the reduction in performance associated with linearization or decoupling of the kinematic equations, such approximations should not be considered at any point during the derivation of the guidance scheme. This will make the guidance law viable, even at large heading angles and in the presence of strong coupling.

\section{Main Results}
\label{Sec:Main_Results}
In this section, we design a three-dimensional nonlinear guidance law to ensure target interception at a prespecified impact time that will obey the FOV constraints. Towards that, we first present some key concepts used in this guidance design. Subsequently, we analyze the derived guidance laws, focusing on error convergence, the range of achievable impact times, and other relevant factors.
 
It is well known that a target interception is guaranteed if the relative speed between the interceptor and the target is negative and their LOS rate is maintained at zero. For a stationary target, we get the geometric rule, in terms of heading angle, given by $\sigma = 0$, which is essentially a homing rule that makes the interceptor point toward the target while flying toward it in a straight line. The impact time for an interceptor tracking a collision course is given by
\begin{equation}\label{eq:t_imp_CC}
    t_{\rm im} = \frac{r(0)}{V_{\rm m}},
\end{equation}
where $r(0)$ is the initial separation between the adversaries and $V_{\rm m}$ is the speed of the interceptor. To achieve an impact time smaller than that given in \eqref{eq:t_imp_CC}, the interceptor would need to fly at higher speeds. However, such interception is not possible due to the assumption of constant speed. Hence, \eqref{eq:t_imp_CC} represents the shortest attainable impact time for an interceptor moving at a constant speed. The assumption of constant speed also implies that tracing a longer trajectory leads to a higher impact time. The same can be achieved by initially maintaining a non-zero lead angle and later converging to the collision course as soon as the time-to-go matches its desired value. In the guidance law in \cite{He20193D}, such compensation was achieved by introducing an additional feedback term in the conventional PNG law. Alternatively, a backstepping approach could be taken where the desired lead angle, $\sigma_{\rm d}$, is employed as a virtual control input to minimize the error in impact time. The main advantage of the backstepping-based approach is that the idea can easily be extended from a planar engagement to a three-dimensional(3D) scenario. The same will be exercised in the upcoming sections to derive the proposed guidance law. Some existing results that aid in guidance design will be useful are presented next.
\begin{lemma}
\label{lem:lyapunov}
    \cite{Khalil2002Ch4} Let $\pmb{x}=0$ be an equilibrium point for the system $\dot{\pmb{x}}=\pmb{f}(\pmb{x})$. Let $V:\mathbb{R}^n\rightarrow \mathbb{R}$ be a continuously differentiable function such that
    \begin{align*}
        V(\pmb{0})=0; \ V(\pmb{x})&>0\,\ \forall\ \pmb{x}\neq \pmb{0},~~
        ||\pmb{x}||\rightarrow \infty \\
        &\implies V(\pmb{x}) \rightarrow \infty
    \end{align*}
    \begin{itemize}
        \item If $\dot{V}(\pmb{x})<0\,\ \forall\ \pmb{x}\neq \pmb{0}$, then $\pmb{x}=\pmb{0}$ is globally asymptotically stable.
        \item If $\dot{V}(\pmb{x})<-kV^\rho(\pmb{x})\,\ \forall\ \pmb{x}\neq \pmb{0}$, where $k>0$ and $\rho \in (0,1)$, then $\pmb{x}=\pmb{0}$ is finite time stable.
    \end{itemize}
\end{lemma}

\subsection{Derivation of Guidance Strategy using Effective Lead Angle}
\label{SubSec:Backstep_Eff_Lead_Angle}
To derive the guidance command for achieving the desired objective, we define the range error, $e_1$, as the difference between the distance covered by the interceptor ($V_{\rm m}t_{\rm go}^{\rm d}$) (assuming the interceptor to be traveling in a straight line) and the actual interceptor-target separation $r$. One may also interpret this error as a time-to-go error for a constant speed interceptor.

Next, to achieve the same, we further define the lead angle error, $e_2$, as the difference between the actual lead angle $\sigma$ and its desired value, $\sigma_{\rm d}$. These errors are expressed as follows 
\begin{align}\label{eq:e1_e2}
 e_1 &= V_{\rm m} t_{\rm go}^{\rm d} - r,~~e_2 = \sigma - \sigma_{\rm d}.
\end{align} 

On differentiating the range error, $e_1$, given in \eqref{eq:e1_e2}, with respect to time, we obtain
\begin{equation}
    \dot{e}_1 = -V_{\rm m} + V_{\rm m} \cos{\sigma} .\label{eq:e_1_dot}
\end{equation}
From \eqref{eq:e_1_dot} and \eqref{sigma bound}, it may be observed that a positive initial range error ($e_1 > 0$), and a non-zero lead angle ensures that the range error decreases, as $\cos{\sigma}<1$ for $|\sigma|\in(0,\pi/2)$. Therefore, one may choose a desired heading angle $\sigma_{\rm d}$ such that $\sigma_{\rm d} \neq 0$ whenever the range error is non-zero and gradually converges to zero as the range error tends to zero. An immediate option for the desired lead angle could be selected as a scaled version of the range error, that is, $\sigma_{\rm d} = C_0 e_1$.
However, choosing a $C_0$ that guarantees the convergence in $e_1$ requires further analysis of the range error rate, $\dot{e}_1$. Therefore, as an alternative, we utilize the dynamic inversion on $\dot{e}_1$ to derive an expression for $\sigma_{\rm d}$. On inverting the dynamics of $e_1$ using \eqref{eq:e_1_dot}, we derive the desired lead angle, $\sigma_{\rm d}$, such that $\dot{e}_1 = -k_1V_{\rm m}\,\sign(e_1)$,  ensuring linear convergence of $e_1$. 
The corresponding expression for the lead angle, which also serves as a virtual input, may be obtained as
\begin{equation}\label{eq:sigmad}
    \sigma_{\rm d} = \cos^{-1}[1-k_1\sign(e_1)],
\end{equation}
where $\sign(\cdot)$ denotes the signum function, defined as 
\begin{equation}
{\sign}(x) = 
  \begin{cases}
    1,&  x>0\\
    -1,& x<0         
  \end{cases}.
  \label{sgn}
\end{equation}
As the function $\sign(\cdot)$ is discontinuous at the origin; therefore, it can be replaced by a differentiable function $\sgmf(\cdot)$, an approximation to $\sign(\cdot)$, \cite{H_J_Kim}, which is given by
\begin{equation}
{\sgmf}(x) = 
    \begin{cases}
     -\dfrac{x^3}{2\phi^3}+\dfrac{3x}{2\phi},&|x|\le\phi\\
      \sign(x),& |x|>\phi
     \end{cases}.
     \label{sgmf}
\end{equation}
Following this, the final expression for the desired lead angle of the interceptor becomes
\begin{equation}\label{sigma_d_kim}
\sigma_{\rm d} = f(e_1) = \cos^{-1}[1-k_1{\sgmf}(e_1)].
\end{equation}
Using the expression in \eqref{sigma_d_kim}, the derivative of desired lead angle, $\dot{\sigma}_{\rm d}$, may be obtained as 
\begin{equation*}
\begin{split}
     \dot{\sigma}_{\rm d} =
    \begin{cases}
        \dfrac{k_1(-V_{\rm m}+V_{\rm m} \cos{\sigma})}{\sin{\sigma_{\rm d}}} \left[\dfrac{-3e^2_1}{2\phi^3} + \dfrac{3}{2\phi}\right];
       |e_1|\le\phi\\
    0; ~~~~~~~~~~~~~~~~~~~~\mathrm{else}
    \end{cases}
\end{split}
\end{equation*}  
\begin{remark}
One may observe that when the range error is zero ($e_1=0$), from \eqref{sigma_d_kim}, we obtain $\sigma_{\rm d}=2n\pi~\forall n\in \mathbb{Z}$. However, due to the field-of-view (FOV) constraints, $\sigma\in(-\pi/2, \pi/2)$, the only feasible solution is $\sigma_{\rm d} = 0$. On the other hand, $\sigma_{\rm d}$ is zero only if $e_1=0$. Therefore, we establish the equivalence: $\sigma_{\rm d}=0 \iff e_1=0$.
Furthermore, the virtual control input in \eqref{sigma_d_kim} ensures that $e_1$ converge linearly when $|e_1|>\phi$ and asymptotically when $|e_1|<\phi$, provided these constraints are satisfied.
\end{remark}
Next, we aim to design the interceptor lateral acceleration commands such that the lead angle errors converge to zero. Once the error $e_2$ converges to zero, the effective lead angle tracks the virtual control input, $\sigma = \sigma_{\rm d}$. The following theorem presents a guidance law that guarantees linear convergence of the lead angle error of the interceptor to its desired values.
\begin{theorem}
\label{thm:thm_1}
Consider the three-dimensional interceptor-target engagement whose kinematics is governed by \eqref{eq:3D_Kine_eqns} with the desired lead angle, $\sigma_{\rm d}$ given by
\begin{equation*}
    \sigma_{\rm d} = \cos^{-1}[1-k_1\sgmf(e_1)].
\end{equation*}
If the interceptor lateral acceleration is designed as 
\begin{align}
\label{eq:a_accel}
    a_{\rm y} &= \dfrac{V_{\rm m} \sin{\sigma}}{\sin{\psi_{\rm m}}} \left( \dot{\sigma}_{\rm d} -k_2~{\sign}(e_2) \right),\\
    a_{\rm z} &= \dfrac{V_{\rm m}}{\sin{\theta_{\rm m}}\cos{\psi_{\rm m}}} \left( \dfrac{-V_{\rm m} \sin^2{\sigma}}{r} \right),
\end{align}
with $k_2>0$, then the interceptor lead angle error, $e_2$, converges to zero within a finite time. Subsequently, target interception will be achieved at the desired impact time. 
\end{theorem}
\begin{proof}
Consider a Lyapunov function candidate as $\mathcal{V} = 0.5e_2^2$. The time derivative of the Lyapunov function candidate $\mathcal{V}$ is obtained as $\dot{\mathcal{V}} = \dot{e}_2e_2$. On differentiating the lead angle error, $e_2$ given in \eqref{eq:e1_e2} with respect to time and substituting it in \eqref{sigma_dot}, the error dynamics of the lead angle error, $\dot e_2$, may be obtained as
\begin{align}
    \nonumber \dot{e}_{2} &= \dfrac{V_{\rm m}\sin{\sigma}}{r} + \dfrac{\sin{\psi_{\rm m}}}{V_{\rm m}\sin{\sigma}}a_{\rm y} + \dfrac{\sin{\theta_{\rm m}}\cos{\psi_{\rm m}}}{V_{\rm m}\sin{\sigma}} a_{\rm z} - \dot{\sigma}_{\rm d},\\
\nonumber     &= \dfrac{1}{\sin{\sigma}}\left[\left(\dfrac{V_{\rm m}\sin^2{\sigma}}{r} + \dfrac{\sin{\theta_{\rm m}}\cos{\psi_{\rm m}}}{V_{\rm m}}a_{\rm z}\right)\right.\\ 
     &\left. + \left(\dfrac{\sin{\psi_{\rm m}}}{V_{\rm m}}a_{\rm y} - \dot{\sigma}_{\rm d} \sin{\sigma}\right)\right].\label{eq:e2_dot_1}
\end{align}
With the interceptor's lateral acceleration command given in \eqref{eq:a_accel}, the expression of $\dot{e}_2$ becomes
\begin{equation}
    \dot{e}_{2} = -k_2~{\sign}(e_2).
    \label{eq:e2_dot_2}
\end{equation}  
On may substitute \eqref{eq:e2_dot_2} in the expression of $\dot{\mathcal{V}}$, to obtain $\dot{\mathcal{V}} = -k_2~|e_2|$, which further implies that $\dot{\mathcal{V}} = -\sqrt{2}k_2 \mathcal{V}^{0.5}$. Hence, for any chosen value of $k_2>0$, $\dot{\mathcal{V}}$ is negative definite. Therefore, the error $e_2$ will converge to zero within a finite time by \Cref{lem:lyapunov}. Once the error $e_2$ converges to zero, the expression for lead angle is given by \eqref{sigma_d_kim}. Substituting \eqref{sigma_d_kim} into \eqref{eq:e_1_dot}, one may observe that the range error converges linearly for $|e_1|>\phi$ and asymptotically for $|e_1|<\phi$. This completes the proof. 
\end{proof}
\begin{remark}
Note that while implementing the proposed guidance law, the parameter $k_2$ needs to be selected. The magnitude of interceptor lateral acceleration and the rate of convergence of error $e_2$ are proportional to the parameter $k_2$. Hence, there is a trade-off between the time of convergence and the interceptor's lateral acceleration demand. 
\end{remark}

The presence of the term $k_1\sgmf(e_1)$ in  \eqref{eq:a_accel} indicates that variations in the lead angle error $\sigma$ influence the interceptor's lateral acceleration command $a_{\rm y}$. However, from \eqref{eq:3D_Kine_eqns}, one can observe that $a_{\rm y}$ can only affect one of the heading angles ($\psi_{\rm m}$). Due to the coupling between $\psi_{\rm m}$ and $\theta_{\rm m}$, choosing $a_{\rm y}$ as in \eqref{eq:a_accel} might result in an oscillatory behavior in acceleration components. To avoid this, one may use heading angles, $\theta_{\rm m}$ and $\psi_{\rm m}$ to design the guidance command. The same is discussed in the upcoming subsection. 

\subsection{Derivation of Guidance Strategy using Heading Angles}
\label{SubSec:Backstep_Heading_Angles}
In this subsection, we propose a guidance strategy wherein the range error and its rate will be regulated to zero by considering the desired heading angles, $\psi_{\rm m_d}$ and $\theta_{\rm m_d}$, as virtual control inputs. This approach offers greater design flexibility, as it allows the selection of two independent functions instead of one, unlike the previous approach.
 
Additionally, since the lateral accelerations, $a_{\rm z}$ and $a_{\rm y}$, directly affect $\theta_{\rm m}$ and $\psi_{\rm m}$, respectively, the oscillatory behavior that was observed in the backstepping based design (using velocity lead angle) can also be prevented. Analogous to the lead angle error, heading errors are defined as follows: 
\begin{align}
e_3 = \theta_{\rm m} - \theta_{\rm m_d},~~
e_4 = \psi_{\rm m} - \psi_{\rm m_d}.\label{eq:e3_e4}
\end{align}
Note that there can be multiple possible desired heading angles corresponding to the same desired lead angle, as seen from \eqref{eq:sigma_psi_m_theta_m}. We have chosen $\psi_{\rm m_d}$ and $\theta_{\rm m_d}$ to be $\mathcal{C}^1$-continuous functions such that $\dot{e}_3$ and $\dot{e}_4$ exist. 
In the following theorem, we present the lateral acceleration commands that will drive the error signals $e_3$ and $e_4$ to zero.

\begin{theorem}
\label{thm:thm_2}
For the errors $e_3$ and $e_4$, as defined in \eqref{eq:e3_e4}, if $\theta_{\rm m_d}$ and $\psi_{\rm m_d}$ are $\mathcal{C}^1$-continuous and the interceptor lateral acceleration commands are chosen as
\begin{subequations}\label{eq:head_angle_accel}
\begin{align}
\nonumber     a_{\rm y} &= V_{\rm m}\,\cos{\theta_{\rm m}} \left[ -\dot{\psi}\,\tan{\theta_{\rm m}}\cos{\psi_{\rm m}}\sin{\theta}+ \dot{\psi}\,\cos{\theta} \right.\\ 
     &\left.
     + \dot{\theta}\,\tan{\theta_{\rm m}}\sin{\psi_{\rm m}} + \dot{\psi}_{\rm m_d} -k_4\,{\sign}(e_4)\right],\label{eq:head_angle_accel_ay}\\
    a_{\rm z} &= V_{\rm m}\left[ \dot{\psi}\,\sin{\theta}\sin{\psi_{\rm m}} + \dot{\theta}\,\cos{\psi_{\rm m}} + \dot{\theta}_{\rm m_d} \nonumber\right.\\
    &\left.-k_3\,{\sign}(e_3)\right]\label{eq:head_angle_accel_az},
\end{align}
\end{subequations}
where $k_3$ and $k_4$ are positive constants, then the errors $e_3$ and $e_4$ will converge to zero in finite time.
\end{theorem}
\begin{proof}
Note that the proof is similar to that of Theorem \ref{thm:thm_1}, so for conciseness, we provide it in compact form to maintain focus on the key technical contributions. Differentiating \eqref{eq:e3_e4} with respect to time and using the relation in \eqref{eq:3D_Kine_eqns_psi_m_dot} and \eqref{eq:3D_Kine_eqns_theta_m_dot}, one may obtain
\begin{subequations}
    \begin{align}
        \dot{e}_3 =& \frac{a_{\rm z}}{V_{\rm m}}-\dot{\psi} \sin \theta \sin \psi_{\rm m}-\dot{\theta} \cos \psi_{\rm m} - \dot{\theta}_{\rm m_d},
    \label{e_3_dot_2}
   \\ 
       \dot{e}_4 =& \frac{a_{\rm y}}{V_{\rm m} \cos \theta_{\rm m}}+\dot{\psi} \tan \theta_{\rm m} \cos \psi_{\rm m} \sin \theta-\dot{\psi} \cos \theta \nonumber\\
       &-\dot{\theta} \tan \theta_{\rm m} \sin \psi_{\rm m} - \dot{\psi}_{\rm m_d}. 
    \label{e_4_dot_2}
    \end{align}
\label{eq:e3dot2_e4dot2}
\end{subequations}
On substituting the value lateral acceleration commands from \eqref{eq:head_angle_accel} into \eqref{eq:e3dot2_e4dot2}, we get
\begin{align}
    \dot{e}_3 & = -k_3 ~{\sign}(e_3),
    \dot{e}_4  = -k_4 ~{\sign}(e_4).\label{eq:e3_e4_desired_rates}
\end{align}
Consider the Lyapunov function candidates, $\mathcal{V}_j = 0.5e_j^2~\forall~j\in\{3,4\}$. The time derivative of $\mathcal{V}_j$ is obtained as  $\dot{\mathcal{V}}_j = -\sqrt{2}k_j \mathcal{V}_j^{0.5}$, which is negative definite $\forall \, e_{j}\neq 0$, therefore the errors $e_j\,\forall~j~\in\{3,4\}$, will converge within a finite time by  \Cref{lem:lyapunov}. This completes the proof. 
\end{proof} 
\begin{remark}
The presence of the rate terms $\dot{\theta}_{\rm m_d}$ and $\dot{\psi}_{\rm m_d}$ in \eqref{eq:head_angle_accel} necessitates that $\theta_{\rm m_d}$ and $\psi_{\rm m_d}$ must be $\mathcal{C}^1$-continuous to ensure that the interceptor's lateral accelerations remain finite. One can observe that the interceptor lateral acceleration commands $a_{\rm z}$ and $a_{\rm y}$ in \eqref{eq:head_angle_accel} cancel out the dynamics of $\theta_{\rm m}$ and $\psi_{\rm m}$, respectively.  Hence, the interceptor lateral acceleration commands in \eqref{eq:head_angle_accel} will not lead to any coupling effect, as opposed to \eqref{eq:a_accel}.    
\end{remark}
 
Note that the virtual inputs need to be chosen such that the desired lead angle remains the same, that is, $\sigma_{\rm d} = \cos^{-1}[1-k_1\sgmf(e_1)]$. One such choice can be obtained by imposing an additional constraint of $\theta_{\rm m_d} = \psi_{\rm m_d}$. With this assumption,  \eqref{eq:sigma_psi_m_theta_m} reduces to the form, $\cos{\sigma} = \cos^2{\psi_{\rm m}} = \cos^2{\theta_{\rm m}}$. 
Under this assumption, in the following proposition, we derive the expressions for $\theta_{\rm m_d}$ and $\psi_{\rm m_d}$.
\begin{proposition}
Consider the desired lead angle $\sigma_{\rm d}$ be defined as in \eqref{sigma_d_kim}, and suppose the desired heading angles satisfies $\theta_{\rm m_d} = \psi_{\rm m_d}$. Under this condition, the desired heading angles admit the following closed-form expression
    \begin{equation}
        \psi_{\rm m_d} = \theta_{\rm m_d} = \frac{1}{2}\cos^{-1}\left[2\cos(\sigma_{\rm d})-1\right].
    \end{equation} 
\end{proposition}
\begin{proof}
With $\sigma_{\rm d}$ given in \eqref{eq:sigmad} and the relation $\theta_{\rm m_d} = \psi_{\rm m_d}$, the expression in \eqref{eq:sigma_psi_m_theta_m} becomes
\begin{equation}\label{eq:prepositioneqThetamd}
    [1-k_1\sgmf(e_1)] = \cos^2{\theta_{\rm m_d}}.
\end{equation}
Using the trigonometric relation, $\cos{2\theta_{\rm m_d}} = 2\cos^2{\theta_{\rm m_d}} - 1$, the expression in \eqref{eq:prepositioneqThetamd} further simplifies to
\begin{equation}
 \nonumber   [1-k_1\sgmf(e_1)] = \frac{1}{2}\left(\cos{2\theta_{\rm m_d}}+1\right),
\end{equation}
which further implies
\begin{equation}\label{eq:thetaDpsiD}
\psi_{\rm m_d} = \theta_{\rm m_d} = \frac{1}{2}\cos^{-1}[1-2k_1\sgmf(e_1)].
\end{equation}
Using the relation $[1-k_1\sgmf(e_1)] = \cos{\sigma_{\rm d}}$, one may write \eqref{eq:thetaDpsiD} as
\begin{equation}
    \label{eq:bk_step_psi_theta_m}
    \psi_{\rm m_d} = \theta_{\rm m_d} = \frac{1}{2}\cos^{-1}\left[2\cos(\sigma_{\rm d})-1\right].
\end{equation}
\end{proof}
We remark that, if the tracking errors $e_3(t)$ and $e_4(t)$ converge to zero, the field-of-view constraints will be obeyed throughout the engagement.

It can be observed from \eqref{eq:head_angle_accel} that the implementation of the guidance commands requires the time derivatives of the desired heading angles, that is, $\dot{\theta}_{\rm m_d}$ and $\dot{\psi}_{\rm m_d}$. One may differentiate the expressions of $\theta_{\rm m_d}$ and $\psi_{\rm m_d}$ from \eqref{eq:thetaDpsiD} to obtain 
\begin{equation}\label{eq:thetaD_dot}
    \dot{\theta}_{\rm m_d} = 
        \begin{cases}
        \dfrac{k_1(-V_{\rm m}+V_{\rm m} \cos{\sigma})}{\sin{2\theta_{\rm m_d}}} \left[\dfrac{-3e^2_1}{2\phi^3} + \dfrac{3}{2\phi}\right]&
       |e_1|\le\phi\\
    0& \phantom{a b c d}{\mathrm{else}}
    \end{cases}
    \end{equation}
    \begin{equation}\label{eq:psiD_dot}
    \dot{\psi}_{\rm m_d} =     \begin{cases}
        \dfrac{k_1(-V_{\rm m}+V_{\rm m} \cos{\sigma})}{\sin{2\psi_{\rm m_d}}} \left[\dfrac{-3e^2_1}{2\phi^3} + \dfrac{3}{2\phi}\right]&
       |e_1|\le\phi\\
    0& \phantom{a b c d}{\mathrm{else}}
    \end{cases}
\end{equation}
\begin{remark}
 From \eqref{eq:thetaDpsiD}, it follows that when the range error is zero, we obtain $\theta_{\rm m_d} = \psi_{\rm m_d}= \dfrac{n\pi}{2}~\forall n\in \mathbb{Z}$. However, due to the field-of-view (FOV) constraints, $\sigma_{\rm d}\in(-\pi/2, \pi/2)$, therefore both $\theta_{\rm m_d}$ and $\psi_{\rm m_d}$ will have values in $\left(-\dfrac{n\pi}{2}, \dfrac{n\pi}{2} \right)$ such that relation in \eqref{eq:sigma_psi_m_theta_m} is satisfied.   
\end{remark}

\begin{corollary}\label{col:corollary_SigmaAcc}
    With the proposed guidance strategy, both the velocity lead angle, $\sigma$, and the lateral acceleration components, $a_{\rm y}$ and $a_{\rm z}$, converge in the neighborhood of zero near target interception. 
\end{corollary}
\begin{proof}
    Consider the Lyapunov function candidates $\bar{\mathcal{V}}_1 = 0.5\psi_{\rm m}^2$ and $\bar{\mathcal{V}}_2 = 0.5\theta_{\rm m}^2$. Differentiating $\bar{\mathcal{V}}_1$ and $\bar{\mathcal{V}}_2$, one may obtain
    \begin{equation*}
        \dot{\bar{\mathcal{V}}}_1 = \psi_{\rm m}\dot{\psi}_{\rm m},\,
        \dot{\bar{\mathcal{V}}}_2 = \theta_{\rm m}\dot{\theta}_{\rm m}.
    \end{equation*}
    On substituting for \eqref{eq:3D_Kine_eqns_psi_m_dot} and \eqref{eq:3D_Kine_eqns_theta_m_dot} in the expression of $ \dot{\bar{\mathcal{V}}}_1$ and $ \dot{\bar{\mathcal{V}}}_2$, one may obtain
   \begin{align}
    \nonumber    \dot{\bar{\mathcal{V}}}_1 &= \psi_{\rm m}\left(\dfrac{a_{\rm y}}{V_{\rm m} \cos \theta_{\rm m}}+\dot{\psi} \tan \theta_{\rm m} \cos \psi_{\rm m} \sin \theta \label{eq:V1DotAy}\right.\\
    &\left.-\dot{\psi} \cos \theta-\dot{\theta} \tan \theta_{\rm m} \sin \psi_{\rm m}\right),\\
        \dot{\bar{\mathcal{V}}}_2 &= \theta_{\rm m}\left(\dfrac{a_{\rm z}}{V_{\rm m}}-\dot{\psi} \sin \theta \sin \psi_{\rm m}-\dot{\theta} \cos \psi_{\rm m}\right).\label{eq:V1DotAz}
    \end{align}
 One may substitute for $a_{\rm y}$ and $a_{\rm z}$ from \Cref{thm:thm_2} in \eqref{eq:V1DotAy} and \eqref{eq:V1DotAz} to obtain
 \begin{align}
      \dot{\bar{\mathcal{V}}}_1 &= \psi_{\rm m}\left(\dot{\psi}_{\rm m_d} - k_4\sign(e_4)\right),\\
      \dot{\bar{\mathcal{V}}}_2 &= \theta_{\rm m}\left(\dot{\theta}_{\rm m_d} - k_3\sign(e_3)\right).
 \end{align}
 One may substitute for $\dot{\theta}_{\rm m_d}$ and $\dot{\psi}_{\rm m_d}$ from \eqref{eq:thetaD_dot} and \eqref{eq:psiD_dot} to get
 \begin{align}
\nonumber     \dot{\bar{\mathcal{V}}}_1 &= \psi_{\rm m}\left(\dfrac{k_1(-V_{\rm m}+V_{\rm m} \cos{\sigma})}{\sin{2\psi_{\rm m_d}}} \left[\dfrac{-3e^2_1}{2\phi^3} + \dfrac{3}{2\phi}\right] \right.\\
     &\left.- k_4\sign(e_4)\right),\label{eq:V1dotpsiMd}\\
 \nonumber    \dot{\bar{\mathcal{V}}}_2 &= \theta_{\rm m}\left(\dfrac{k_1(-V_{\rm m}+V_{\rm m} \cos{\sigma})}{\sin{2\theta_{\rm m_d}}} \left[\dfrac{-3e^2_1}{2\phi^3} + \dfrac{3}{2\phi}\right] \right.\\ 
     &\left.- k_3\sign(e_3)\right),\label{eq:V1dotthetaMd}
 \end{align}
 for $|e_1| \leq \phi$. Furthermore, once $e_3$ and $e_4$ converge to zero, we have $\theta_{\rm m} = \theta_{\rm m_d}$ and $\psi_{\rm m} = \psi_{\rm m_d}$. Without loss of generality, we assume $\theta_{\rm m} = \psi_{\rm m} = \theta_{\rm m_d} = \psi_{\rm m_d}$. Substituting this into the expressions in \eqref{eq:V1dotpsiMd} and \eqref{eq:V1dotthetaMd}, we obtain
  \begin{align}
    \dot{\bar{\mathcal{V}}}_1 &= \psi_{\rm m_d}\left(\dfrac{k_1(-V_{\rm m}+V_{\rm m} \cos^2{\psi_{\rm m_d}})}{\sin{2\psi_{\rm m_d}}} \left[\dfrac{-3e^2_1}{2\phi^3} + \dfrac{3}{2\phi}\right]\right),\label{eq:V1dotpsiMd1}\\
 \dot{\bar{\mathcal{V}}}_2 &= \theta_{\rm m_d}\left(\dfrac{k_1(-V_{\rm m}+V_{\rm m} \cos^2{\theta_{\rm m_d}})}{\sin{2\theta_{\rm m_d}}} \left[\dfrac{-3e^2_1}{2\phi^3} + \dfrac{3}{2\phi}\right]\right).\label{eq:V1dotthetaMd1}
 \end{align}
 For the case when $|e_1| \leq \phi$, we have $\dot{\bar{\mathcal{V}}}_1 \leq 0$ and $\dot{\bar{\mathcal{V}}}_2 \leq 0$. This follows from the fact that the terms $\dfrac{\theta_{\rm m_d}}{\sin{2\theta_{\rm m_d}}}$ and $\dfrac{\psi_{\rm m_d}}{\sin{2\psi_{\rm m_d}}}$ are positive for $\theta_{\rm m_d}, \psi_{\rm m_d} \in \left[-\dfrac{n\pi}{2}, \dfrac{n\pi}{2}\right]\setminus{\{0\}}$, and the inequality 
$-V_{\rm m} + V_{\rm m}\cos^2{D} \leq 0$ holds for $D = \theta_{\rm m_d}, \psi_{\rm m_d}$. This further implies that both the terms $\psi_{\rm m_d}$ and $\theta_{\rm m_d}$ are asymptoticaly convergent. Additionaly, once $e_3 = e_4 = 0$, $\psi_{\rm m}$ and $\theta_{\rm m}$ will tend to zero, which results in  $\sigma \rightarrow 0$.  Using the expressions of $a_{\rm y}$ and $a_{\rm z}$ from \eqref{eq:head_angle_accel}, along with the dynamics of $\dot{\psi}$ and $\dot{\theta}$, it can be readily shown that $\lim_{\{\theta_{\rm m},\psi_{\rm m}\} \to 0} a_{\rm y} = 0$ and $\lim_{\{\theta_{\rm m},\psi_{\rm m}\} \to 0} a_{\rm z} = 0$ hold once $e_3 = e_4 = 0$.
\end{proof}

\subsection{Analysis of the Guidance Law}
\label{Sec:Analysis}
This subsection presents further analysis of the guidance law proposed in Theorem \ref{thm:thm_2} and the desired heading angles in \eqref{eq:bk_step_psi_theta_m}. Firstly, the asymptotic convergence of the range error is examined. Subsequently, an expression for the convergence time of the range error is derived. Next, the range of achievable impact time for the proposed guidance law is investigated. Finally, expressions for the limiting value of $k_1$ based on the FOV bound are obtained. At any instant during the engagement, the minimum time-to-go corresponds to the collision course and is given by $r/V_{\rm m}$. Hence, $t_{\rm go}\ge r/V_{\rm m}$ must hold throughout the engagement. Thus, a negative range error indicates that the corresponding impact time constraint cannot be satisfied. The below lemma gives the sufficient condition for the range error, $e_1$, to remain non-negative throughout the engagement.
\begin{lemma}
    \label{lem:lemma_2}
    ~\citep{H_J_Kim} Let $\eta$ be the maximum allowable range error at interception. Then for all initial conditions satisfying the relation, $|e_1(0)-\eta|/(2V_{\rm m})>t_{e_2=0}$, the error variable $e_1(t)$ is always larger than or equal to zero, that is, $e_1(t)\ge0~\forall~t\ge0$.
\end{lemma}

\begin{remark}
 From here, one can make an interesting observation that the condition for range error staying non-negative is related to the time taken for the lead angle error to vanish `$t_{e_{2}=0}$'. The initial departure from the collision course is longer for larger impact times, and hence, the time taken for the lead angle error to vanish is higher. Thus, ultimately, the condition for range error to stay non-negative depends on the choice of impact time and the constant $k_2$.   
\end{remark}

\subsubsection{Convergence of Range Error} 
To prove the convergence of the range error $e_1$, consider the interval after which both errors $e_3$ and $e_4$ have converged to zero, which further implies $\psi_{\rm m} = \psi_{\rm m_{d}}$ and $\theta_{\rm m} = \theta_{\rm m_{d}}$. From this and the relation in \eqref{eq:sigma_psi_m_theta_m},  it follows that $e_2=0$, as $\sigma$ = $\sigma_{\rm d}$.  
The asymptotic convergence of the range error is presented in the following theorem.
\begin{theorem}
\label{thm:thm_3}
Consider the heading angle relations ($\theta_{\rm m_d}$ and $\psi_{\rm m_d}$) in \eqref{eq:bk_step_psi_theta_m}. Once the errors $e_3$ and $e_4$ vanish, the range error $e_1$ asymptotically converges to zero.
\end{theorem}
\begin{proof} \label{proof_3}
After the heading errors have vanished, we have the following equality: 
\begin{align*}
    e_3 = e_4 &= 0 \implies \theta_{\rm m} = \theta_{\rm m_d} ,~~ \psi_{\rm m} = \psi_{\rm m_d}.
\end{align*}
Hence, by substituting \eqref{eq:bk_step_psi_theta_m} into \eqref{eq:sigma_psi_m_theta_m}, the effective lead angle can be obtained to be the same as in \eqref{sigma_d_kim}. Now, substituting \eqref{sigma_d_kim} in \eqref{eq:e_1_dot} yields
\begin{equation}\label{eq:e1_dot_pf3}
    \dot{e}_1 = -k_1V_{\rm m}{\sgmf}(e_1).
\end{equation}
Consider a Lyapunov function candidate $\mathcal{W} = 0.5e_1^2$, the time derivative of which is given by $\dot{\mathcal{W}}=e_1\dot{e}_1$. On substituting for $\dot{e}_1$ from \eqref{eq:e1_dot_pf3}, we get $\dot{\mathcal{W}}= -k_1V_{\rm m}e_1{\sgmf}(e_1)$. Using the expression for ${\sgmf}(\cdot)$ from \eqref{sgmf}, one may obtain
\begin{align*}
\dot{\mathcal{W}} &=  \begin{cases}
        -k_1V_{\rm m}\left[-\dfrac{e_1^4}{2\phi^3}+\dfrac{3e_1^2}{2\phi}\right]& |e_1|\le\phi\\
        -k_1V_{\rm m}|e_1|& \mathrm{else}
      \end{cases}.
\end{align*}
It is evident that when error $|e_1|>\phi$, then $\dot{\mathcal{W}}<0$. For the case when $|e_1|\le\phi$,  we define $|e_1|=\alpha\phi$, where $\alpha\in[0,1]$ since $\phi>0$. It is evident that $e_1=0$ $\iff$ $\alpha=0$. On substitution $e_1 = \alpha\phi$, the expression of $\dot{\mathcal{W}}$ becomes
\begin{align}
    \dot{\mathcal{W}} = & -k_1\phi\left(-\frac{\alpha^4}{2}+\frac{3\alpha^2}{2}\right), \mathrm{if}~~|e_1|\le\phi,
\end{align}
For a non-zero $\alpha$, $\dot{\mathcal{W}}$ is negative if and only if 
\begin{align}
\left(-\frac{\alpha^4}{2}+\frac{3\alpha^2}{2}\right)>0 \implies \alpha^2\left(-\frac{\alpha^2}{2}+\frac{3}{2}\right)>0.\label{eq:e1alphaRange}
\end{align}
Furthermore, since $\alpha^2>0$ for any $\alpha\in\mathbb{R}$, therefore from \eqref{eq:e1alphaRange}, we get $ -\sqrt{3}<\alpha<\sqrt{3}$.
This condition on $\alpha$ is readily satisfied as $\alpha\in[0,1]$. Thus, the rate of Lyapunov function candidate $\mathcal{W}$ is negative definite, and hence by \Cref{lem:lyapunov}, the error $e_1$ converges to zero asymptotically. This completes the proof.
\end{proof} 
From \eqref{eq:e1_dot_pf3}, it is evident that the range error converges to zero at a rate of $k_1$. However, any value of $k_1$ will not guarantee the interception. To ensure that the proposed guidance command is well-defined, we present the following proposition to provide a range of the gain $k_1$.
\begin{proposition}
  Consider \Cref{lem:lemma_2} and the fact that $\cos{\sigma_{\rm d}}$ must remain in the interval $[-1,1]$, it follows that the gain $k_1$ will lie in the range $\left[0, 2 \right]$.    
\end{proposition}
\begin{proof}    
The range error rate is given by  
\begin{equation*}
    \dot{e}_1 = -V_{\rm m} + V_{\rm m} \cos{\sigma_{\rm d}}.
\end{equation*}
Furthermore, as $\cos{\sigma_{\rm d}} \in \left[-1, 1\right]$ and maximum value of $\sgmf(e_1)$ is $1$, therefore
\begin{align*}
\sigma_{\rm d} &= \cos^{-1}[1-k_1{\sgmf}(e_1)]\\
\implies&
\cos{\sigma_{\rm d}} = 1-k_1{\sgmf}(e_1)\\
\implies& -1 \le 1-k_1{\sgmf}(e_1) \le 1\\
\implies& 0\le k_1{\sgmf}(e_1) \le 2.
\end{align*}
\end{proof}
In the backstepping approach taken, the range error converges after the heading errors have vanished.
\begin{remark}\label{rem:conv_t_e2_e3_e4}
One may observe from \eqref{eq:e3_e4_desired_rates} that the magnitudes of errors $e_3$ and $e_4$ decrease at a rate of $k_3$ and $k_4$, respectively. Therefore, the time taken for the heading errors to vanish is finite and is given by $t_{e_j=0} = {|e_j(0)|}/{k_j}~\forall~j\in\{3,4\}$. For a proper choice of $k_3$ and $k_4$, the heading errors converge much earlier than the range error. 
\end{remark}
In addition to the asymptotic convergence that was proven earlier in this section, it is necessary that the errors converge to a value as low as possible during target interception. The lethal radius of the interceptor provides the maximum permissible range error at the time of interception. This upper bound on the range error can be translated into an upper bound on the convergence time of the errors, as stated in the following theorem.
\begin{theorem}
\label{thm:thm_4}
For the guidance law  and the desired heading angles given by \Cref{thm:thm_2} and \eqref{eq:bk_step_psi_theta_m}, respectively, the range error, $e_1(t)$, converges to a finite interval $[-\eta, \eta]$, within a finite time $t_{\rm c}$, where $\eta\in(0,\phi)$, with the upper bound on convergence time given as
\begin{align}
   t_{\rm c} \le&~ \mathrm{max}\bigg(\frac{|e_3(0)|}{k_3}, \frac{|e_4(0)|}{k_4}\bigg) + \frac{V_{\rm m}t_{\rm f}-r(0)-\phi}{k_1V_{\rm m}}\nonumber \\
   &+ \frac{\phi}{3k_1V_{\rm m}}\log\bigg(\frac{3\phi^2-\eta^2}{2\eta^2}\bigg).\label{eq:tc1_case_b_e}
\end{align}
\end{theorem}
\begin{proof}
Consider the engagement scenario at $t=0$, when all the errors have a non-zero finite value. By virtue of the guidance law in \eqref{eq:head_angle_accel}, the magnitude of the heading angle errors, $e_3$ and $e_4$, linearly converge to zero. Let $t_{e_3=0}$ and $t_{e_4=0}$ be the time taken for the errors $e_3$ and $e_4$ to vanish as noted in \Cref{rem:conv_t_e2_e3_e4}. Once $e_3$ and $e_4$ have vanished, the effective lead angle tracks their desired value, $\sigma_{\rm d}$, ensuring $e_2 = 0$. However, $e_1$ remains positive. The worst case occurs when the magnitude of range error is greater than $\phi$ at the start of the engagement and persists to be greater than $\phi$ even after $e_3$ and $e_4$ have converged to zero, where $\phi$ is a small positive constant in \eqref{sgmf}. From  \eqref{eq:e1_dot_pf3}, it is evident that the range error converges linearly for $e_1>\phi$ and asymptotically for $e_1<\phi$. Therefore, the time interval of engagement $[0,t_{e_1=\eta}]$ is divided into three regimes as follows:
\begin{align*}
    \begin{cases}
        e_1>\phi, e_3\neq0 \ \mathrm{or}\ e_4\neq0 & \mathrm{for}~0\le t<t_{e_2=0},\\
        e_1>\phi, e_3=0 \ \mathrm{and}\ e_4=0 & \mathrm{for}~t_{e_2=0}\le t<t_{e_1=\phi},\\
        e_1\le\phi, e_3=0 \ \mathrm{and}\ e_4=0 & \mathrm{for}~t_{e_1=\phi}\le t\le t_{e_1=\eta},
    \end{cases}
\end{align*}
where $t_{e_1=\eta}$ is the time taken for the range error to decrease to small magnitude, $\eta$, such that $0<\eta<<\phi$. In the first regime, $t\in[0,t_{e_2=0})$, the heading errors, $e_3$ and $e_4$ reduce from their initial values of $e_3(0)$ and $e_4(0)$ to zero. However, the lead angle error, $e_2$, is zero if and only if both $e_3$ and $e_4$ are zero. This further implies
\begin{equation}\label{eq:t_e2=0_eq1}
    t_{e_2=0} = \mathrm{max}\bigg(t_{e_3=0}, t_{e_4=0}\bigg).
\end{equation}
Using \Cref{rem:conv_t_e2_e3_e4}, one may modify \eqref{eq:t_e2=0_eq1} to get
\begin{equation}\label{eq:t_e2=0_e3=e4=0}
    t_{e_2=0} = \mathrm{max}\bigg(\frac{|e_3(0)|}{k_3}, \frac{|e_4(0)|}{k_4}\bigg).
\end{equation}
Now that the heading errors have converged, the lead angle follows the virtual input, $\sigma_{\rm d}$. The additional time taken for the range error to converge to $\phi$ can be calculated in the second regime, $t\in[t_{e_2=0},t_{e_1=\phi})$. In this interval, the $\sgmf(\cdot)$ function \eqref{sgmf} behaves as the signum function, and from \eqref{eq:e1_dot_pf3} the expression for the range rate is given by
\begin{equation*}
    \dot{e}_1 = -k_1V_{\rm m},
\end{equation*}
which on integrating within the time interval $[t_{e_2=0},t_{e_1=\phi}]$ yields
\begin{align}
    \nonumber e_1(t_{e_2=0})-\phi &= -k_1V_{\rm m}(t_{e_2=0} - t_{e_1=\phi}),\\
    \implies t_{e_1=\phi} - t_{e_2=0} &= \frac{e_1(t_{e_2=0})-\phi}{k_1V_{\rm m}}.\label{eq:thm4_t_phi_0_1}
\end{align}
As established in \Cref{thm:thm_3}, the range error is asymptotically convergent and hence monotonically decreasing. Consequently substituting the relation, $e_1(t_{e_2=0}) < e_1(0)$ into \eqref{eq:thm4_t_phi_0_1} results in
\begin{align}
 t_{e_1=\phi} &- t_{e_2=0} \le \frac{e_1(0)-\phi}{k_1V_{\rm m}}\implies 
    t_{e_1=\phi} - t_{e_2=0} \nonumber \\ 
    &\le \frac{V_{\rm m}t_{\rm f}-r(0)-\phi}{k_1V_{\rm m}}. \label{eq:thm4_t_phi_0_2}
\end{align}
Note that the equality in \eqref{eq:thm4_t_phi_0_2} occurs when the range error remained constant in the first regime $t\in[0,t_{e_2=0})$. For the third regime, $t\in[t_{e_1=\phi},t_{e_1=\eta}]$. Once the range error converges to $\phi$, its dynamics can be obtained from \eqref{eq:e1_dot_pf3} and \eqref{sgmf}, and is given by
\begin{align}
\nonumber \dot{e}_1 &= -k_1V_{\rm m}\left[-\dfrac{e^3_1}{2\phi^3} + \dfrac{3e_1}{2\phi}\right]\\
 &\implies \dfrac{de_1}{\left[\dfrac{e^3_1}{2\phi^3} - \dfrac{3e_1}{2\phi}\right]} 
 = k_1V_{\rm m}{dt}.\label{eq:kv1dt}
\end{align}
Integrating \eqref{eq:kv1dt} in the interval $[t_{e_1=\phi},t_{e_1=\eta}]$, lead us to arrive at
\begin{equation}
    t_{e_1=\eta}-t_{e_1=\phi} = \frac{\phi}{3k_1V_{\rm m}}\log\bigg(\frac{3\phi^2-\eta^2}{2\eta^2}\bigg). \label{eq:thm4_t_eta_phi}
\end{equation}
Therefore, the total time taken for the range error to converge to $\eta$ is obtained by taking the sum of \eqref{eq:t_e2=0_e3=e4=0}, \eqref{eq:thm4_t_phi_0_2}, and \eqref{eq:thm4_t_eta_phi}, and is given by
\begin{align}
    t_{\rm c} &\le \mathrm{max}\bigg(\dfrac{|e_3(0)|}{k_3}, \frac{|e_4(0)|}{k_4}\bigg) + \frac{V_{\rm m}t_{\rm f}-r(0)-\phi}{k_1V_{\rm m}}\nonumber \\
    &+ \dfrac{\phi}{3k_1V_{\rm m}}\log\bigg(\frac{3\phi^2-\eta^2}{2\eta^2}\bigg).\label{tc1 case a in pf}
\end{align}
This completes the proof.
\end{proof} 
\subsubsection{Lead Angle Analysis}
In this subsection, we analyze whether the effective velocity lead angle under the proposed control law remains within the prescribed field-of-view (FOV) limit.

To that end, let us assume $e_3 = e_4 = 0$, then $\theta_{\rm m} = \theta_{\rm{m_d}}$ and $\psi_{\rm m} = \psi_{\rm{m_d}}$. Consider the case when, $\theta_{\rm{m_d}} = \psi_{\rm{m_d}}$. Under this condition, the relation 
$\cos{\sigma} = \cos{\psi_{\rm{m_d}}}\cos{\theta_{\rm{m_d}}} $
simplifies to $ \cos{\sigma} = \left({\cos^{2}{\psi_{\rm{m_d}}}}\right)$.
Using the relation in \eqref{eq:prepositioneqThetamd}, one may obtain
$$ \sigma = \cos^{-1}\left[1 - k_1\sgmf\left(e_1\right)\right] = \sigma_{\rm d}.$$ According to the result in \cref{col:corollary_SigmaAcc}, as $\sigma \rightarrow 0$, it follows that, $\sigma_{\rm{d}}\rightarrow 0$. Therefore, as long as $\sigma_{\rm{d}} \leq \sigma_{\max}$, the effective velocity lead angle satisfies $\sigma \leq \sigma_{\max}$, and the FOV constraint is maintained.
\begin{proposition}
If the gain $k_1$ satisfies the relation  
\begin{equation}\label{eq:FOV_Bound}
    0\leq k_1 \leq 1 - \cos{\sigma_{\max}} - \epsilon,
\end{equation}
where $\epsilon$ is the positive constant smaller than $1 - \cos{\sigma_{\max}}$, then $\sigma_{\rm d} \leq \sigma_{\max}$ for all $t \geq 0$, and the FOV constraint is maintained.
\end{proposition}
\begin{proof}
  Since $e_1$ is positive, as shown in \cref{lem:lemma_2}, and using the expression for the desired velocity lead angle, $\sigma_{\rm d}$ as in \eqref{sigma_d_kim}, one may get
  \begin{subequations}\label{eq:FOV_Bound1}
    \begin{align}
    \cos{\sigma_{\rm d}} &= 1 - k_1\sgmf\left(e_1\right) \leq 1\\
    \cos{\sigma_{\rm d}} &= 1 - k_1\sgmf\left(e_1\right) \geq 1 - k_1,  
  \end{align}
  \end{subequations}
where $0\leq\sgmf\left(e_1\right)\leq 1$ for $e_1> 0$. 
Using \eqref{eq:FOV_Bound} and \eqref{eq:FOV_Bound1}, we get
\begin{equation}
    \cos{\sigma_{\max}} \leq \epsilon \leq \cos{\sigma_{\rm{d}}} \leq 1,
\end{equation}
which further leads to $\cos^{-1}{1} \leq |\sigma_{\rm{d}}\leq \cos^{-1}\left(\cos{\sigma_{\rm{d}}} _ \epsilon\right)$. Since $\epsilon$ is chosen to satisfy the relation $0 \leq \epsilon + \cos{\sigma_{\max}} < 1$, therefore,
\begin{align*}
    0 \leq |\sigma_{\rm d}| &\leq \cos^{-1}\left(\cos{\sigma_{\max} + \epsilon}\right),\\
                            &< \cos^{-1}\left(\cos{\sigma_{\max}}\right) = \sigma_{\max}.
\end{align*}
Hence, one may conclude that $\sigma_{\rm d} \leq \sigma_{\max}$. This completes the proof.
\end{proof}
\subsubsection{Achievable Impact Time}
\label{Subsec:achieve_tf}
For an interceptor moving at a constant speed, the lowest achievable impact time is the one corresponding to the collision course. The lowest possible closing speed is $V_{\rm m}\cos{\sigma_{\rm max}}$, where $\sigma_{\rm max}$ is the lead angle bound. These two conditions together provide lower and upper bounds on achievable impact time for engagement geometry, which may be obtained as
\begin{equation}\label{tf_Ness}
    \frac{r(0)}{V_{\rm m}} \le t_{\rm f} \le \frac{r(0)}{V_{\rm m} \cos{\sigma_{\rm max}}}.
\end{equation}
The condition in \eqref{tf_Ness} is necessary for target interception irrespective of the guidance law, while obeying FOV constraints. To derive a sufficient condition, the guidance law must be taken into account. In the remainder of this subsection, we derive the minimum and maximum achievable impact time.

The lower bound on the impact time, $r(0)/V_{\rm m}$, is equivalent to the condition $e_1(t)\ge0$ by the definition of $e_1$. Thus, a sufficient condition corresponding to the minimum achievable impact time is given by \Cref{lem:lemma_2}. Accordingly, for the lower bound of the impact time, $t_{\rm f}$, we obtain
\begin{align}\label{VmR0Eta}
    \frac{|e_1(0)-\eta|}{2V_{\rm m}}>t_{e_2=0} 
    \implies \frac{|V_{\rm m}t_{\rm f}-r(0)-\eta|}{2V_{\rm m}}>t_{e_2=0}.
\end{align}
Substituting the value of $t_{e_2=0}$ from \eqref{eq:t_e2=0_e3=e4=0} in \eqref{VmR0Eta}, results in 
\begin{align} 
    \nonumber \frac{|V_{\rm m}t_{\rm f}-r(0)-\eta|}{2V_{\rm m}}>\mathrm{max}\bigg(\frac{|e_3(0)|}{k_3}, \frac{|e_4(0)|}{k_4}\bigg)\\
    \implies t_{\rm f}>\frac{r(0)+\eta}{V_{\rm m}}+2\mathrm{max}\bigg(\frac{|e_3(0)|}{k_3}, \frac{|e_4(0)|}{k_4}\bigg)\label{eq:td_min_bce}.
\end{align}
It is important to note that the time of convergence has to be lower than the desired impact time for a successful interception. Therefore, the constraint, $t_{\rm c} < t_{\rm f}$ together with \Cref{thm:thm_4}, implies 
\begin{align}
    t_{\rm c} &\le \mathrm{max}\bigg(\dfrac{|e_3(0)|}{k_3}, \frac{|e_4(0)|}{k_4}\bigg) + \frac{V_{\rm m}t_{\rm f}-r(0)-\phi}{k_1V_{\rm m}}\nonumber \\
    &+ \dfrac{\phi}{3k_1V_{\rm m}}\log\bigg(\frac{3\phi^2-\eta^2}{2\eta^2}\bigg) < t_{\rm f},
\end{align}
which may be further rewritten as
\begin{align}
 \nonumber  t_{\rm f} < &~ \frac{k_1}{1-k_1}\bigg[ \frac{r(0)+\phi}{k_1V_{\rm m}} - \mathrm{max}\bigg(\frac{|e_3(0)|}{k_3}, \frac{|e_4(0)|}{k_4}\bigg) \\
   &- \frac{\phi}{3k_1V_{\rm m}}\log\bigg(\frac{3\phi^2-\eta^2}{2\eta^2}\bigg)\bigg].\label{eq:td_max_case_e}
\end{align}
The expression in \eqref{eq:td_max_case_e} provides the upper bound on the achievable impact time. Consequently, the set pertaining to achievable impact times is given by 
\begin{align}
    \nonumber t_{\rm f} \in&~ \left(\dfrac{r(0)+\eta}{V_{\rm m}}+\mathrm{max}\left(\dfrac{|e_3(0)|}{k_3},
\dfrac{|e_4(0)|}{k_4}\right), \right.\\
&~\left.\dfrac{k_1}{1-k_1}\left[\dfrac{r(0)+\phi}{k_1V_{\rm m}} - \mathrm{max}\left(\dfrac{|e_3(0)|}{k_3}, \frac{|e_4(0)|}{k_4}\right) \nonumber \right.\right.\\
&\left.\left.- \dfrac{\phi}{3k_1V_{\rm m}}\log\left(\dfrac{3\phi^2-\eta^2}{2\eta^2}\right)\right]\right). \label{eq:tf_min_max}
\end{align} 
Now that the achievable range of impact time has been derived, it is necessary to ensure that the lead angle remains within the upper bound, $\sigma_{\rm max}$ throughout the engagement.
\begin{remark}\label{rem:rem_2}
Substituting the lead angle bound, $\sigma_{\max}$, into the $\sigma_{\rm d}$ given in \eqref{sigma_d_kim}, the upper bound on the value of $k_1$ can be obtained as follows
\begin{align}
    k_1 < 1 - \cos{\sigma_{\max}}.\label{eq:case_e_k1_FOV_limit}
\end{align}
\end{remark}
Any value of $k_1$ satisfying \eqref{eq:case_e_k1_FOV_limit} ensures that the lead angle remains within the permissible FOV range. Moreover, once the constant $k_1$ is selected in accordance with \eqref{eq:case_e_k1_FOV_limit}, the maximum lead angle that can be attained during the engagement is $\sigma = \cos^{-1}(1-k_1)$. 

\subsection{Choice of Virtual Inputs}
\label{Subsec:Vir_Inp}
This subsection summarizes a few possible choices of virtual inputs, including the one already discussed in \Cref{SubSec:Backstep_Heading_Angles}. To meet the collision conditions, the desired heading angles ($\theta_{\rm m_d}$ and $\psi_{\rm m_d}$) can be chosen appropriately, and the interceptor lateral acceleration commands stated in Theorem \ref{thm:thm_2} would guide the interceptor to track the desired value. As there is no constraint on the impact angle, infinitely many possible choices exist for the virtual inputs. The different choices are made such that the desired lead angle is similar to \eqref{sigma_d_kim}, and they are tabulated in \Cref{tab:Vir_Inp}. Here, the function $f(e_1)$ is the same as \eqref{sigma_d_kim}.
\begin{table*}[!ht]
    \caption{Choice of Virtual Inputs}
    \label{tab:Vir_Inp}
    \centering
    \begin{tabular}{cccc}
    \hline
    \hline
    Case & $\theta_{\rm m_d}$ & $\psi_{\rm m_d}$ & $\dot{e}_1$ at $\sigma = \sigma_{\rm d}$\\
    \hline 
    (a) & 0 & $f(e_1)$ & $-k_1V_{\rm m}{\sgmf}(e_1)$\\
    (b) & $f(e_1)/2$ & $f(e_1)/2$ & $-\dfrac{k_1V_{\rm m}}{2}{\sgmf}(e_1)$\\
    (c) & $f(e_1)$ & $f(e_1)$ & $k_1^2V_{\rm m}[{\sgmf}(e_1)]^2 -2 k_1V_{\rm m}{\sgmf}(e_1)$\\
    (d) & $\dfrac{\cos^{-1}[2\cos(f(e_1))-1]}{2}$ & $\dfrac{\cos^{-1}[2\cos(f(e_1))-1]}{2}$ & $-k_1V_{\rm m}{\sgmf}(e_1)$\\[1.8pt]
    \hline
    \hline
    \end{tabular}
\end{table*}

The dynamics of range error after the heading errors have converged are also noted in \Cref{tab:Vir_Inp}. The virtual inputs of Cases (d) are the same as the ones introduced in \Cref{SubSec:Backstep_Heading_Angles}. The main idea of the guidance scheme is to maintain a non-zero lead angle until the impact time constraint is met and then converge into the collision course as the range error converges to zero. Therefore, the desired heading angles can be chosen to be a scalar multiple of \eqref{sigma_d_kim} as in the cases (b) and (c). The lead angle bound is taken into account by a suitable selection of parameter $k_1$. The permissible bound of $k_1$ can be obtained analogous to \eqref{eq:case_e_k1_FOV_limit}.
\begin{align}
    k_1 < \begin{cases}
        1 - \cos{\sigma_{\max}} & ,\text{Cases (a), (d)}\\
        2\,[1 - \cos{\sigma_{\max}}] & ,\text{Case (b)}\\
        1 - \sqrt{\cos{\sigma_{\max}}} & ,\text{Case (c)}
    \end{cases}
\label{eq:k1_FOV_limit}
\end{align}
The range of achievable impact time for the different choices of virtual inputs are also derived analogous to \eqref{eq:td_max_case_e} and listed below. For Cases (a) and (d), we have 
\begin{subequations}
\begin{align}
    t_{\rm f}&>\frac{r(0)+\eta}{V_{\rm m}}+2\mathrm{max}\bigg(\frac{|e_3(0)|}{k_3}, \frac{|e_4(0)|}{k_4}\bigg)\label{eq:td_min_bcde}\\
    t_{\rm f} & < \frac{k_1}{1-k_1}\bigg[ \frac{r(0)+\phi}{k_1V_{\rm m}} - \mathrm{max}\bigg(\frac{|e_3(0)|}{k_3}, \frac{|e_4(0)|}{k_4}\bigg)\nonumber \\
    &- \frac{\phi}{3k_1V_{\rm m}}\log\bigg(\frac{3\phi^2-\eta^2}{2\eta^2}\bigg)\bigg]\label{eq:td_max_case_be}
\end{align}
\end{subequations}
Equation \eqref{eq:td_min_bcde} forms the lower bound on achievable impact time for Cases (b) and (c) as well. However, the upper bound on achievable impact time for Case (b) is obtained by replacing $k_1$ by $k_1/2$ in \eqref{eq:td_max_case_be}. The derivation of the analytical expression for convergence time being cumbersome for Case (c). The maximum achievable impact time can be obtained by numerically integrating the last term in \eqref{eq:td_max_case_d} as shown below.
\begin{align}
   t_{\rm f} < &~ \frac{k_1}{1-k_1}\bigg[ \frac{r(0)+\phi}{k_1V_{\rm m}} - \mathrm{max}\bigg(\frac{|e_3(0)|}{k_3}, \frac{|e_4(0)|}{k_4}\bigg)\nonumber \\
   &- \int_{\phi}^{\eta}\frac{1}{\dot{e}_1}dt\bigg]\label{eq:td_max_case_d}
\end{align}
Here the term $\dot{e}_1$ corresponds to the range rate expression mentioned for Case (c) in \Cref{tab:Vir_Inp}.

\section{Performance Study}
\label{Sec:Performance_Study}
In this section, we validate the performance of the proposed guidance schemes designed through backstepping using effective lead angle and heading angles via numerical simulations. The first subsection illustrates the coupling effect arising while backstepping using the effective lead angle, for the lateral acceleration commands given in \eqref{eq:a_accel}. The later subsections present simulations and inferences for the lateral acceleration commands in \eqref{eq:head_angle_accel}. 
Note that the function $\cos^{-1}(\cdot)$ in \eqref{eq:bk_step_psi_theta_m} may create an ambiguity in the sign of $\theta_{\rm m_d}$ and $\psi_{\rm m_d}$. However, the same can be exercised to maintain the trajectory of the interceptor within one of the octants. This will be demonstrated in the second subsection. Note that the octants are designated in the initial LOS frame. Simulations are performed for different impact times and initial headings to validate the guidance law and for comparative study.  Furthermore, the derived guidance law is extended for constant velocity targets. The interceptors are assumed to have a constant speed of $250\,$m/s. As the maximum acceleration rate in thrusters and control surface deflections are bounded, the autopilot is assumed to be a first-order system \cite{Nanavati-IEEETAES-57-5-3357-2021}, given by
\begin{equation}
    \frac{a_{\rm z}}{a_{\rm z}^{\rm c}}=\frac{1}{\tau s + 1},~~~~\frac{a_{\rm y}}{a_{\rm y}^{\rm c}}=\frac{1}{\tau s + 1},
    \label{eq:first_order_autopilot}
\end{equation}
where $a_{\rm z}^{\rm c}$ and $a_{\rm y}^{\rm c}$ are the commanded accelerations, and $\tau$ is the time constant of the autopilot, considered as $0.1\,$s here. To alleviate the chattering issue due to presence of the signum function, it is replaced by a continuous sigmoid function $\sgmf_2(x)$ \citep{SRK2014TSMCIACG} with the parameter $a=10$ and is given by
\begin{equation}
   \sgmf_2(x)=2\bigg(\dfrac{1}{1+\exp^{-ax}}-\frac{1}{2}\bigg),~~a>0.
   \label{eq:sgmf_2}
\end{equation}
To respect the physical limitation of the actuators, a bound of $10$ g on the maximum available lateral acceleration in each direction, where $g=9.81$ m/s$^2$ is the acceleration due to gravity, is considered in all simulations.
\renewcommand{\arraystretch}{1.2}
\setlength{\tabcolsep}{2pt}
\begin{table*}[!ht]
\centering
    \caption{Simulation Parameters.}
    \label{tab:Sim_params}
    \begin{tabular}{lcc}
        \hline\hline
        Parameter & Value & Unit\\ \hline
        Interceptor speed ($V_{\rm m}$) & $250$ & $\rm ms^{-1}$\\ 
        Interceptor initial coordinates ($x(0),y(0),z(0)$) & $(0,0,0)$ & $\rm km$\\
        Target coordinates ($x_{\rm T},y_{\rm T},z_{\rm T}$) & $(10,10,0)$ & $\rm km$\\
        $\phi$ & 500 & $-$\\
        $k_1$ & $1-\cos(\sigma_{\rm max})-0.001$ & $-$\\
        $k_2$ & $1$ & $-$\\
        Autopilot time constant ($\tau$) & $1$ & $\rm s$\\
        Acceleration magnitude limit ($a_{\rm max}$) & $10$g & $\rm ms^{-2}$\\
        Seeker FOV constraint ($\sigma_{\rm max}$) & $60$ & $\rm deg$\\ \hline\hline
    \end{tabular}
\end{table*}
Note that for all these simulations, unless stated otherwise, the simulation parameters are the same as in \Cref{tab:Sim_params}. Additionally, the gains $k_3$ and $k_4$ are taken as unity.

\subsection{Performance of Backstepping using Effective Lead Angle}
\label{Subsec:Perf_bk_eff_lead}
\begin{figure*}[!htpb]
\centering
\begin{subfigure}{0.48\linewidth}
\includegraphics[width=\linewidth]{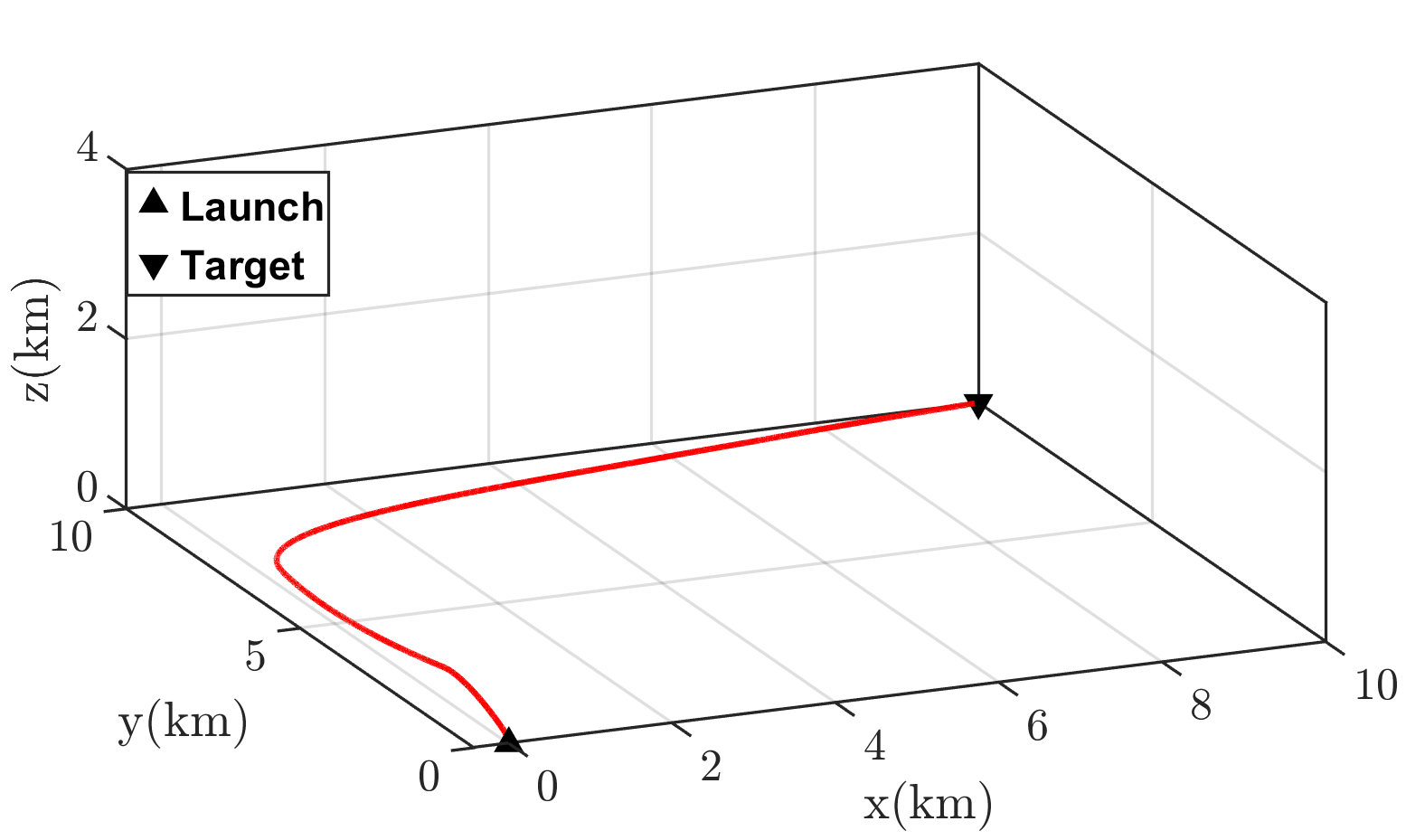}
\caption{Trajectory}\label{fig:one_tf_case_a_xyz}
\end{subfigure}%
\begin{subfigure}{0.48\linewidth}
\includegraphics[width=\linewidth]{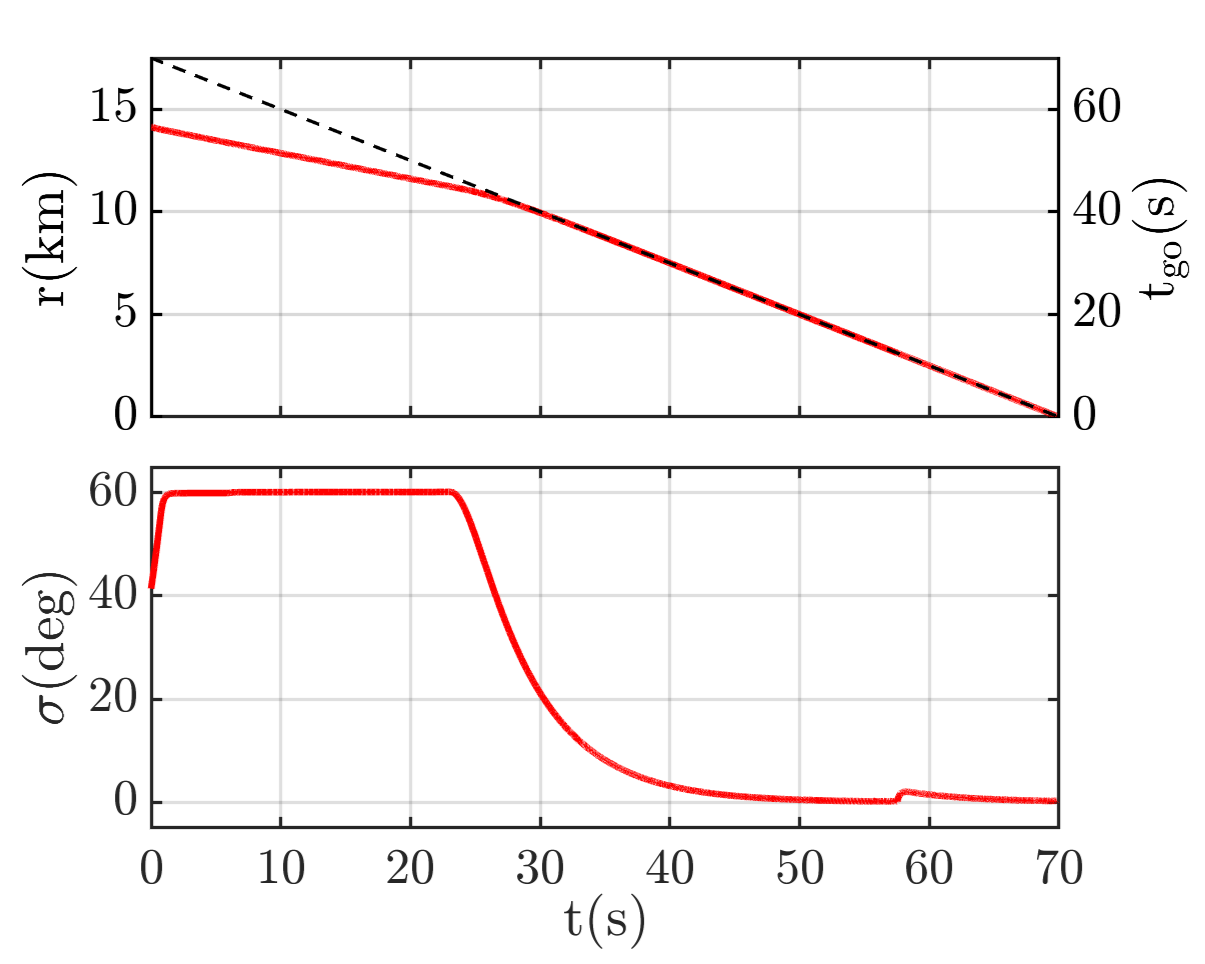}
\caption{Range and Effective Lead Angle}\label{fig:one_tf_case_a_r_sigma}
\end{subfigure}
\begin{subfigure}{0.48\linewidth}
\includegraphics[width=\linewidth]{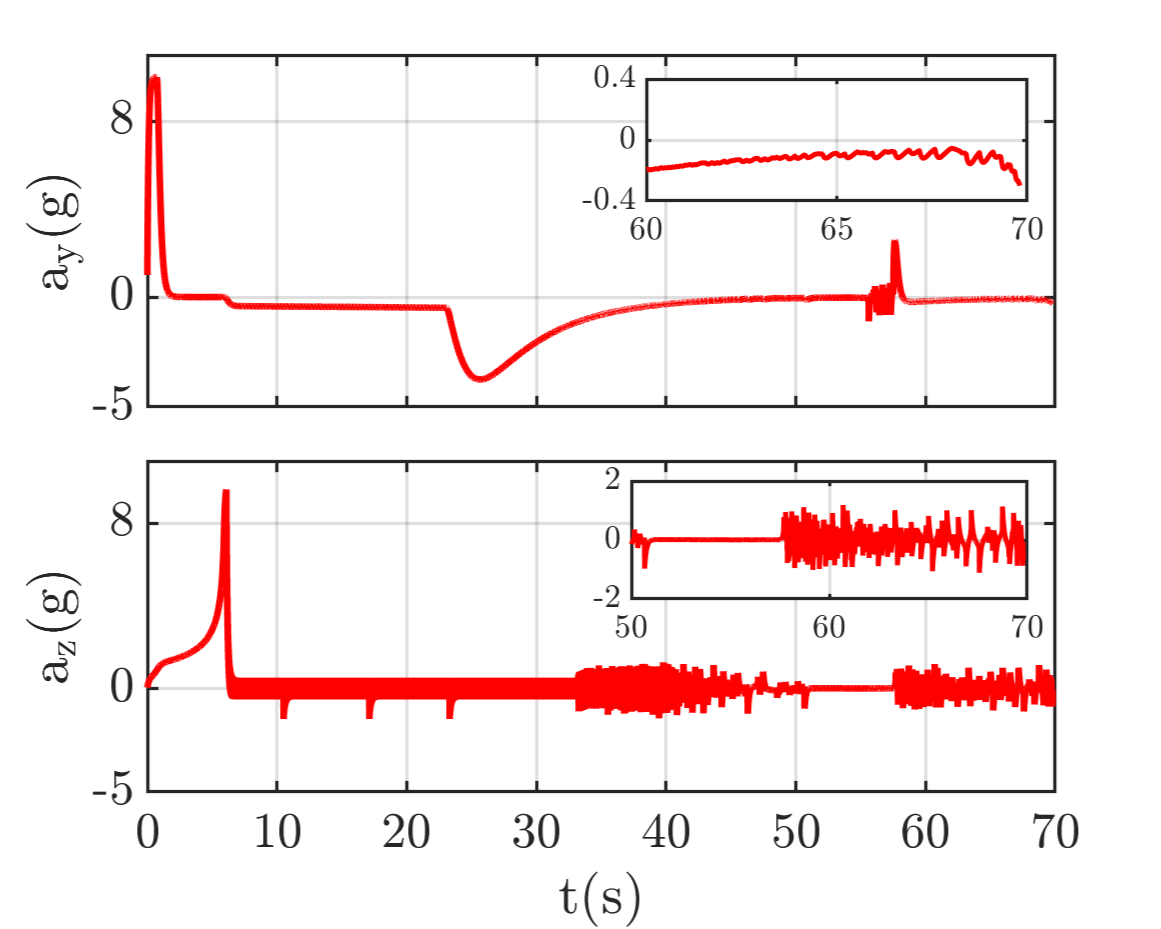}
\caption{Lateral Acceleration}\label{fig:one_tf_case_a_ay_az}
\end{subfigure}%
\begin{subfigure}{0.48\linewidth}
\includegraphics[width=\linewidth]{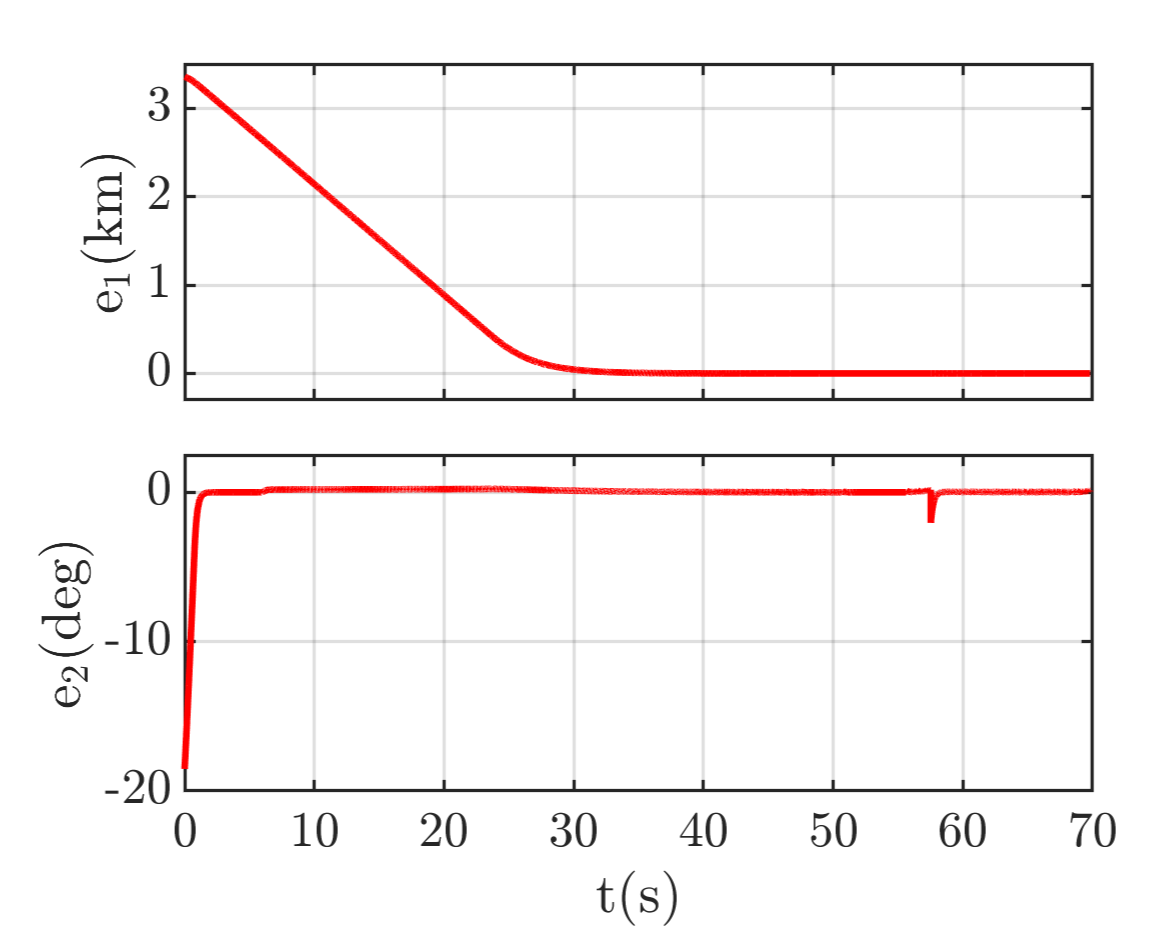}
\caption{Errors $e_1$ and $e_2$}\label{fig:one_tf_case_a_e1_e2}
\end{subfigure}
\caption{Performance for $t_{\rm f}=70\,$s, initial heading angles $\theta_{\rm m}=-30^\circ,\,\psi_{\rm m}=30^\circ$.}
\label{fig:one_tf_case_a}
\end{figure*}

The performance of the guidance law proposed in Theorem \ref{thm:thm_1} is numerically validated in this subsection. 
For this simulation, the initial conditions remain the same as listed in \Cref{tab:Sim_params}, with a desired impact time of 70\,s. The interceptor's states, trajectory, acceleration commands, and errors for this scenario are shown in \Cref{fig:one_tf_case_a}. \Cref{fig:one_tf_case_a_e1_e2} indicates that the lead angle error decreases to zero linearly. However, owing to the replacement of the signum with a sigmoid function \eqref{eq:sgmf_2}, the convergence turns asymptotic as $e_2$ approaches zero. The range error, $e_1$, also shows a similar trend, with its asymptotic behavior being a result of the $\sgmf(\cdot)$ function in \eqref{sigma_d_kim}. 
The dotted black line in \Cref{fig:one_tf_case_a_r_sigma} conveys the variation of the quantity $V_{\rm m}t_{\rm go}$. The vertical separation between the red and black lines indicates the difference between the desired and actual time-to-go values. As the range error reduces, the vertical separation decreases, indicating the time-to-go of the trajectory is approaching its desired value.
The effective lead angle, $\sigma$, is a function of the heading angles ($\psi_{\rm m}$ and $\theta_{\rm m}$). Equation \eqref{eq:a_accel} shows that the lead angle error only affects lateral acceleration $a_y$. However, $a_y$ can only influence the heading angle $\psi_{\rm m}$ directly. Such a choice of acceleration command results in coupling between the two heading angles. As a result, though the target is intercepted successfully, the lateral acceleration command $a_z$ exhibits sharp peaks and oscillations that may not be desirable in practice. 

\subsection{Octant Of Interceptor's Trajectories}
\label{Subsec:Trajec_Oct}
In this subsection, we present the ability to control the octant of the interceptor trajectories during the engagement. Consider the expressions of $\theta_{\rm m_d}$ and $\psi_{\rm m_d}$  in \eqref{eq:bk_step_psi_theta_m}, since the cosine function has the property, $\cos{(-\chi)}=\cos{(\chi)}$, the sign of the heading angles calculated from \eqref{eq:bk_step_psi_theta_m} is ambiguous. However, the signs of $\theta_{\rm m_d}$ and $\psi_{\rm m_d}$ closely connect with the octant in which the interceptor is located in the LOS frame. This relation is shown in \Cref{tab:Octant_signs}. 
\begin{table}[!ht]
\renewcommand{\arraystretch}{1.2}
\setlength{\tabcolsep}{4pt}
\centering
\caption{Relation between LOS frame octants and sign of angles $\theta_{\rm m_d}$ and $\psi_{\rm m_d}$.}
\label{tab:Octant_signs}
\begin{tabular}{lcc}
    \hline
    \hline
    Octant & $\sign(\theta_{\rm m_d})$ & $\sign(\psi_{\rm m_d}$) \\
     \hline
    \Romannum{1} & $-$ & $+$ \\
    \Romannum{4} & $-$ & $-$ \\
    \Romannum{5} & $+$ & $+$ \\
    \Romannum{8} & $+$ & $-$ \\
    \hline
    \hline
\end{tabular}
\end{table}

By choosing the sign of $\theta_{\rm m_d}$ and $\psi_{\rm m_d}$ according to those in \Cref{tab:Octant_signs}, we can easily regulate the octant of the interceptor's trajectory. It is important to note that due to the assumption that the FOV bound is within the interval $(-\pi/2,\pi/2)$, the possible octants in the LOS frame are reduced to only four: octants \Romannum{1}, \Romannum{4}, \Romannum{5} and \Romannum{8}.
\Cref{fig:Trajec_Oct_case_e} demonstrates how the octant of the trajectory changes when the signs of the heading angles ($\theta_{\rm m_d}$ and $\psi_{\rm m_d}$) are varied in accordance with those noted in \Cref{tab:Octant_signs}. The interceptor is assumed to be initially located at the origin, which we also refer to as the launch in the subsequent trajectory plots, and the target is fixed at $(10,0,0)\,$km. The impact time and initial heading angles are set as $(t_{\rm f}, \theta_{\rm m}(0), \psi_{\rm m}(0)) = (60\,\rm s, 30^\circ, 30^\circ)$. The speed of the interceptor is $200$ m/s. The parameter $\phi$ is taken as $200$. It was observed from these plots that the interceptor successfully intercepts the target while adhering to the field-of-view (FOV) constraint. In \Cref{fig:Trajec_Oct_case_e_r_tgo}, the relative separation becomes zero at a time instant of $60\,$s, indicating a successful target interception. Moreover, the lead angle is regulated within the FOV bound as observed from \Cref{fig:Trajec_Oct_case_e_sigma}. \Cref{fig:Trajec_Oct_case_e_xyz} and \Cref{fig:Trajec_Oct_case_e_xyz_x_view} show that the trajectory of the interceptor remained in the expected octant. Adding a negative sign to the original $\theta_{\rm m_d}$ and $\psi_{\rm m_d}$ expressions resulted in an additional peak around $20\,$s in the errors $e_3$ and $e_4$, respectively, as shown in \Cref{fig:Trajec_Oct_case_e_e3_e4}. This peak occurs when the interceptor deviates from its initial trajectory back toward the collision course. The peak in the lateral acceleration commands in \Cref{fig:Trajec_Oct_case_e_ay_az} around $20\,$s is positive when the corresponding virtual input (heading angle) has a negative sign. For instance, the trajectory in $\rm Octant\ \Romannum{1}$ has the following expressions for the virtual inputs
\begin{align*}
    \theta_{\rm m_d} &= -\frac{1}{2}\cos^{-1}\left[2\cos(\sigma_{\rm d})-1\right],~~\\
    \psi_{\rm m_d} &= \frac{1}{2}\cos^{-1}\left[2\cos(\sigma_{\rm d})-1\right],
\end{align*}
where $\sigma_{\rm d}$ is given by $\eqref{sigma_d_kim}$. Based on the expression of $\sigma_{\rm d}$, once the heading errors vanish, a positive $\theta_{\rm m}$ and a negative $\psi_{\rm m}$ are maintained until the range error vanishes. Subsequently, the heading angles are reduced to zero to track the collision course. When the interceptor returns to the collision course, $a_{\rm z}$  initially drops momentarily when the third term in \eqref{eq:head_angle_accel_az} dominates the other terms. Thereafter, the contribution of other terms causes the lateral acceleration to increase. This causes the negative peak in the error $e_3$ around $20\,$s. However, such behavior is not observed in $a_{\rm y}$ owing to the positive sign in $\psi_{\rm m_d}$ expression; hence no peak occurs in error $e_4$ at the same instant. For all the upcoming simulations, the signs of the $\theta_{\rm m_d}$ and $\psi_{\rm m_d}$ expressions are fixed to ensure that the trajectory remains in the first octant of the LOS frame.

\subsection{Performance with Different Impact Times}
\label{Subsec:one_case_diff_t_f}
\begin{figure*}[!htpb]
\centering
\begin{subfigure}{0.32\linewidth}
\includegraphics[width=\linewidth]{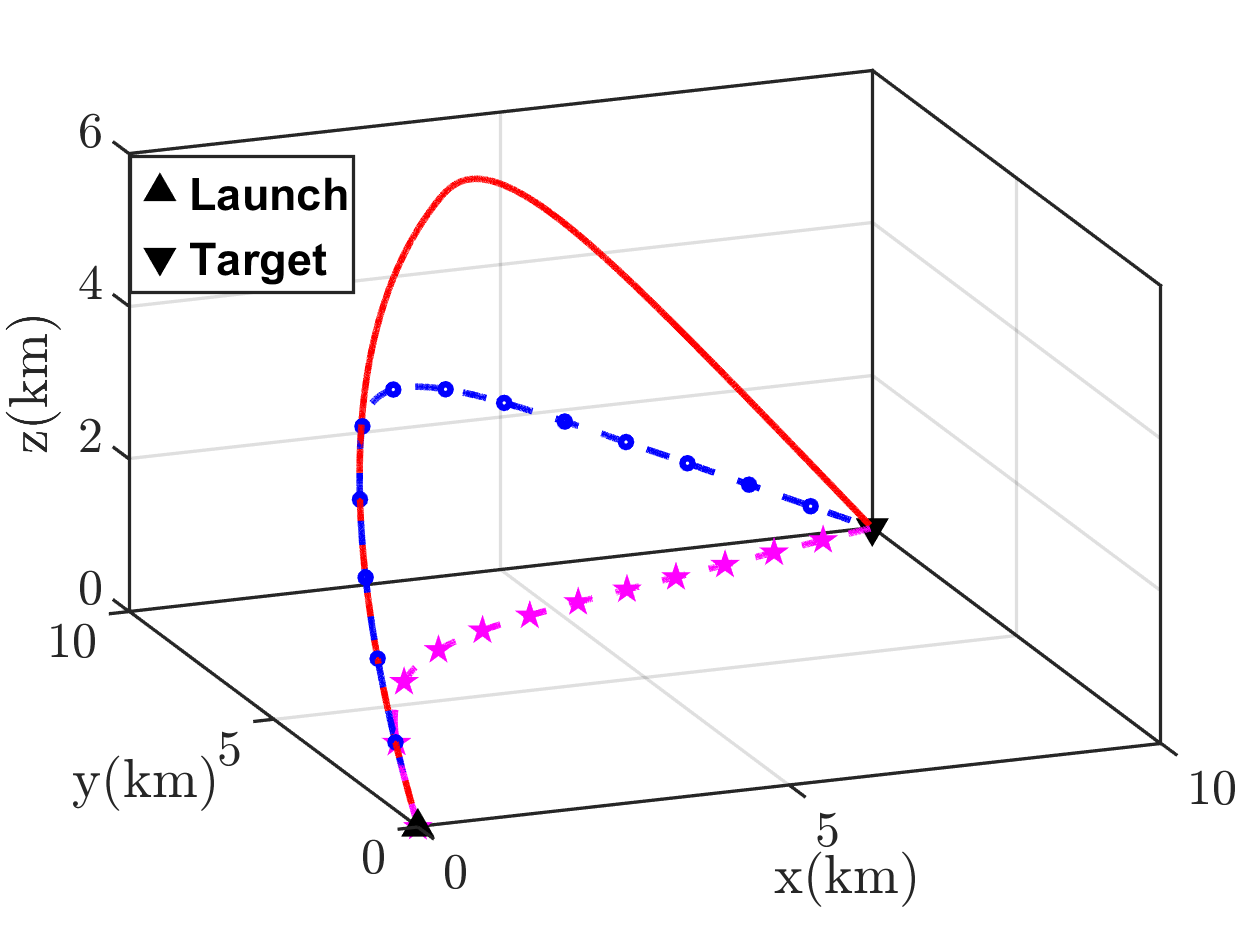}
\caption{Trajectory}\label{fig:3_tf_case_e_xyz}
\end{subfigure}
\begin{subfigure}{0.32\linewidth}
\includegraphics[width=\linewidth]{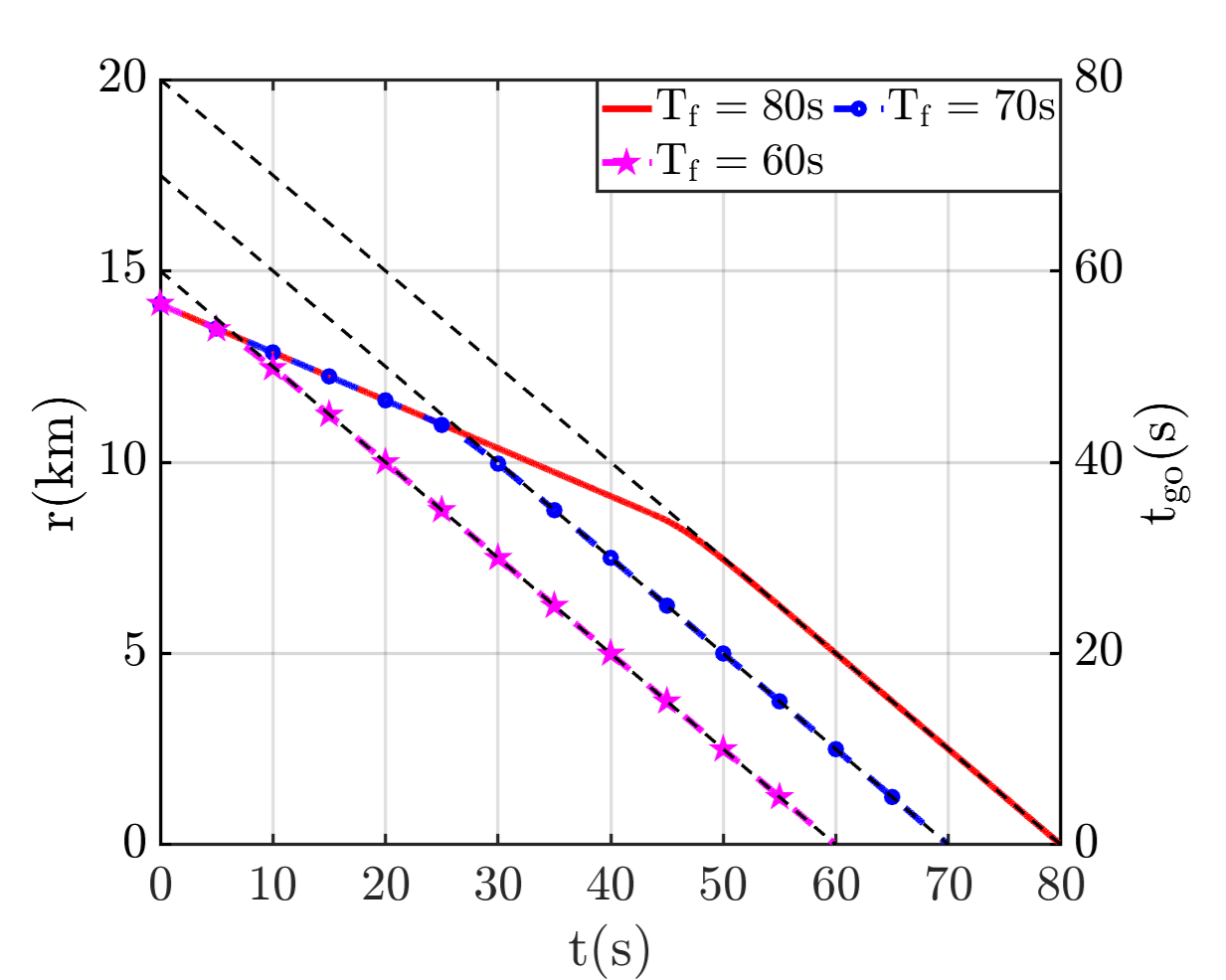}
\caption{Range and Time-to-go}\label{fig:3_tf_case_e_r_tgo}
\end{subfigure}
\begin{subfigure}{0.32\linewidth}
\includegraphics[width=\linewidth]{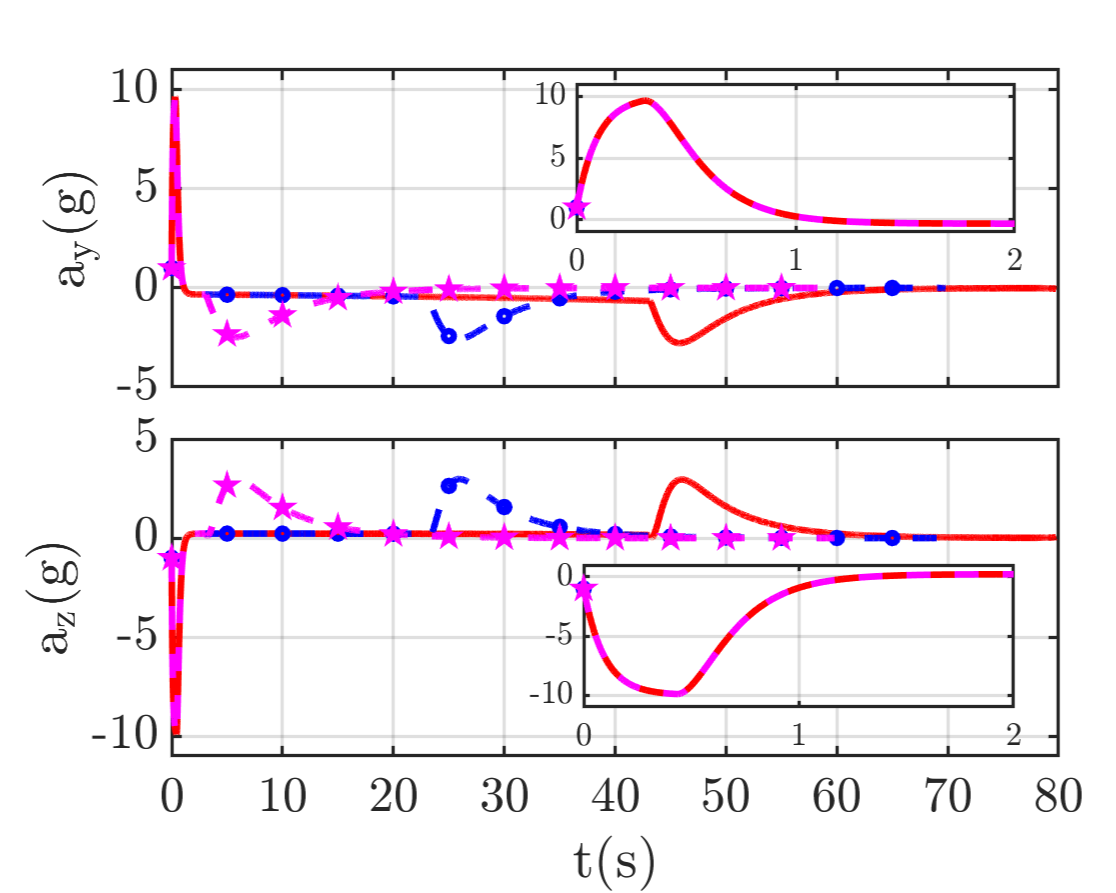}
\caption{Lateral Acceleration}\label{fig:3_tf_case_e_ay_az}
\end{subfigure}
\begin{subfigure}{0.32\linewidth}
\includegraphics[width=\linewidth]{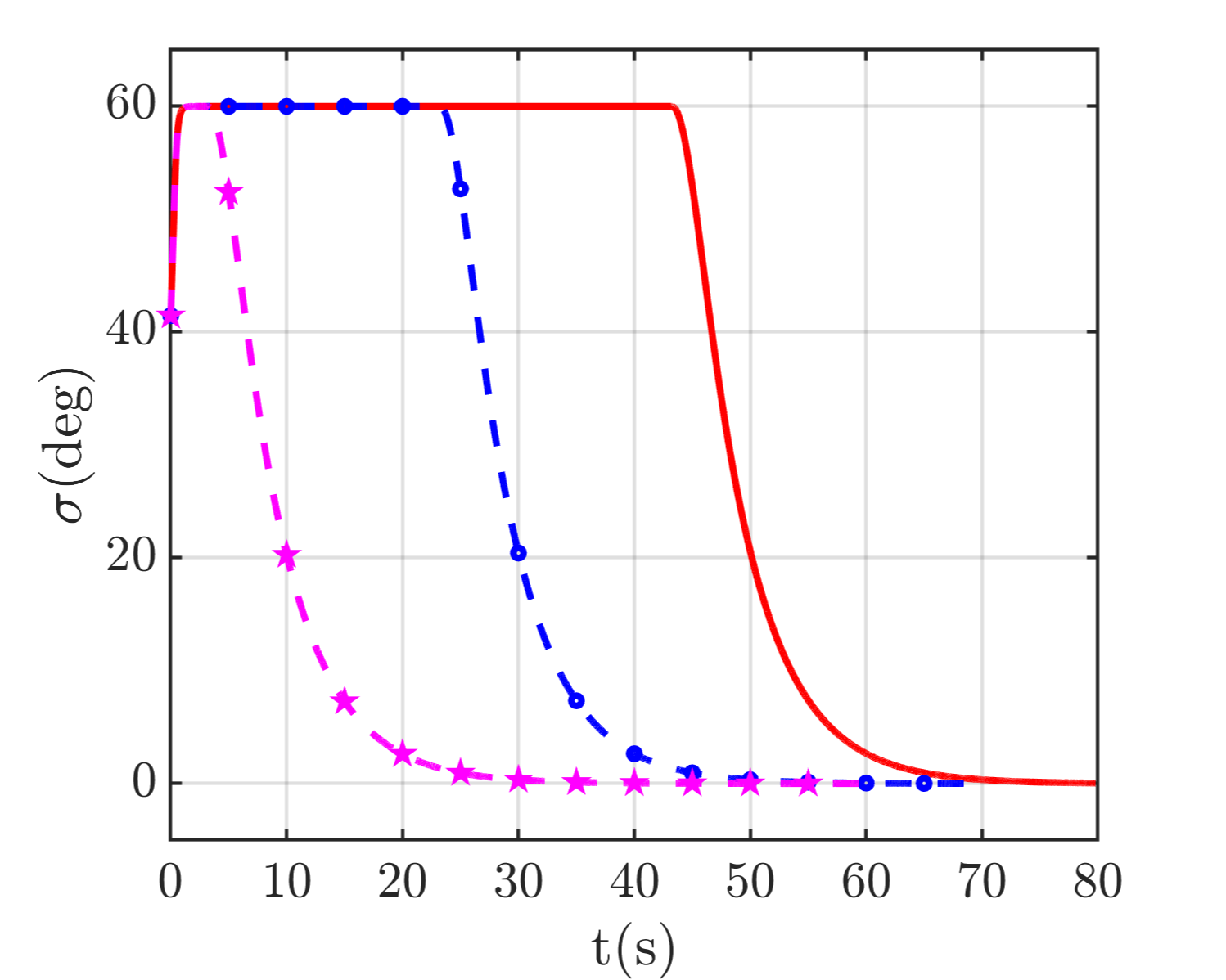}
\caption{Effective Lead Angle}\label{fig:3_tf_case_e_sigma}
\end{subfigure}
\begin{subfigure}{0.32\linewidth}
\includegraphics[width=\linewidth]{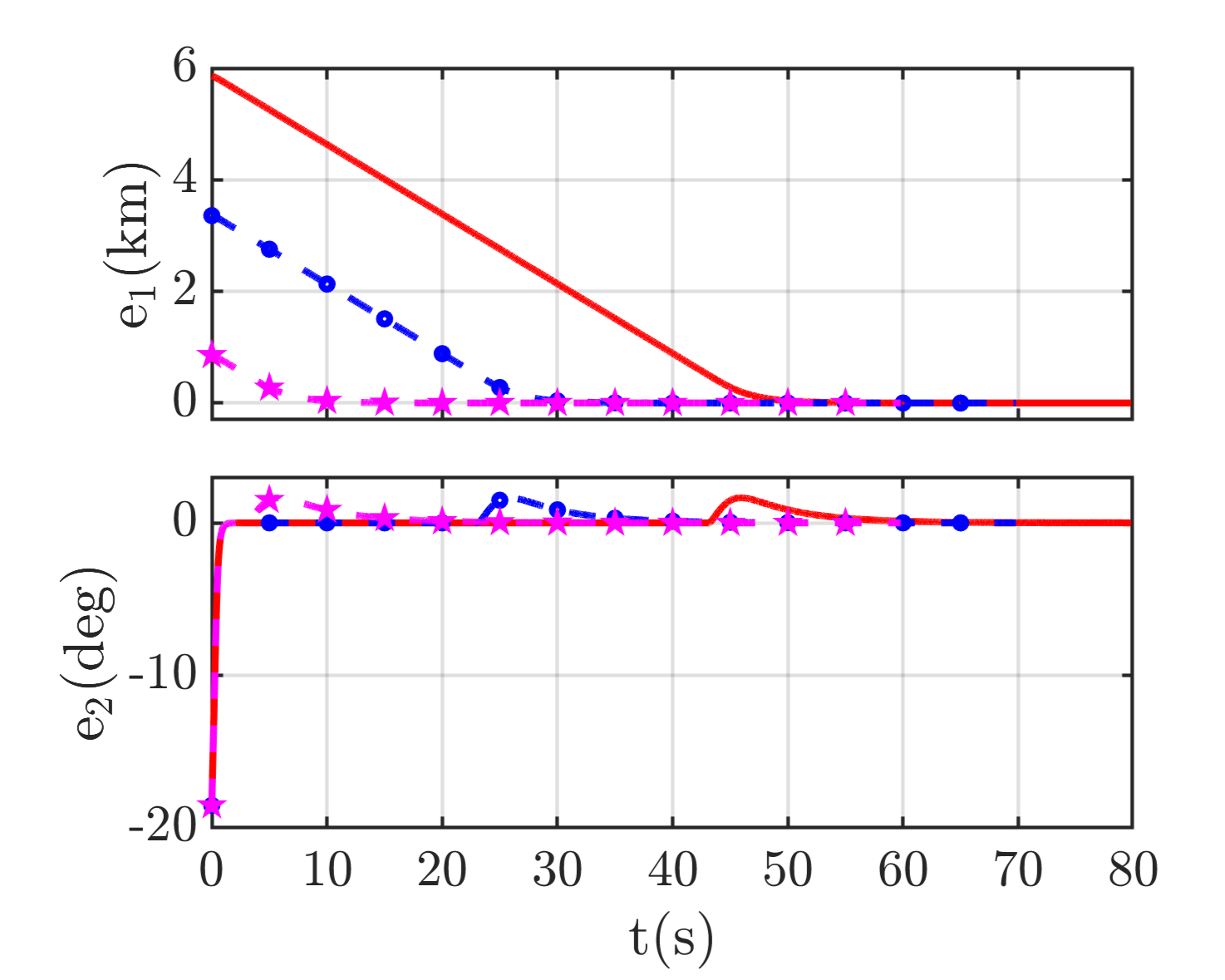}
\caption{Errors $e_1$ and $e_2$}\label{fig:3_tf_case_e_e1_e2}
\end{subfigure}
\begin{subfigure}{0.32\linewidth}
\includegraphics[width=\linewidth]{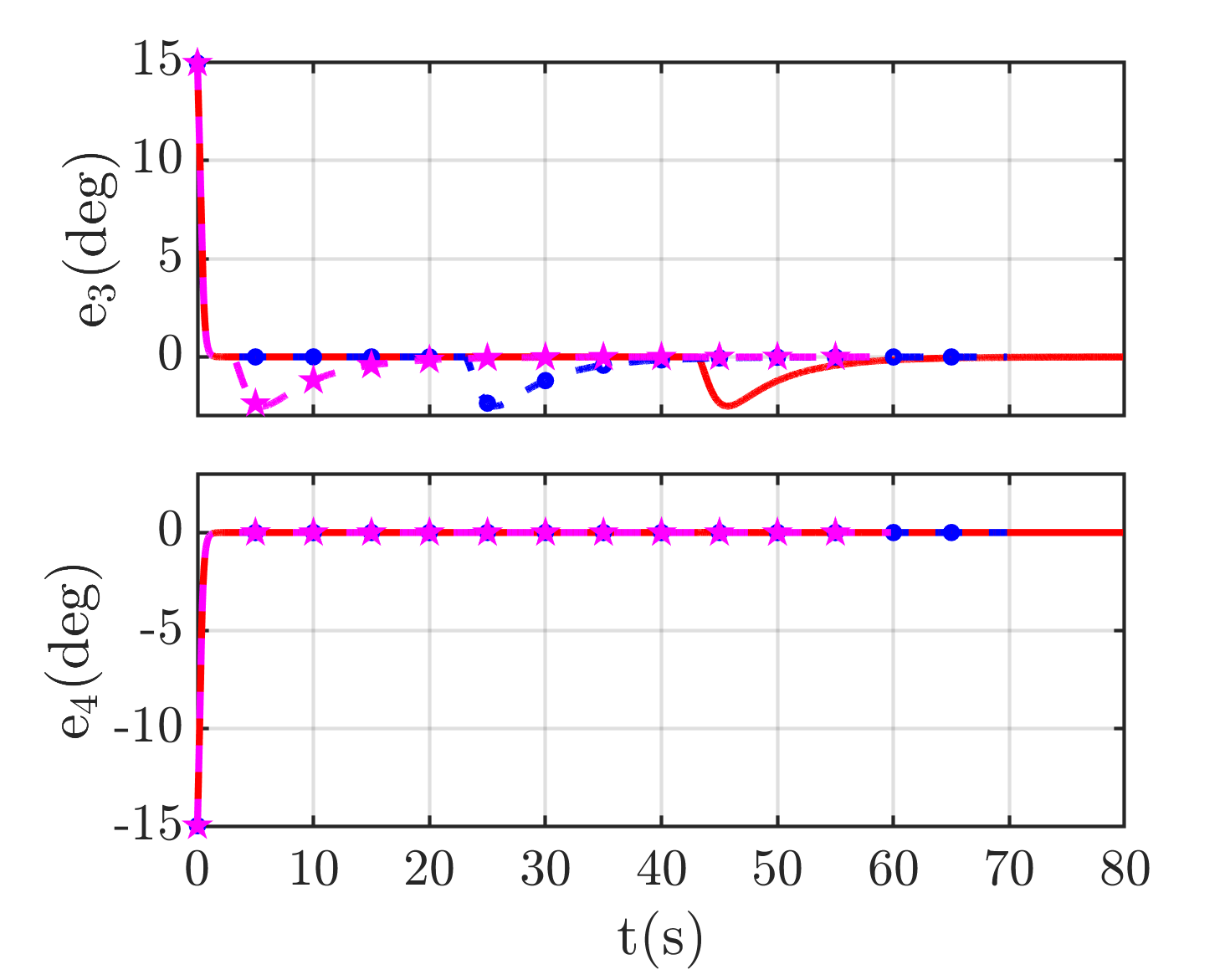}
\caption{Errors $e_3$ and $e_4$}\label{fig:3_tf_case_e_e3_e4}
\end{subfigure}
\caption{Performance for $t_{\rm f}=60\,$s, $70\,$s, $80\,$s, initial heading angles $\theta_{\rm m}=-30^\circ,\,\psi_{\rm m}=30^\circ$.}
\label{fig:3_tf_case_e}
\end{figure*}

In this subsection, the simulation results for three different impact times, $60$, $70$, and $80\,$s, are presented. \Cref{fig:3_tf_case_e} shows the relevant plots for these cases. It can be observed from \Cref{fig:3_tf_case_e} that with our proposed guidance law, the interceptor is able to achieve a successful interception while maintaining its heading angles within the FOV bound. The equations \eqref{eq:td_min_bce} and \eqref{eq:td_max_case_e} yield an achievable impact time range of $(59.19,~86.27)\,$s. 
It is to be noted that the relation in \eqref{eq:td_max_case_e} is only a sufficient condition for interception at the given impact time, but not a necessary one. 
\Cref{fig:3_tf_case_e_xyz} illustrates that the interceptor initially follows the same curved path and later diverts into the collision course. The time at which the interceptor departs from the initial curved path is proportional to the desired impact time $t_{\rm f}$. From \Cref{fig:3_tf_case_e_xyz}, it is evident that after reaching the initial curved path, $a_{\rm z}$, in \Cref{fig:3_tf_case_e_ay_az} remains at zero until the interceptor departs towards the collision course. However, $a_{\rm y}$ in \Cref{fig:3_tf_case_e_ay_az} decreases gradually, reflecting the interceptor's movement towards the target in the $xy$-plane as in \Cref{fig:3_tf_case_e_xyz}. It was observed that all the errors converge within a finite time.

\subsection{Performance with Different Initial Heading Angles}
\label{Subsec:Perf_diff_ini_head}
\begin{figure*}[!htpb]
\centering
\begin{subfigure}{0.32\linewidth}
\includegraphics[width=\linewidth]{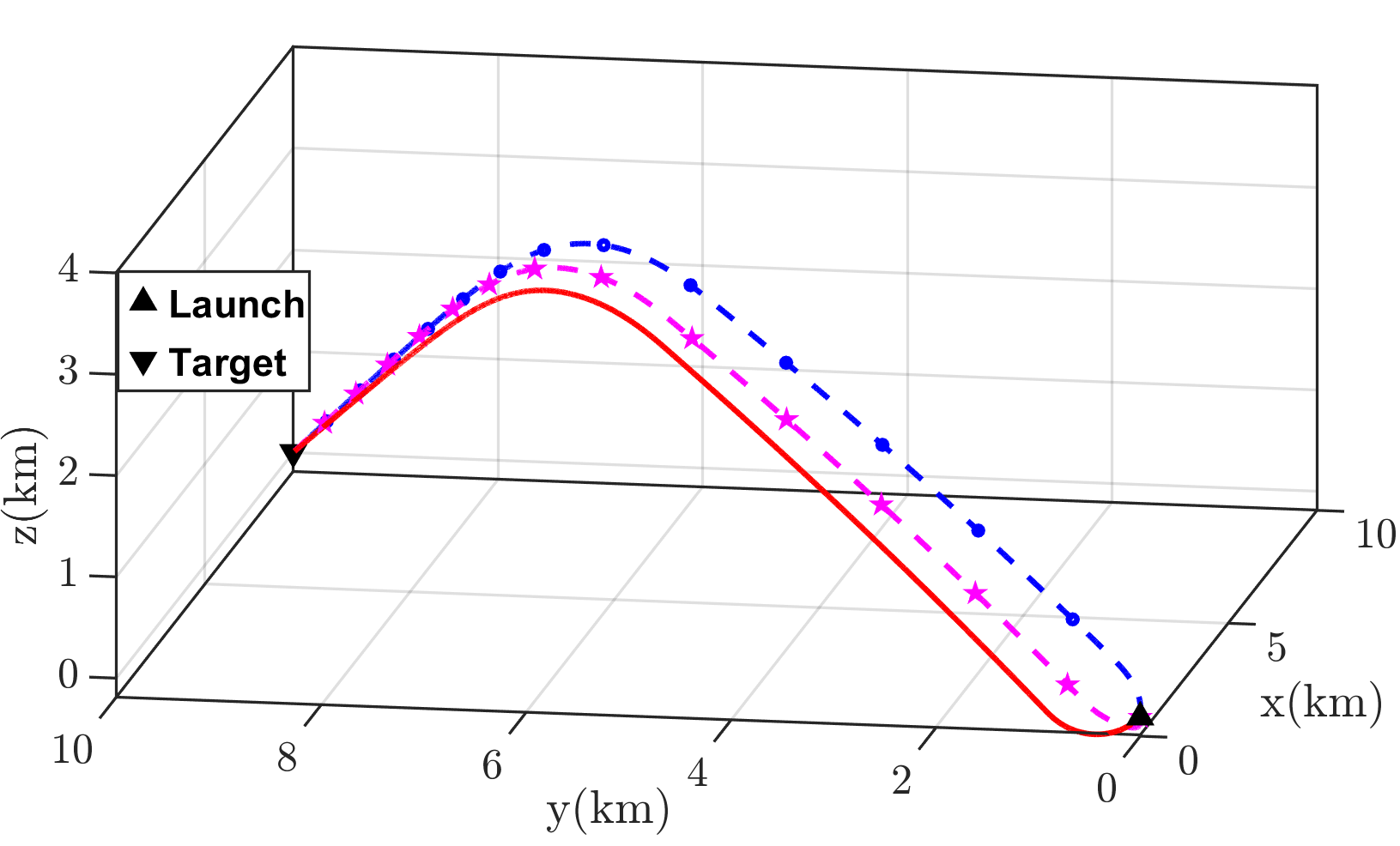}
\caption{Trajectory}\label{fig:3_ini_head_case_e_xyz}
\end{subfigure}
\begin{subfigure}{0.32\linewidth}
\includegraphics[width=\linewidth]{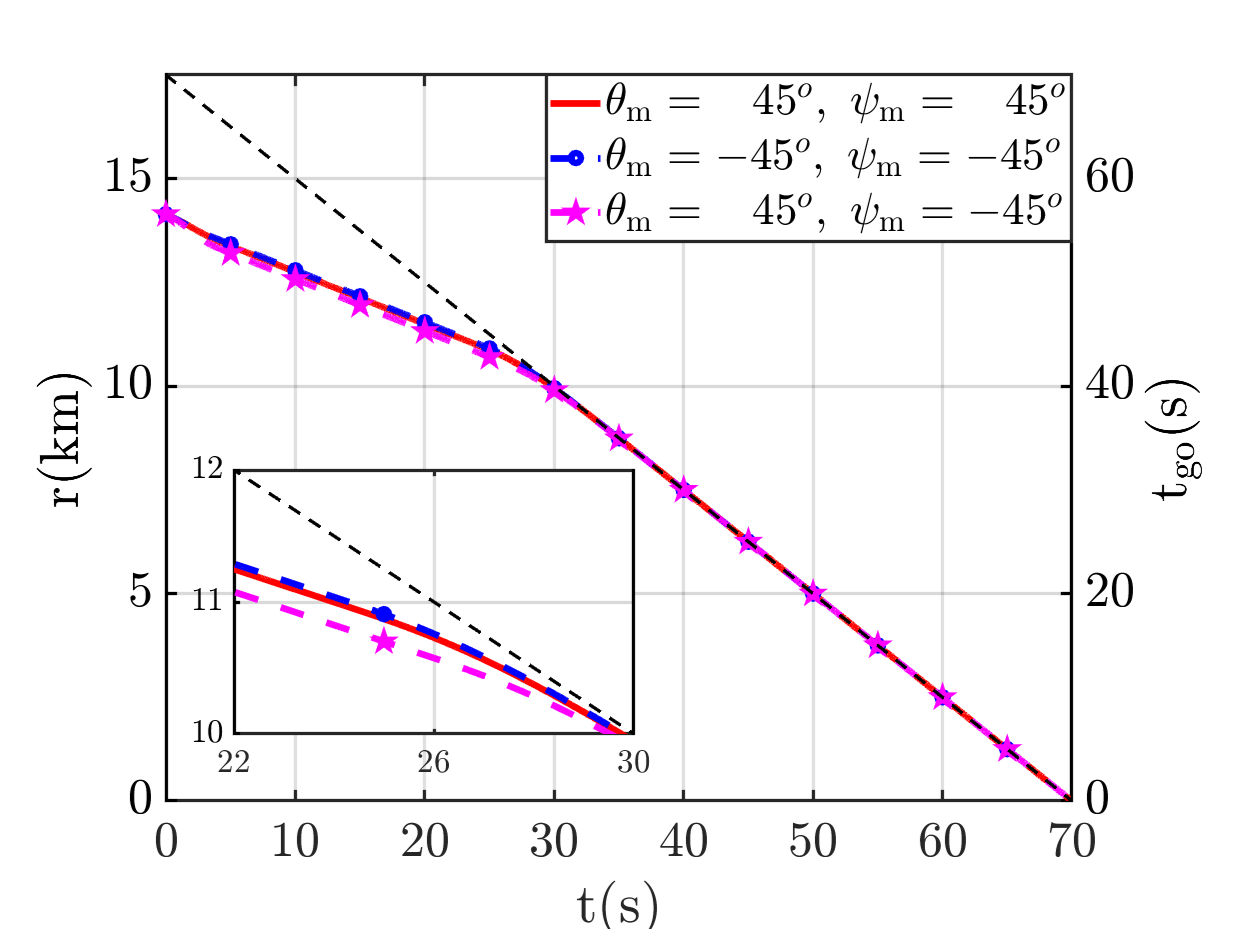}
\caption{Range and Time-to-go}\label{fig:3_ini_head_case_e_r_tgo}
\end{subfigure}
\begin{subfigure}{0.32\linewidth}
\includegraphics[width=\linewidth]{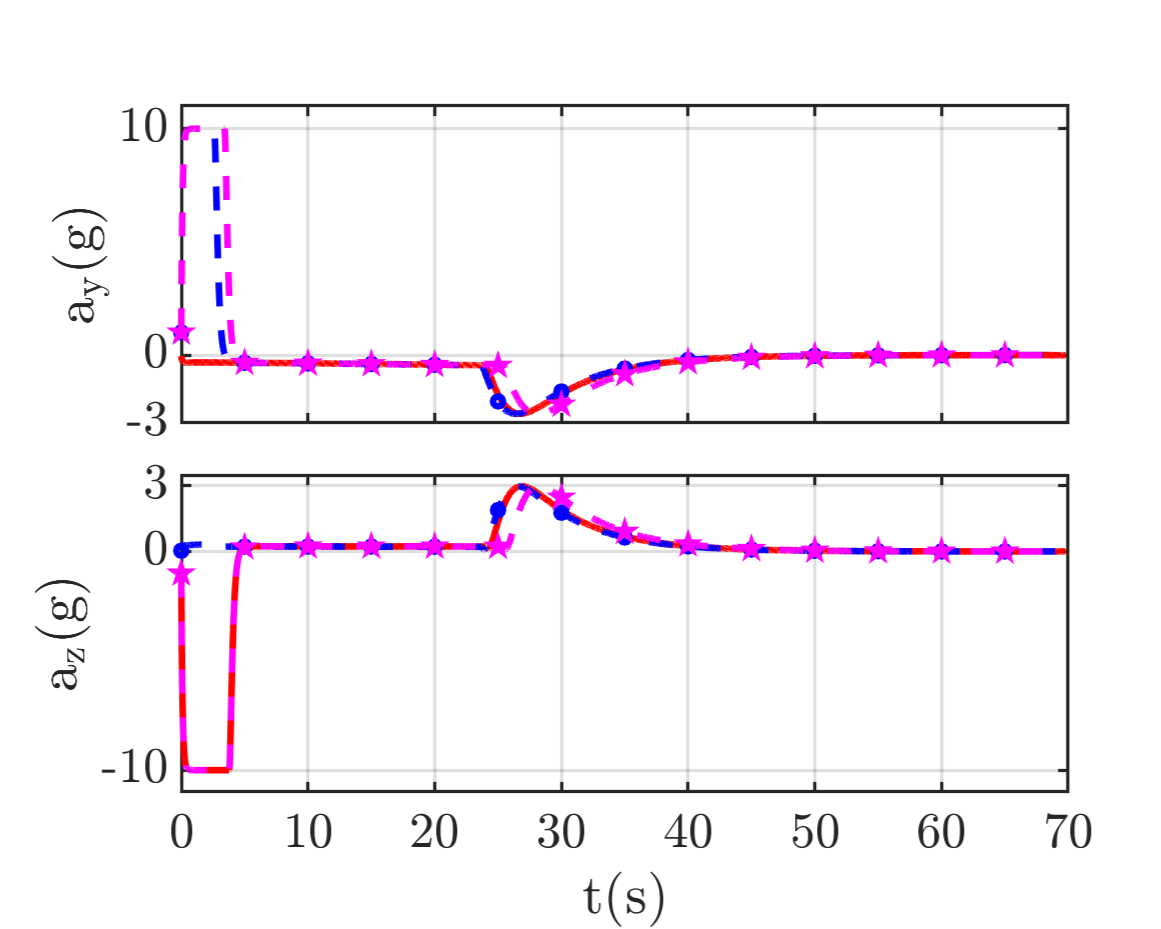}
\caption{Lateral Acceleration}\label{fig:3_ini_head_case_e_ay_az}
\end{subfigure}
\begin{subfigure}{0.32\linewidth}
\includegraphics[width=\linewidth]{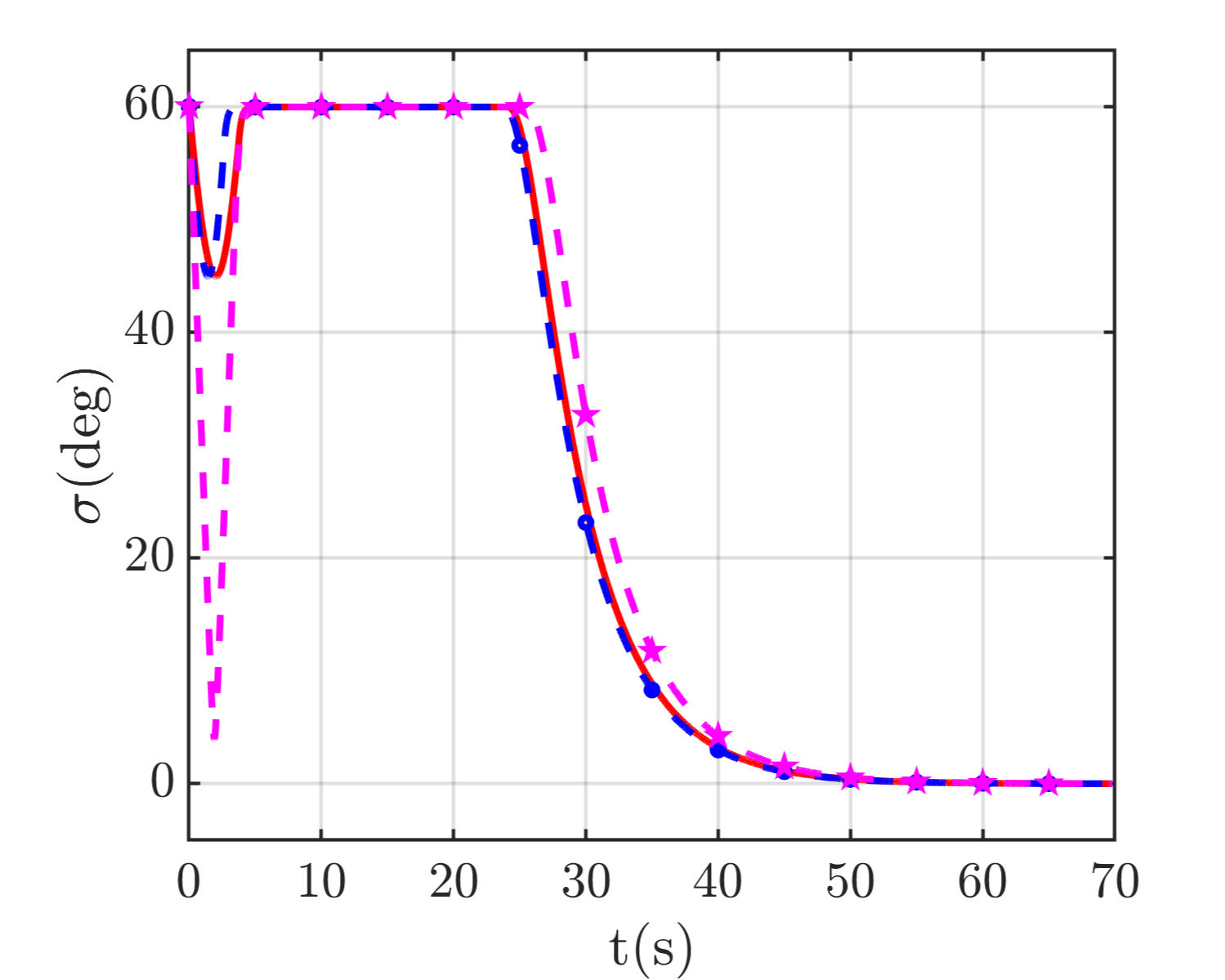}
\caption{Effective Lead Angle}\label{fig:3_ini_head_case_e_sigma}
\end{subfigure}
\begin{subfigure}{0.32\linewidth}
\includegraphics[width=\linewidth]{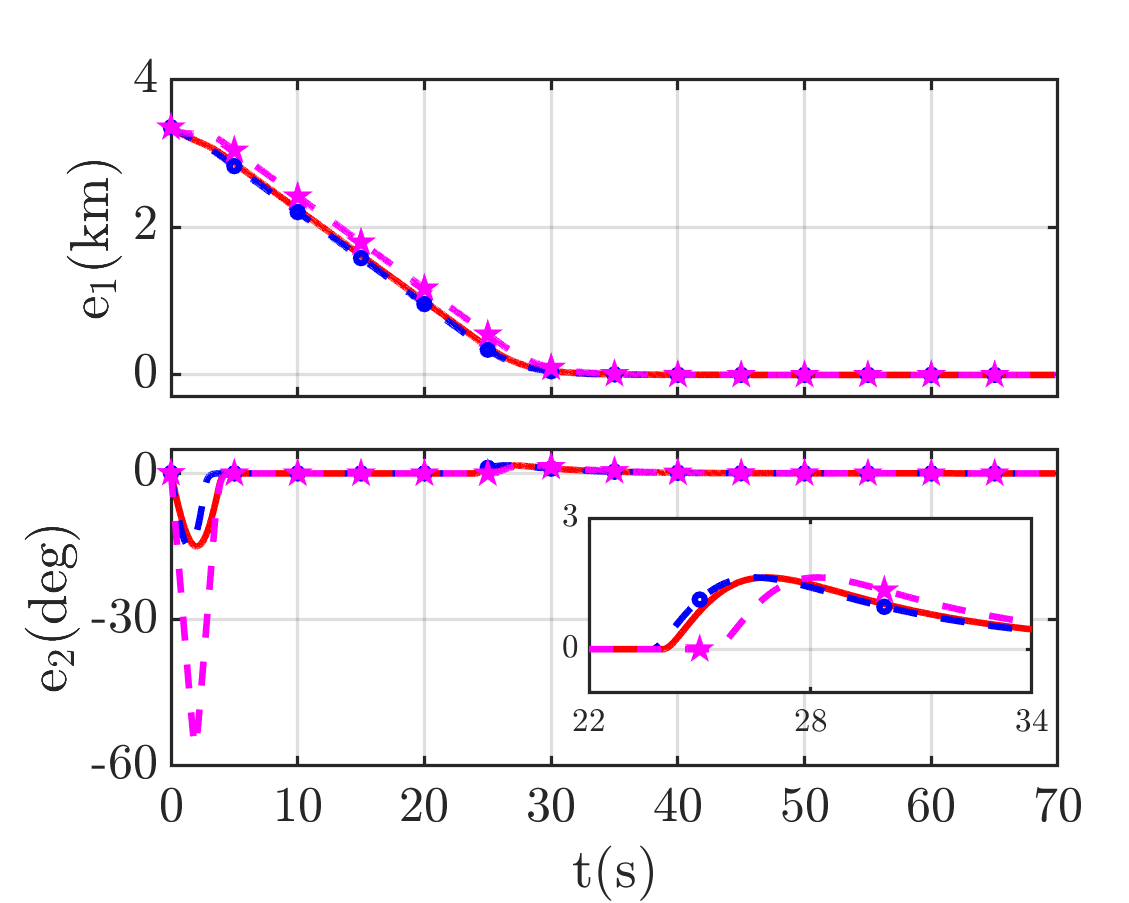}
\caption{Errors $e_1$ and $e_2$}\label{fig:3_ini_head_case_e_e1_e2}
\end{subfigure}
\begin{subfigure}{0.32\linewidth}
\includegraphics[width=\linewidth]{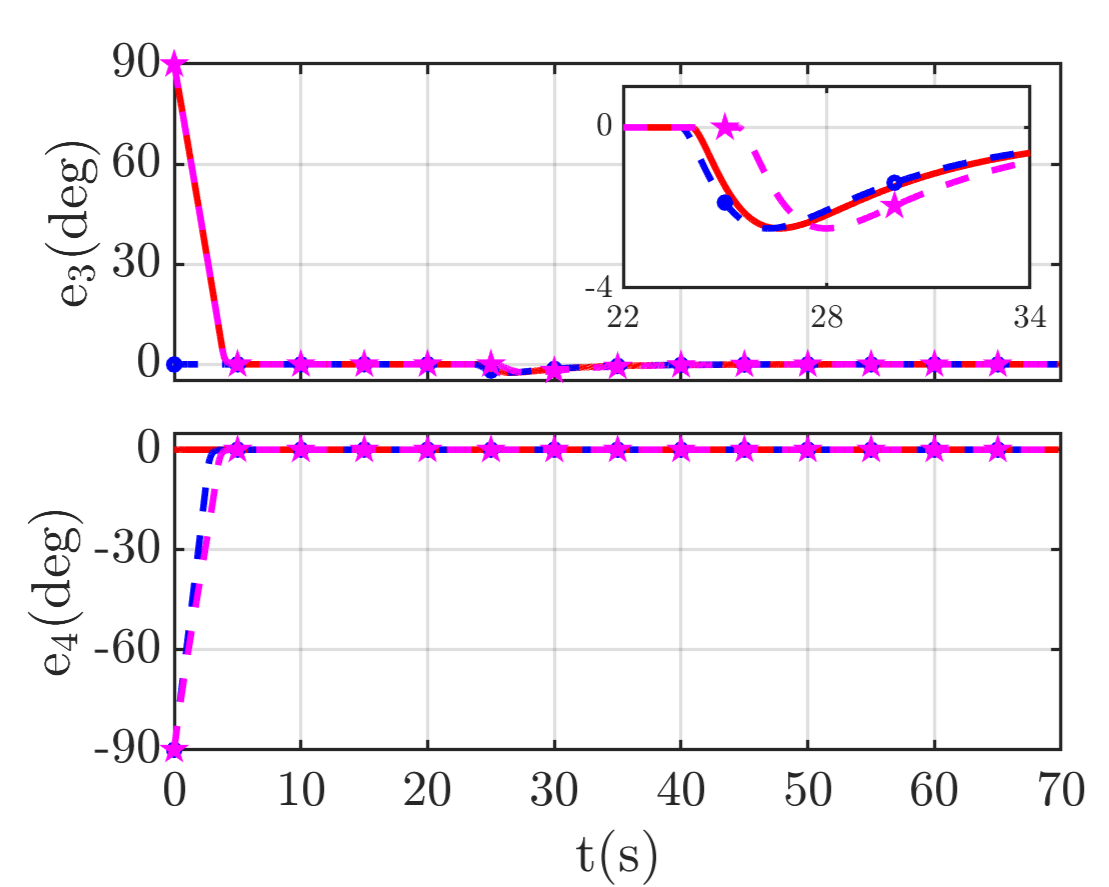}
\caption{Errors $e_3$ and $e_4$}\label{fig:3_ini_head_case_e_e3_e4}
\end{subfigure}
\caption{Performance for a desired impact time $t_{\rm f}=70\,$s, and different initial headings.}
\label{fig:3_ini_head_case_e}
\end{figure*}

The results of the third scenario, featuring three different initial headings angles, are presented in this subsection. The simulations are performed with the initial headings $(\theta_{\rm m},\psi_{\rm m})$, set to $(45^\circ,45^\circ)$, $(-45^\circ,-45^\circ)$, and $(45^\circ,-45^\circ)$, while keeping all other simulation parameters unchanged. \Cref{fig:3_ini_head_case_e} depicts all the simulation results for this scenario. The initial headings, $\theta_{\rm m}$ and $\psi_{\rm m}$, are chosen such that the initial effective lead angle is equal to the FOV bound, allowing us to observe the performance of the guidance law at the verge of the lead angle boundary.

\begin{figure*}[!htpb]
\centering
\begin{subfigure}{0.32\linewidth}
\includegraphics[width=\linewidth]{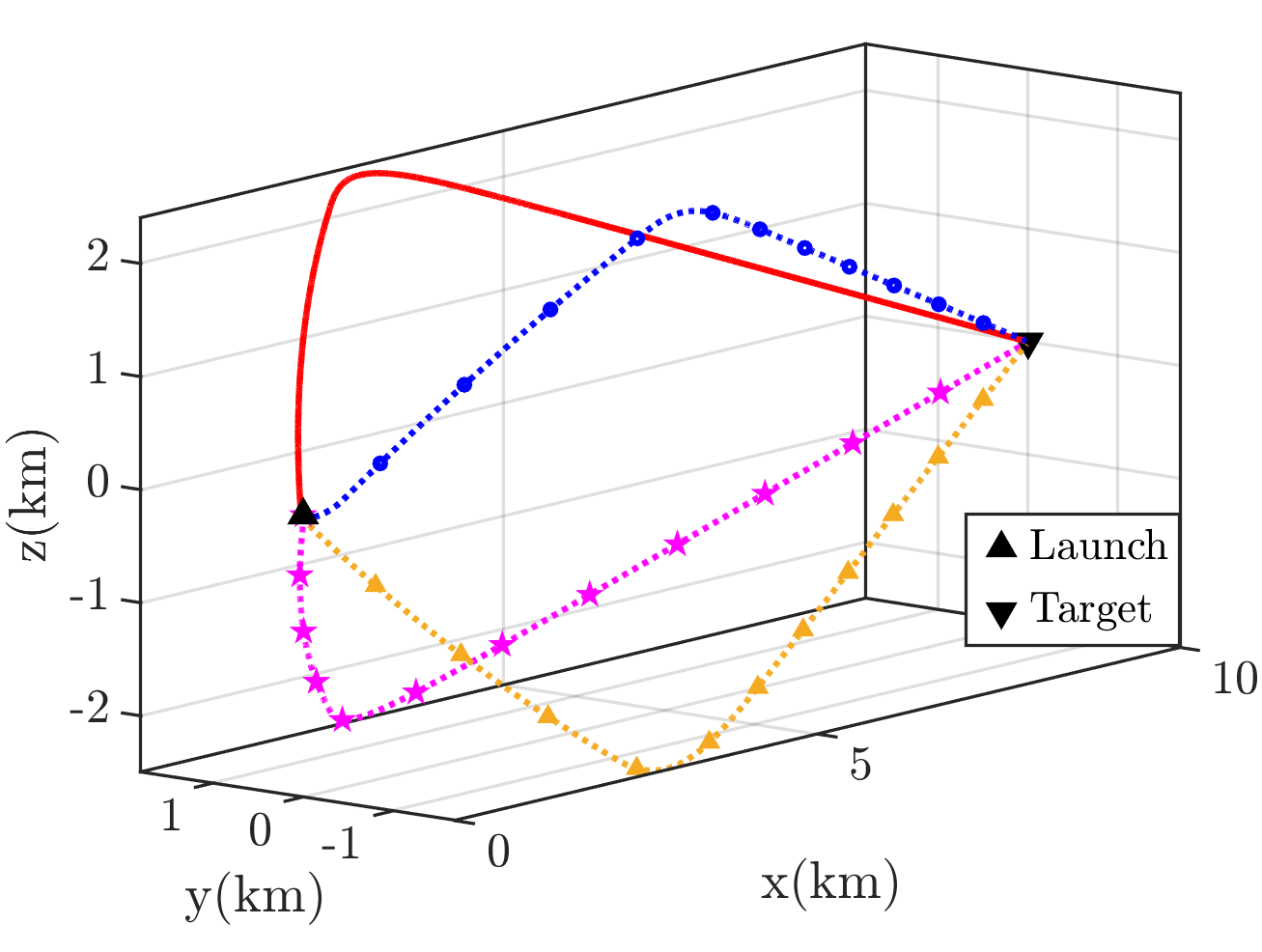}
\caption{Trajectory}\label{fig:Trajec_Oct_case_e_xyz}
\end{subfigure}%
\begin{subfigure}{0.32\linewidth}
\includegraphics[width=\linewidth]{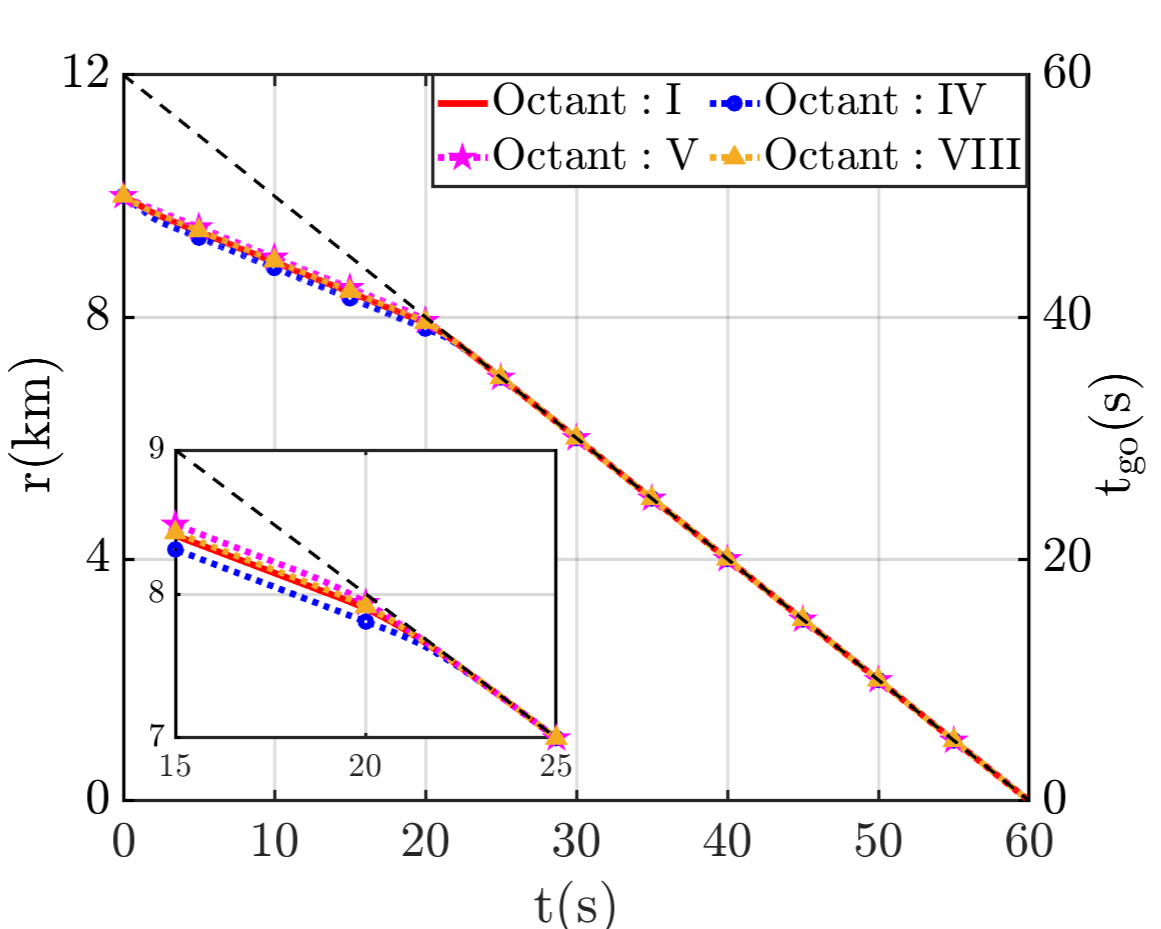}
\caption{Range and Time-to-go}\label{fig:Trajec_Oct_case_e_r_tgo}
\end{subfigure}%
\begin{subfigure}{0.32\linewidth}
\includegraphics[width=\linewidth]{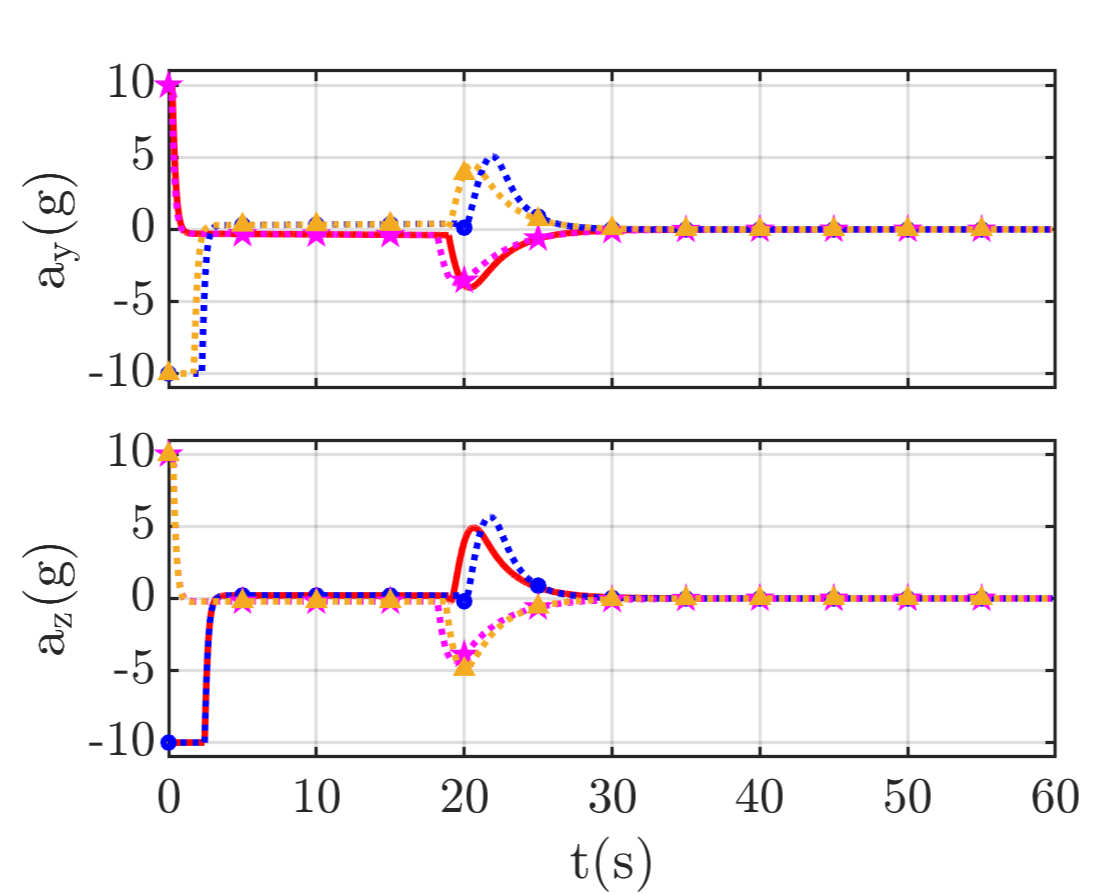}
\caption{Lateral Acceleration}\label{fig:Trajec_Oct_case_e_ay_az}
\end{subfigure}
\begin{subfigure}{0.32\linewidth}
\includegraphics[width=\linewidth]{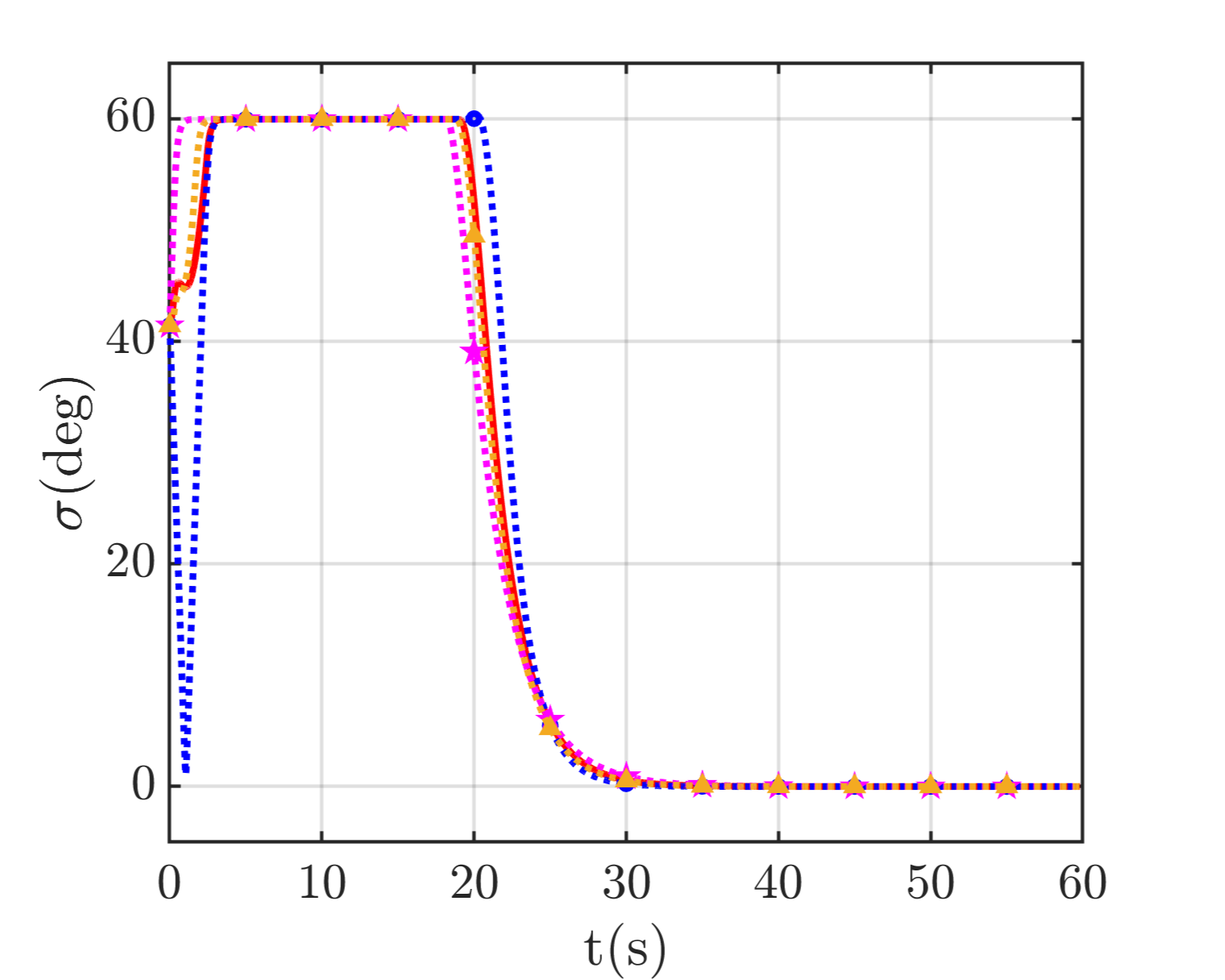}
\caption{Effective Lead Angle}\label{fig:Trajec_Oct_case_e_sigma}
\end{subfigure}%
\begin{subfigure}{0.32\linewidth}
\includegraphics[width=\linewidth]{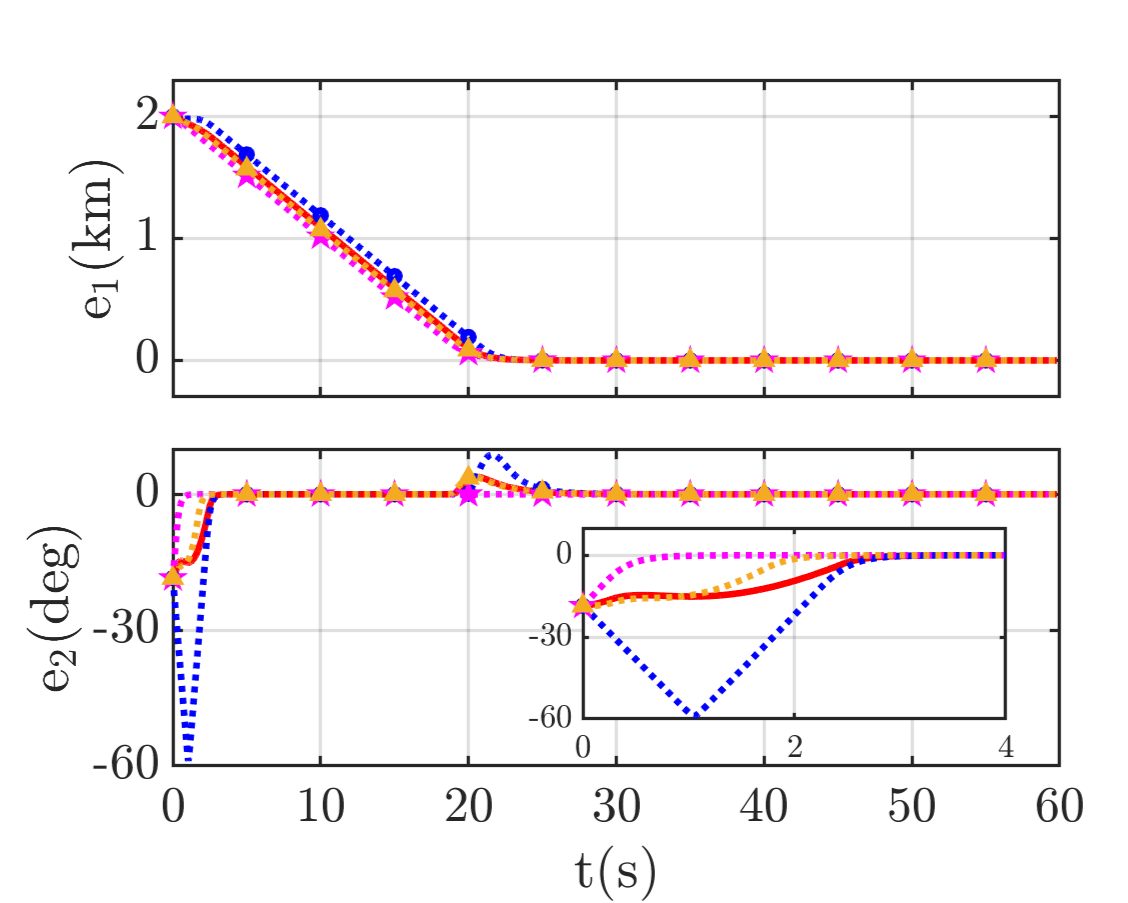}
\caption{Errors $e_1$ and $e_2$}\label{fig:Trajec_Oct_case_e_e1_e2}
\end{subfigure}%
\begin{subfigure}{0.32\linewidth}
\includegraphics[width=\linewidth]{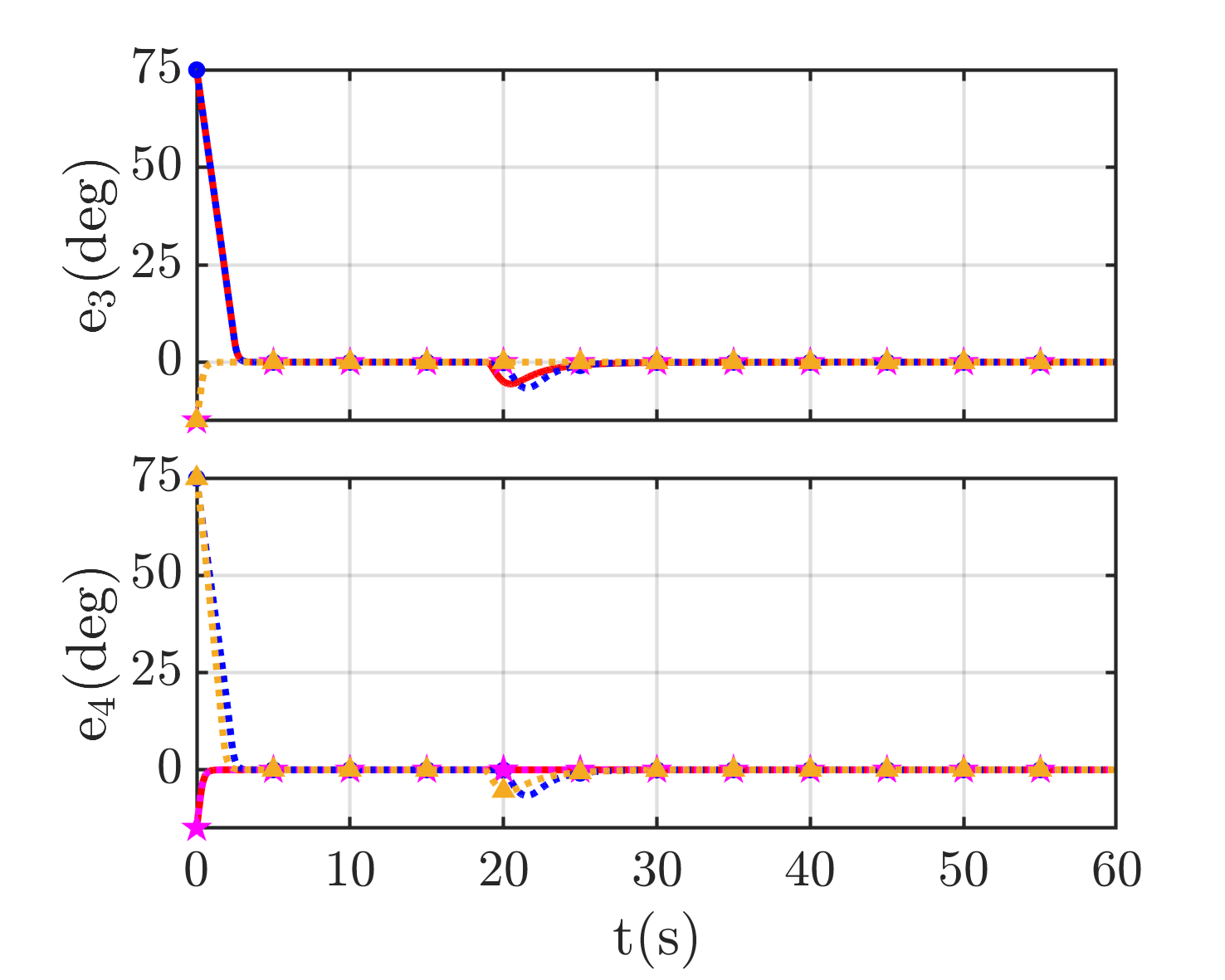}
\caption{Errors $e_3$ and $e_4$}
\label{fig:Trajec_Oct_case_e_e3_e4}
\end{subfigure}
\caption{Control over the Octant of the Interceptor Trajectories.}
\label{fig:Trajec_Oct_case_e}
\end{figure*}
\begin{figure}[!htpb]
    \centering
    \includegraphics[width=0.55\linewidth]{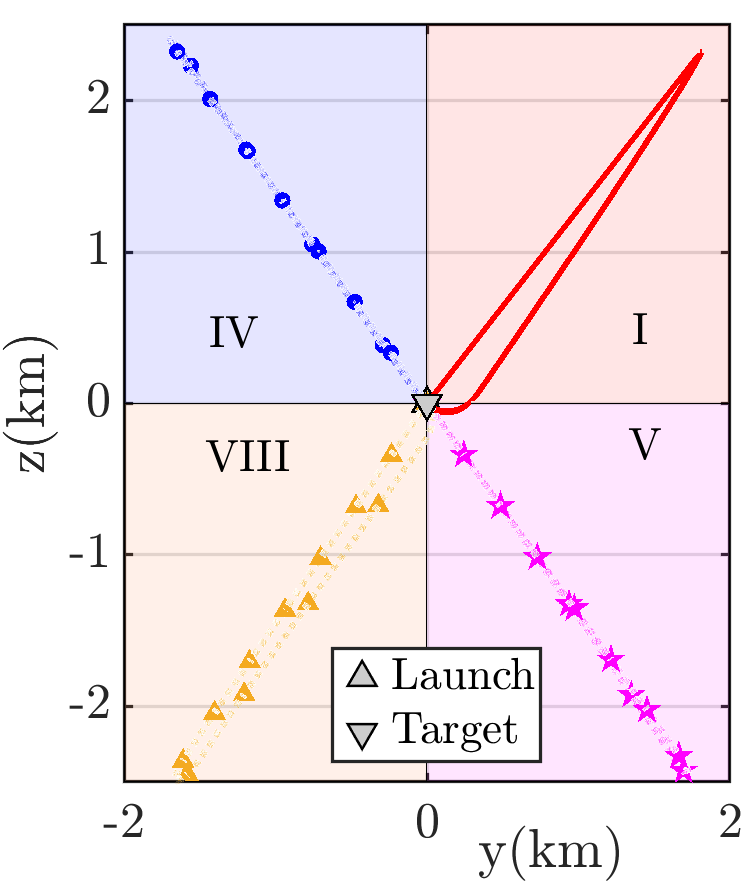}
    \caption{X-axis view of fig.\ref{fig:Trajec_Oct_case_e_xyz}.}
    \label{fig:Trajec_Oct_case_e_xyz_x_view}
\end{figure}

\Cref{fig:3_ini_head_case_e_xyz} demonstrates that a successful target interception at the desired impact time was achieved for all initial heading angles. Also, it was observed that the trajectories return to the first octant of the LOS frame. As we have not considered any small angle assumption in the guidance design, it can be seen to perform well even for large initial heading angles as high as $60^\circ$. Additionally, \Cref{fig:3_ini_head_case_e_sigma} indicates the effective lead angle, which is also the FOV, remained within its upper bound even though the value of the initial lead angle was chosen as $\sigma_{\max}$. As the guidance law brings back the interceptor's trajectory to the first octant, the lead angle might initially drop, as observed earlier in \Cref{fig:3_ini_head_case_e_sigma}, the same behavior was observed in \Cref{fig:Trajec_Oct_case_e_sigma}. However, at the time instant corresponding to this drop, the magnitude of the error $e_2$ in \Cref{fig:3_ini_head_case_e_e1_e2} is positive, and hence the lateral acceleration commands in \Cref{fig:3_ini_head_case_e_ay_az} are non-zero, preventing the interceptor from veering off course. Therefore, even if the interceptor does not align itself with the target momentarily, the non-zero acceleration command will drive the interceptor to a correct trajectory.

\subsection{Performance Against Constant Velocity Target}
Though the guidance law was derived for stationary targets, extension to constant velocity targets is feasible by using the concept of predicted interception point (PIP). The target is assumed to be non-maneuvering and moving at a constant speed. Considering the target's initial coordinates, heading angles, and speed to be ($x_{\rm T}(0), y_{\rm T}(0), z_{\rm T}(0)$), ($\theta_{\rm T}, \psi_{\rm T}$), and $V_{\rm T}$, respectively, the coordinates of the PIP for an impact time of $t_{\rm f}$ are obtained as
\begin{subequations}
\begin{align}
    x_{\rm PIP} &= x_{\rm T}(0) + V_{\rm T}t_{\rm f}\cos{\theta_{\rm T}}\cos{\psi_{\rm T}},\\
    y_{\rm PIP} &= y_{\rm T}(0) + V_{\rm T}t_{\rm f}\cos{\theta_{\rm T}}\sin{\psi_{\rm T}},\\
    z_{\rm PIP} &= z_{\rm T}(0) - V_{\rm T}t_{\rm f}\sin{\theta_{\rm T}}.
\label{eq:PIP_xyz}
\end{align}
\end{subequations}
To analyze the performance of the proposed guidance law against a non-maneuvering constant speed target, we performed simulations with the target and interceptor speeds taken as $100$ m/s and $300$ m/s, respectively. The simulations are conducted for three combinations of ($t_{\rm f},\ \theta_{\rm m},\ \psi_{\rm m}$), i.e., ($80\,\rm s,\ -30^\circ,\ 30^\circ$), ($80\,\rm s,\ 30^\circ,\ 30^\circ$) and ($70\,\rm s,\ -30^\circ,\ 30^\circ$). The value of $\phi$ is taken as $700$ for these simulations. 
\begin{figure*}[!htpb]
\centering
\begin{subfigure}{0.32\linewidth}
\includegraphics[width=\linewidth]{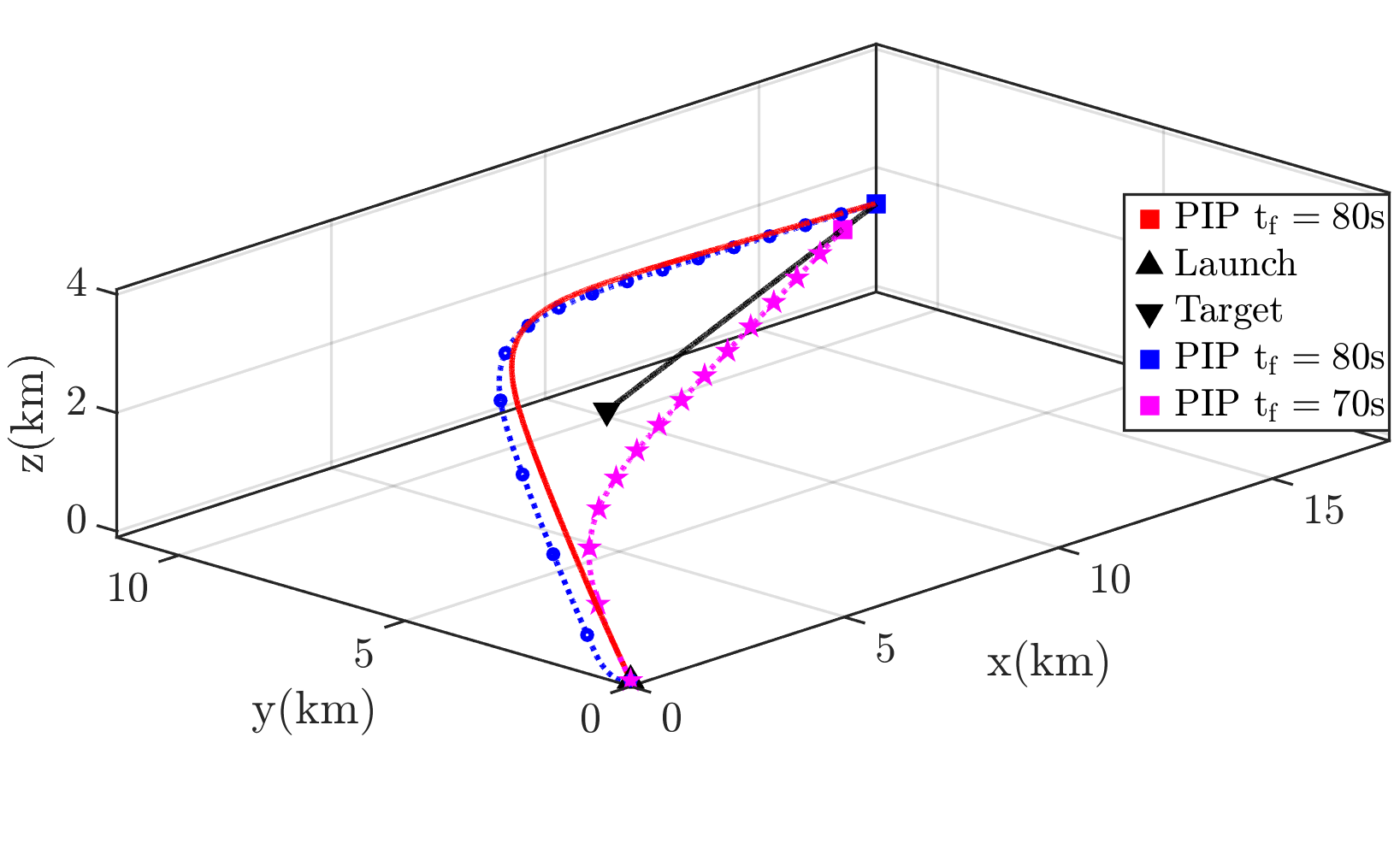}
\caption{Trajectory}\label{fig:PIP_case_e_xyz}
\end{subfigure}
\begin{subfigure}{0.32\linewidth}
\includegraphics[width=\linewidth]{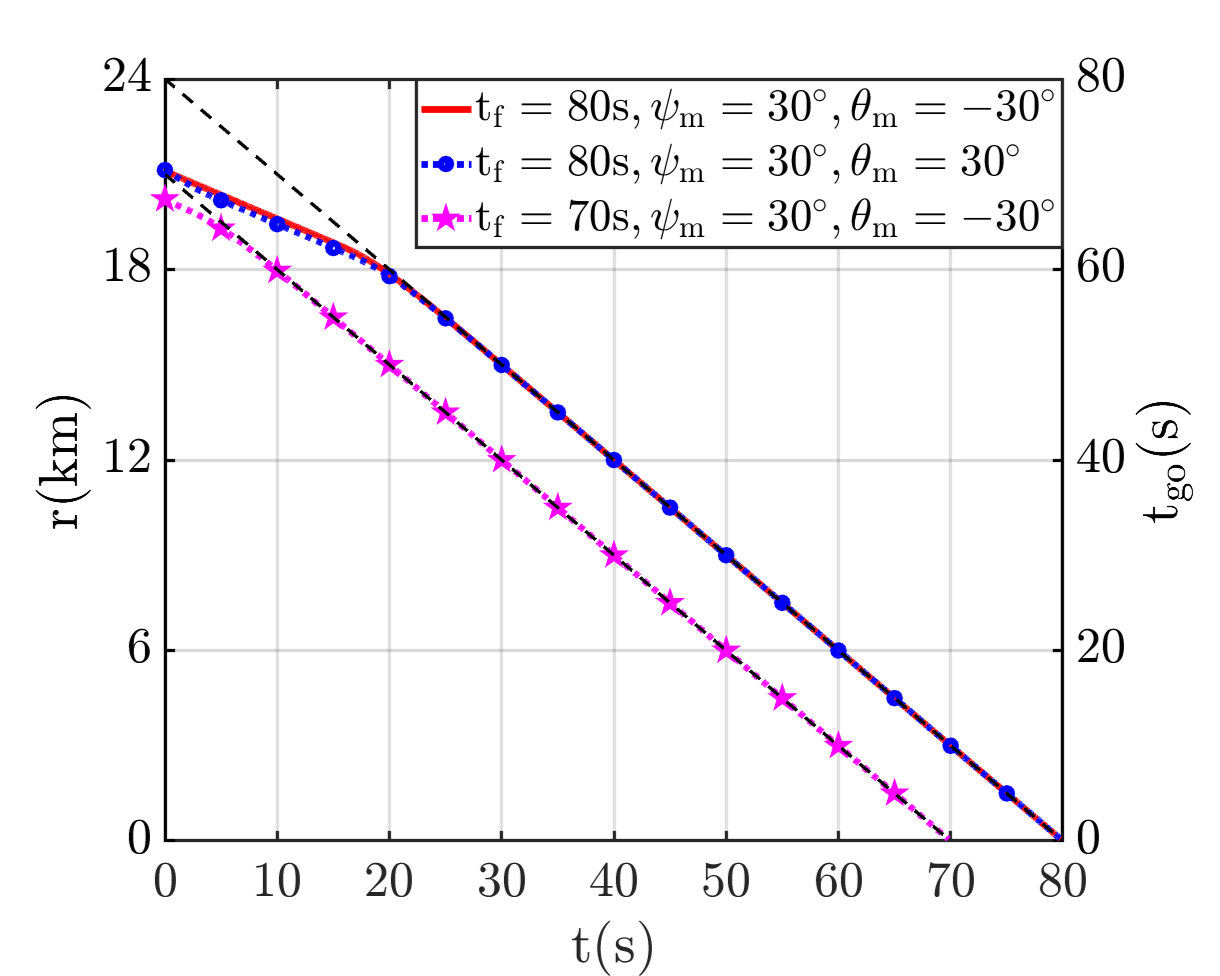}
\caption{Range and Time-to-go}\label{fig:PIP_case_e_r_tgo}
\end{subfigure}
\begin{subfigure}{0.32\linewidth}
\includegraphics[width=\linewidth]{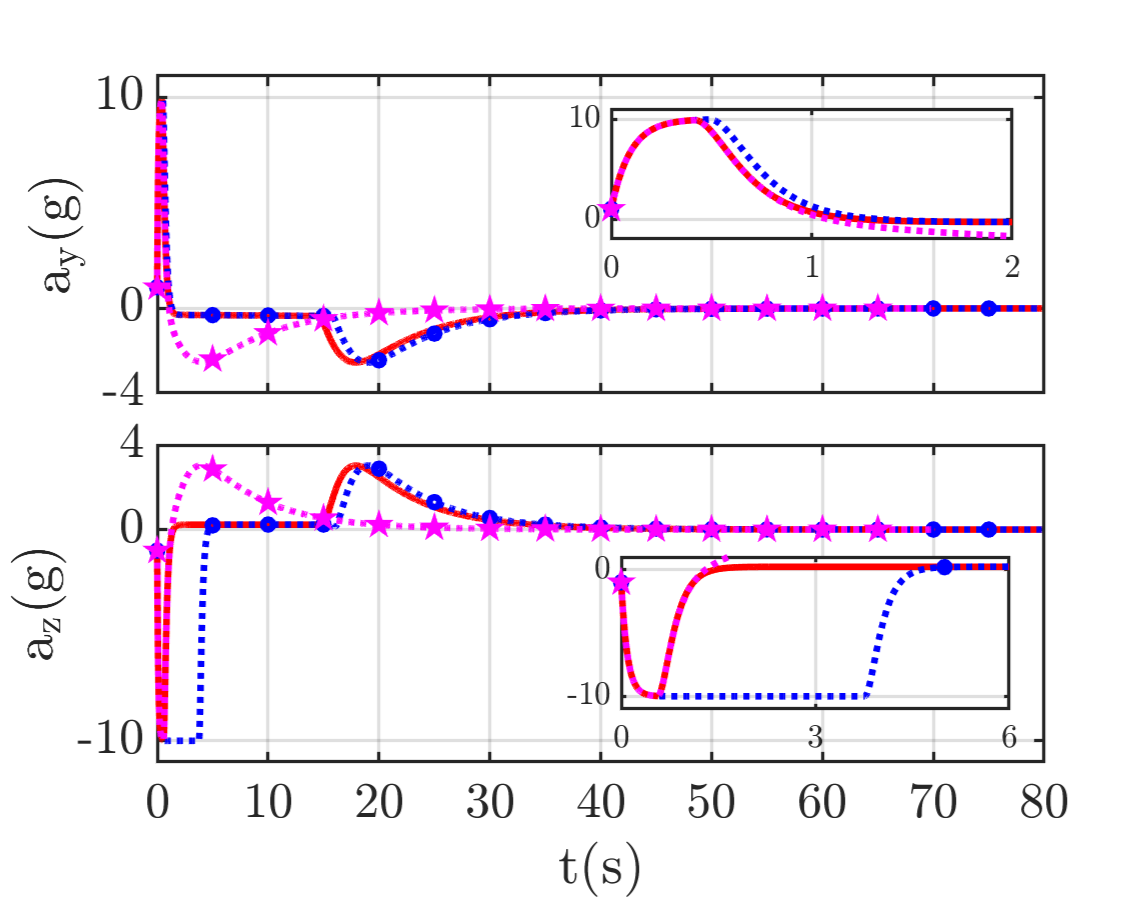}
\caption{Lateral Acceleration}\label{fig:PIP_case_e_ay_az}
\end{subfigure}
\begin{subfigure}{0.32\linewidth}
\includegraphics[width=\linewidth]{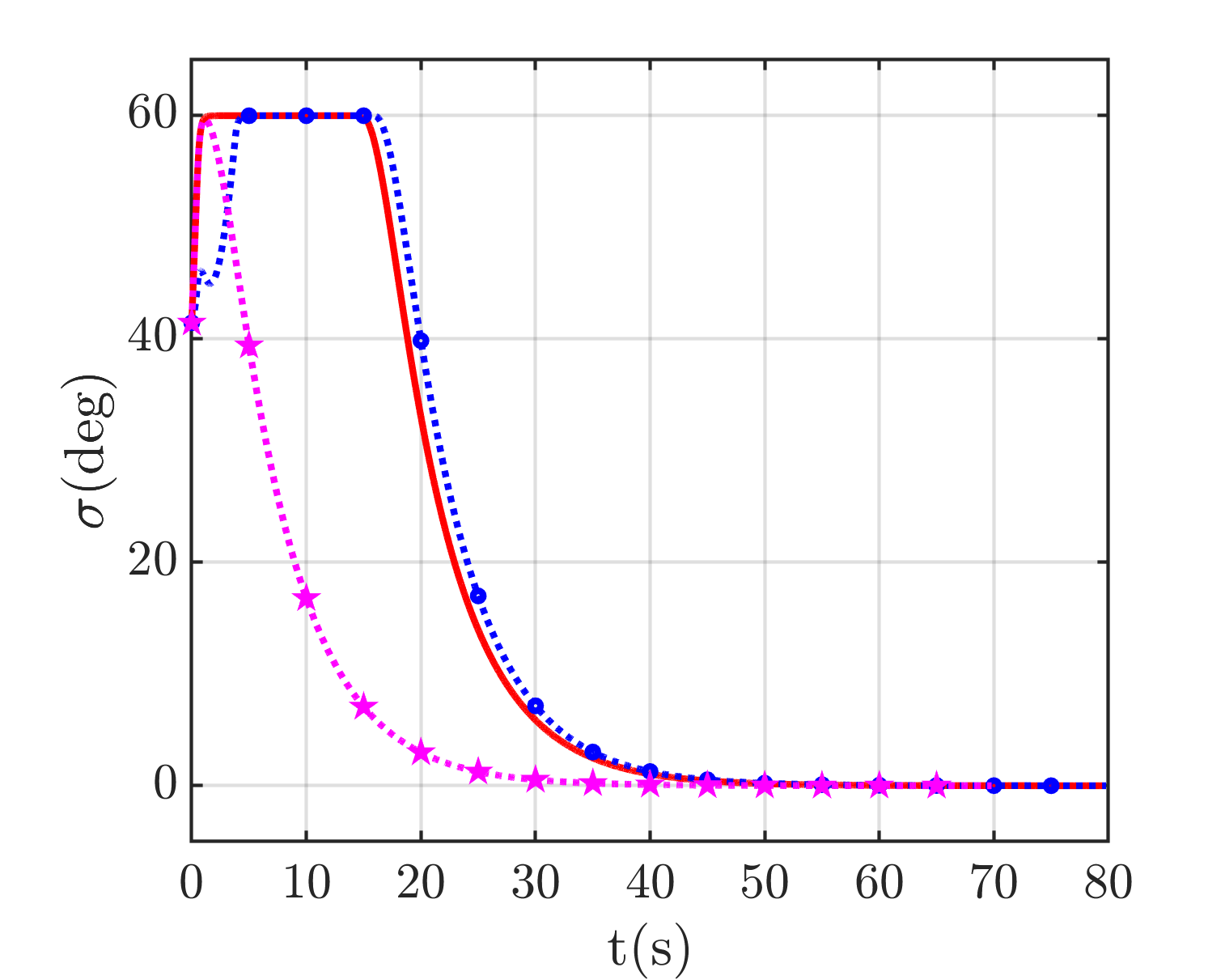}
\caption{Effective Lead Angle}\label{fig:PIP_case_e_sigma}
\end{subfigure}
\begin{subfigure}{0.32\linewidth}
\includegraphics[width=\linewidth]{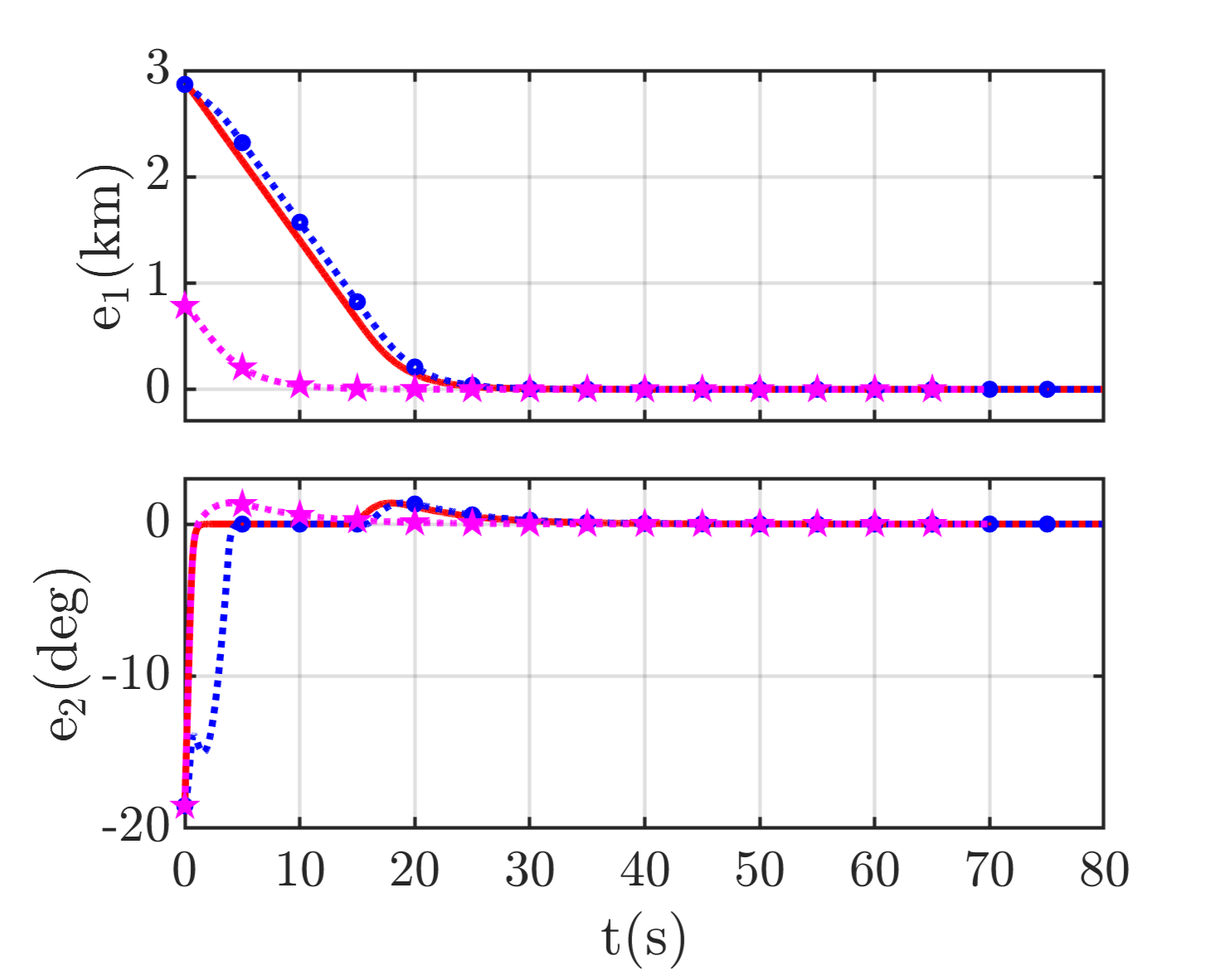}
\caption{Errors $e_1$ and $e_2$}\label{fig:PIP_case_e_e1_e2}
\end{subfigure}
\begin{subfigure}{0.32\linewidth}
\includegraphics[width=\linewidth]{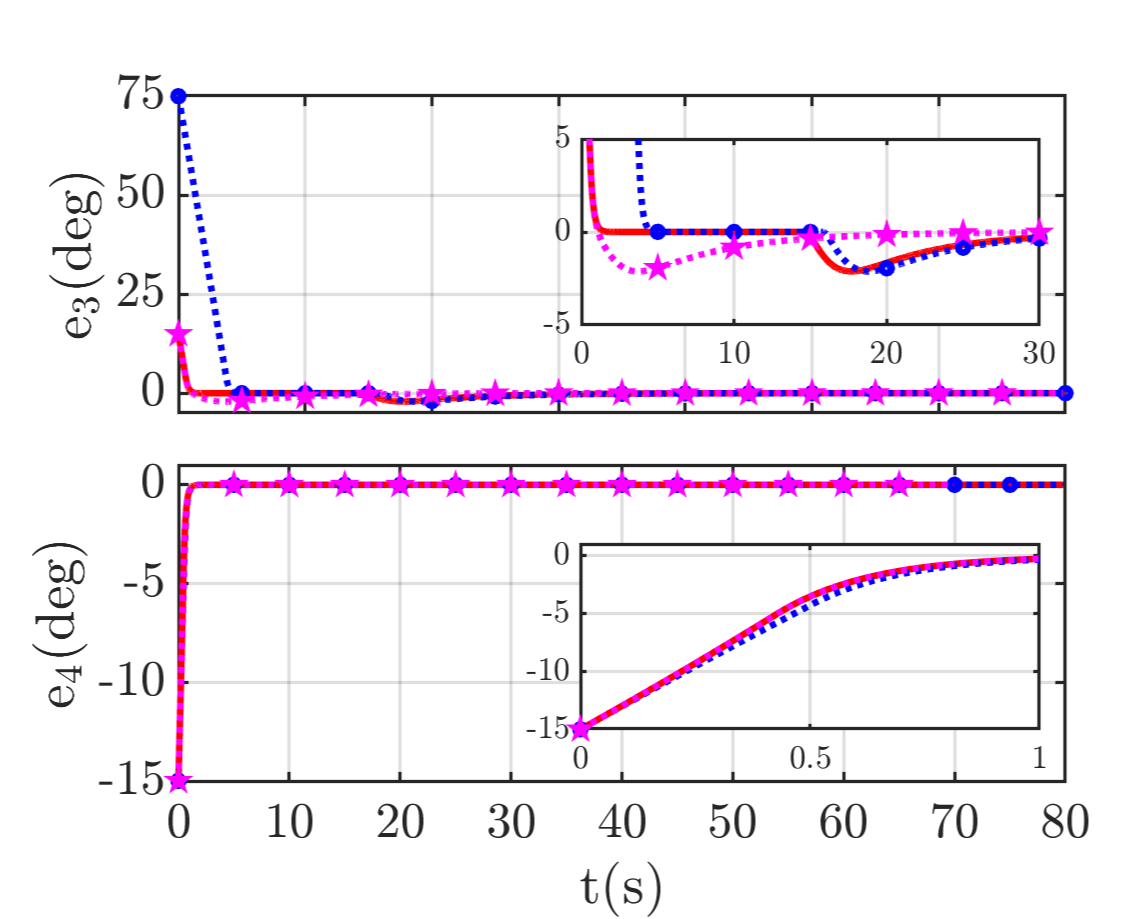}
\caption{Errors $e_3$ and $e_4$}\label{fig:PIP_case_e_e3_e4}
\end{subfigure}
\caption{Performance against constant velocity target.}
\label{fig:PIP_case_e}
\end{figure*}
It can be observed from \Cref{fig:PIP_case_e_xyz} that the interceptor successfully captures the non-maneuvering target while exhibiting similar behavior in the range, errors, and lateral acceleration profiles as depicted in previous scenarios. This further bolsters the claim that the proposed guidance law remains effective for constant velocity moving targets. It is worth mentioning that the interceptor never violates its FOV bound throughout the engagement, which can be seen from \Cref{fig:PIP_case_e_sigma}. 

\subsection{Effect of Tunable Parameters on Performance of Proposed Guidance Laws}
\begin{figure*}[!htpb]
\centering
\begin{subfigure}{0.32\linewidth}
\includegraphics[width=\linewidth]{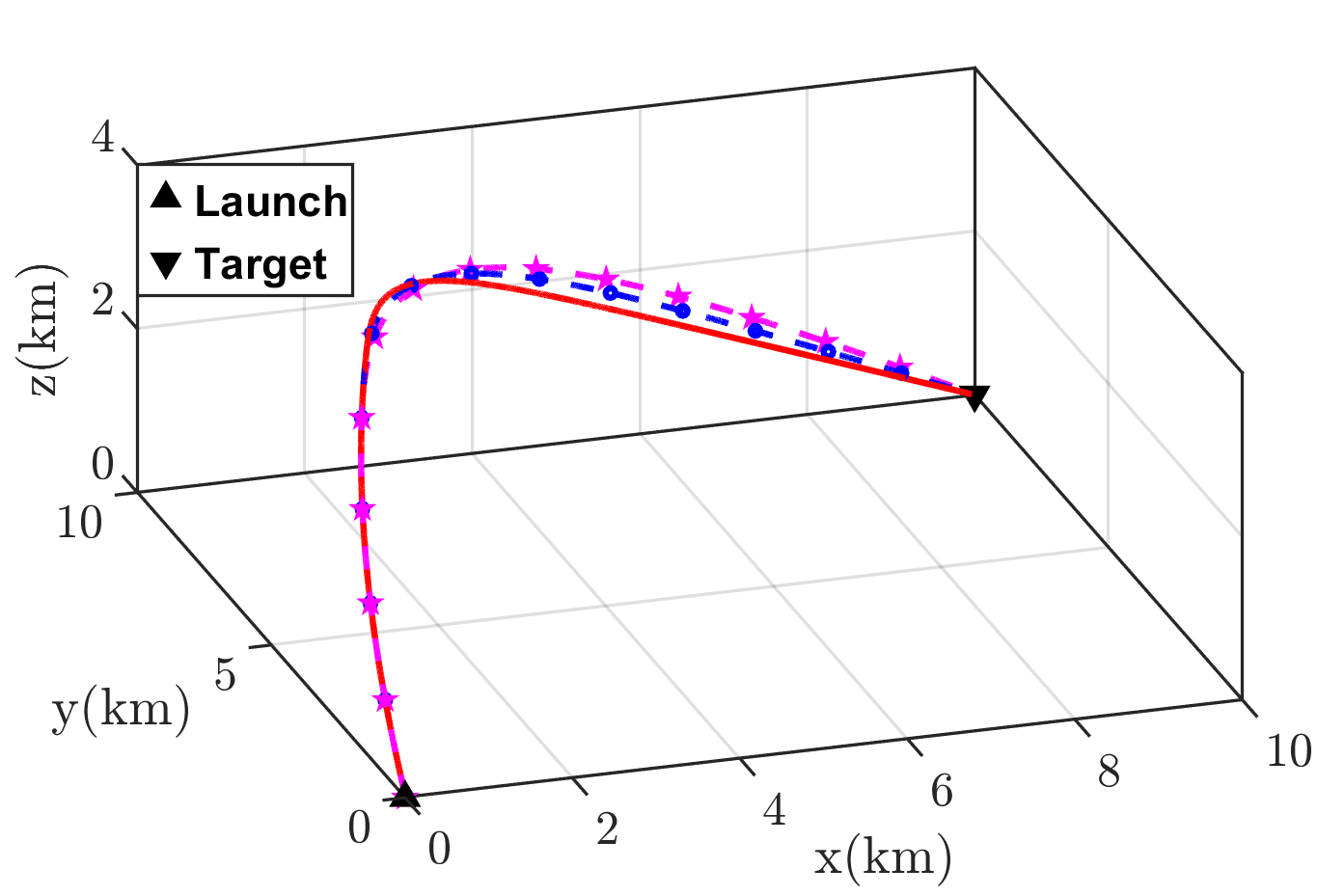}
\caption{Trajectory}\label{fig:Phi_effect_case_e_xyz}
\end{subfigure}
\begin{subfigure}{0.32\linewidth}
\includegraphics[width=\linewidth]{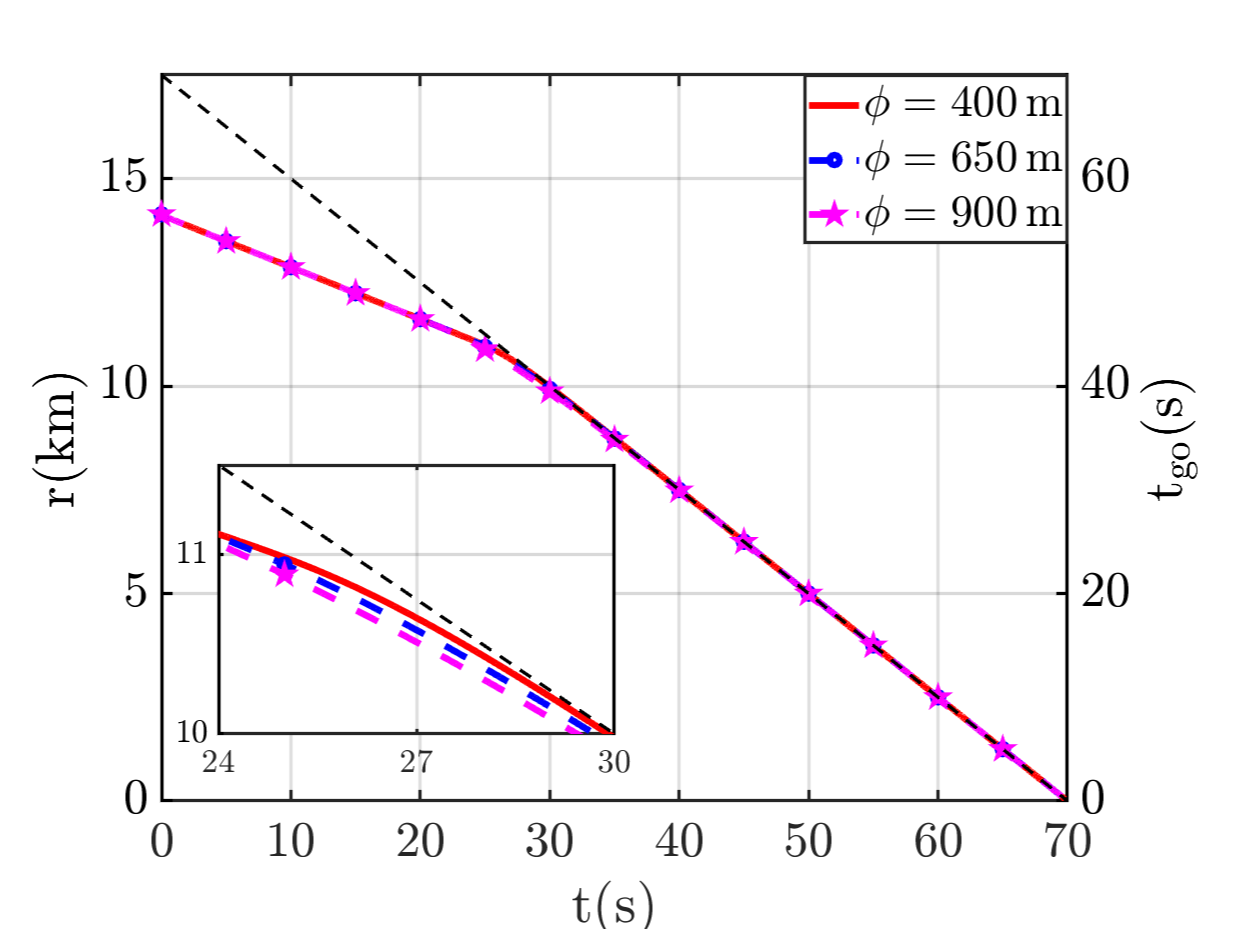}
\caption{Range and Time-to-go}\label{fig:Phi_effect_case_e_r_tgo}
\end{subfigure}
\begin{subfigure}{0.32\linewidth}
\includegraphics[width=\linewidth]{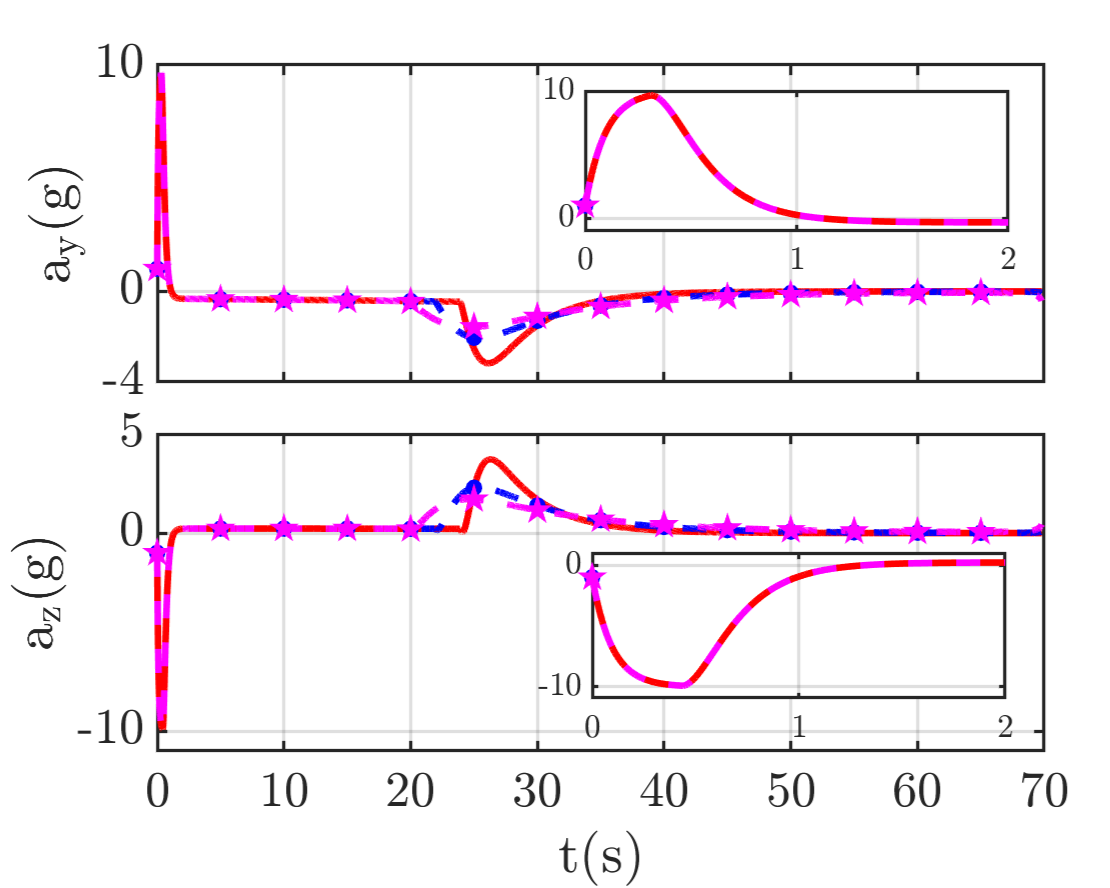}
\caption{Lateral Acceleration}\label{fig:Phi_effect_case_e_ay_az}
\end{subfigure}
\begin{subfigure}{0.32\linewidth}
\includegraphics[width=\linewidth]{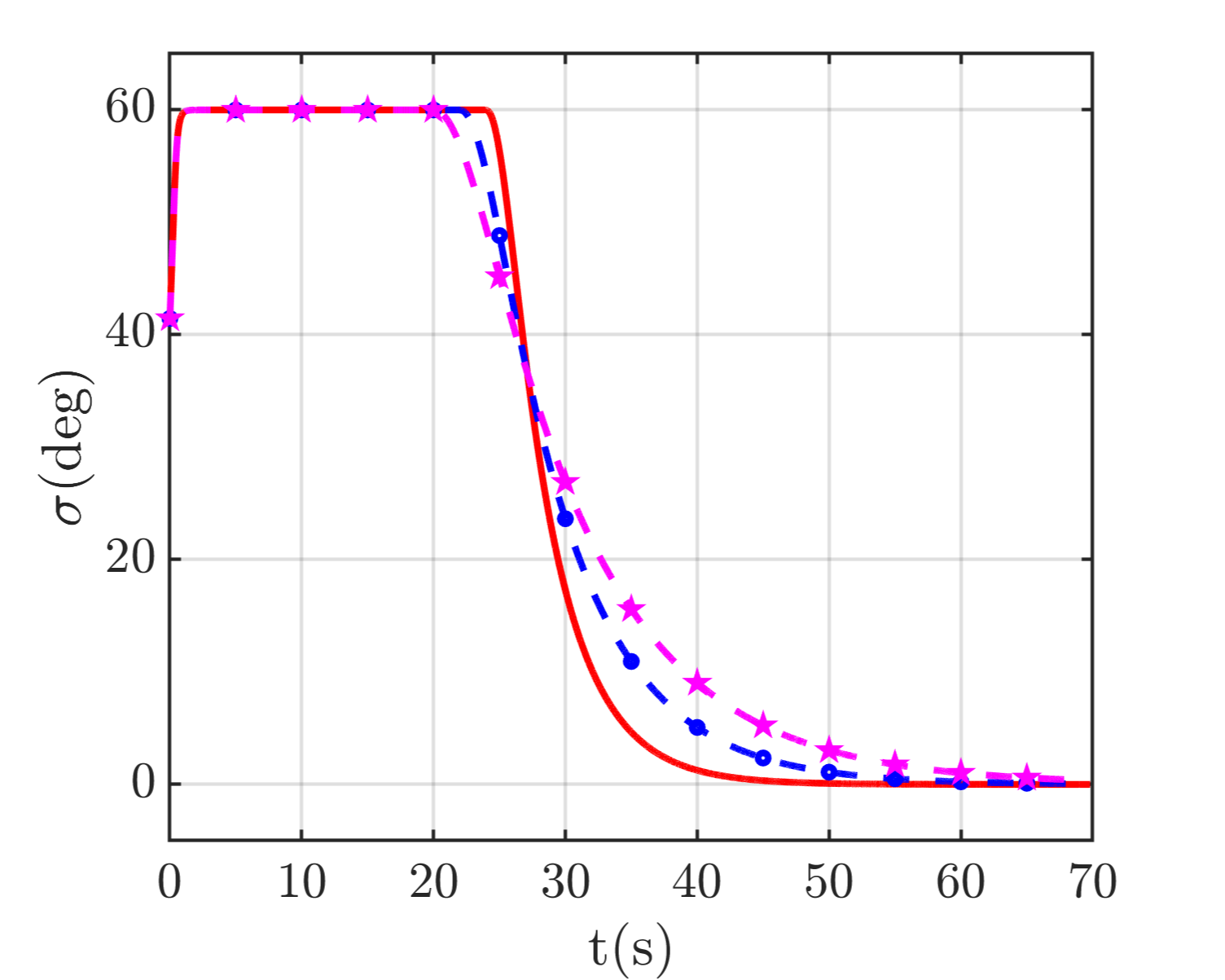}
\caption{Effective Lead Angle}\label{fig:Phi_effect_case_e_sigma}
\end{subfigure}
\begin{subfigure}{0.32\linewidth}
\includegraphics[width=\linewidth]{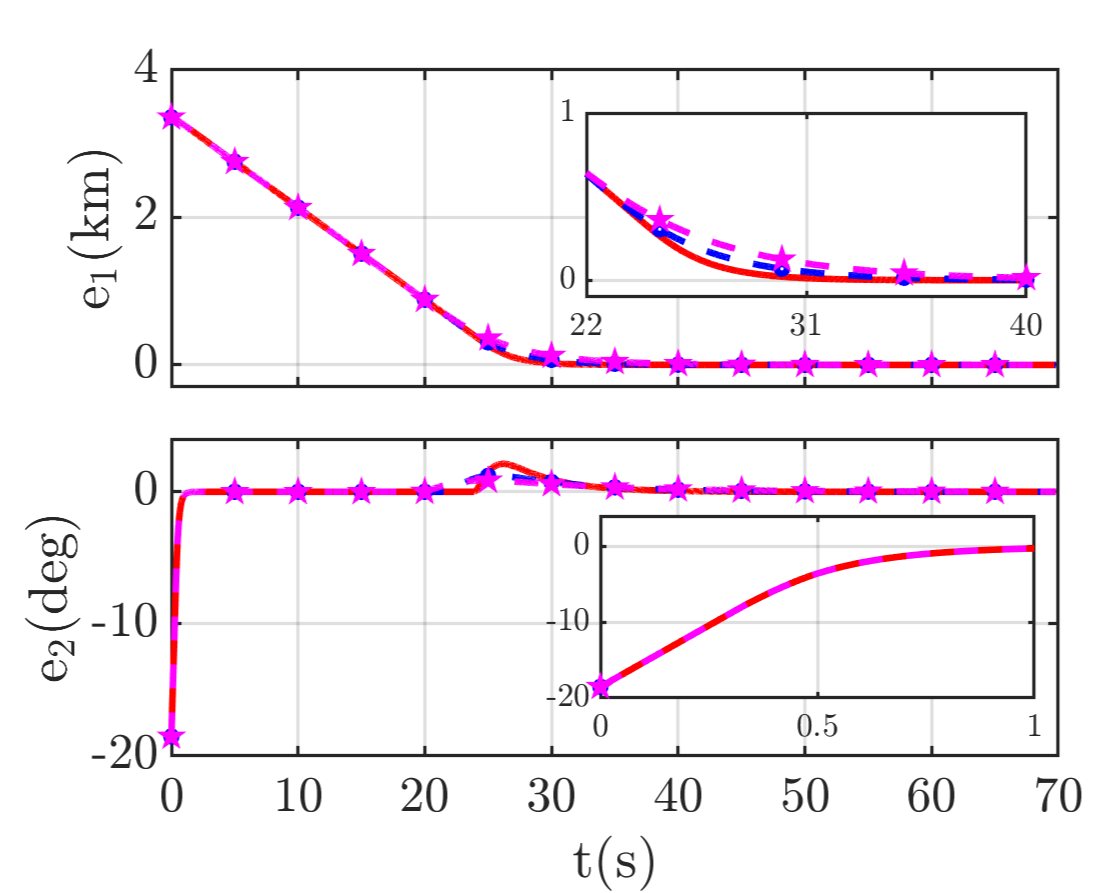}
\caption{Errors $e_1$ and $e_2$}\label{fig:Phi_effect_case_e_e1_e2}
\end{subfigure}
\begin{subfigure}{0.32\linewidth}
\includegraphics[width=\linewidth]{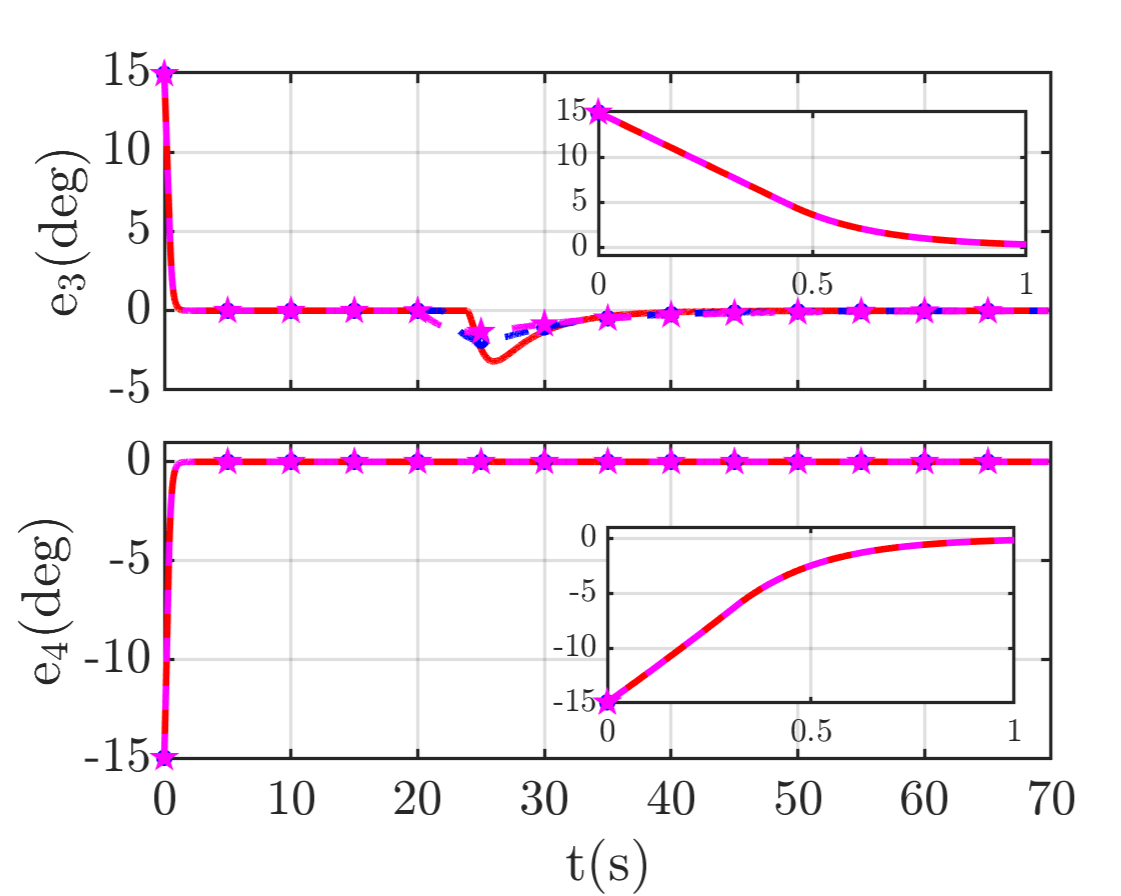}
\caption{Errors $e_3$ and $e_4$}\label{fig:Phi_effect_case_e_e3_e4}
\end{subfigure}
\caption{Effect of Variation in Parameter $\phi$.}
\label{fig:Phi_effect_case_e}
\end{figure*}

\begin{figure*}[!htpb]
\centering
\begin{subfigure}{0.32\linewidth}
\includegraphics[width=\linewidth]{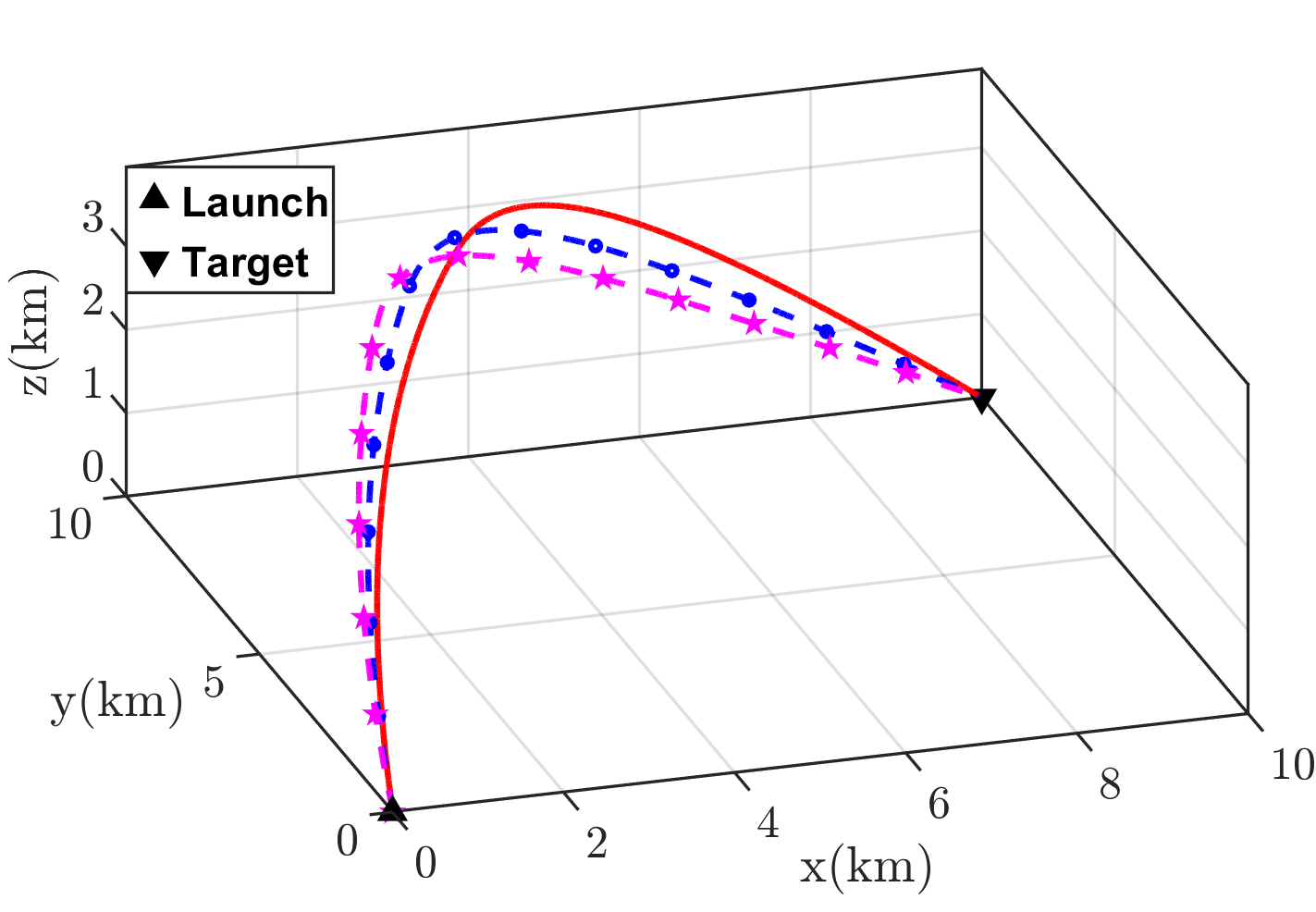}
\caption{Trajectory}\label{fig:k1_effect_case_e_xyz}
\end{subfigure}
\begin{subfigure}{0.32\linewidth}
\includegraphics[width=\linewidth]{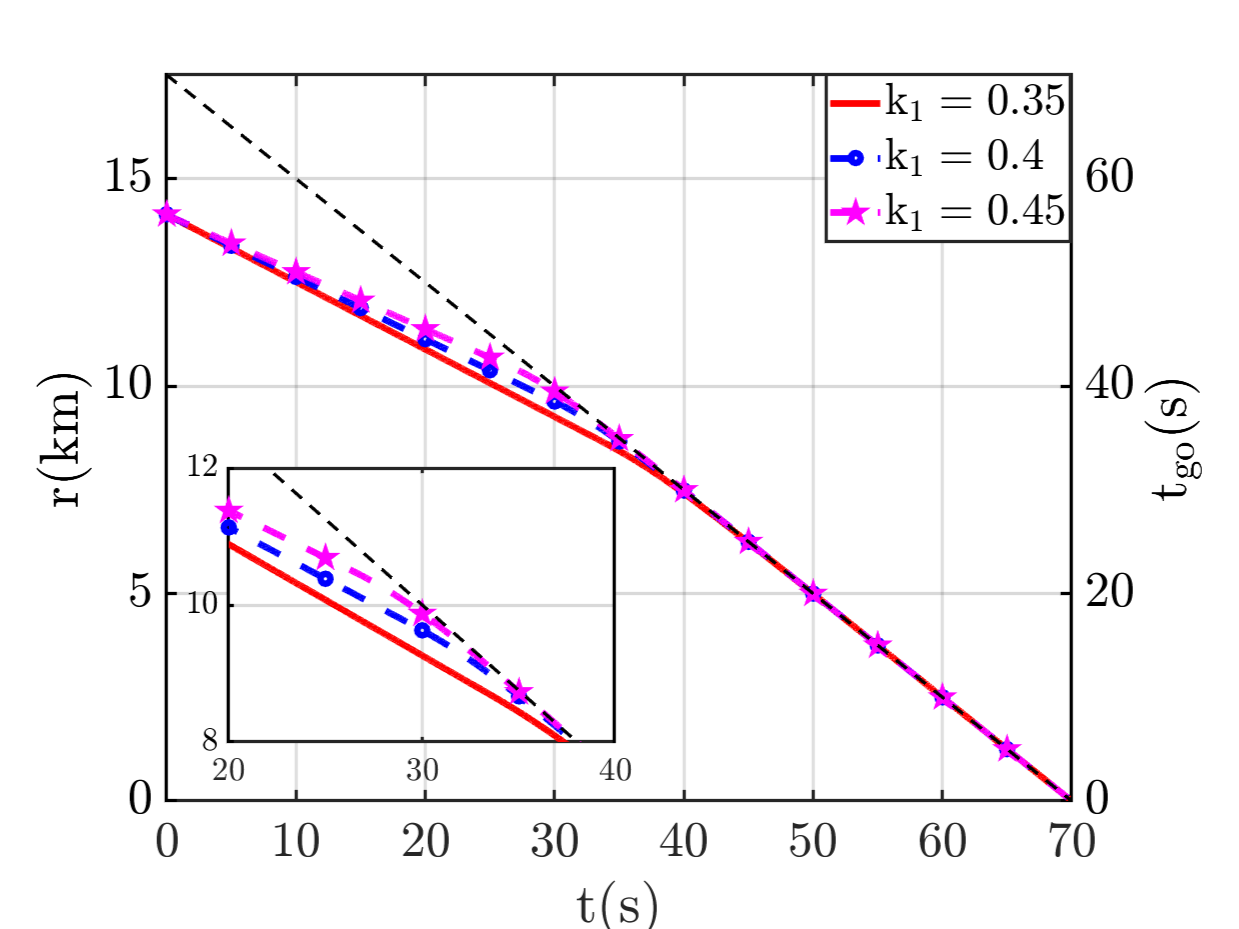}
\caption{Range and Time-to-go}\label{fig:k1_effect_case_e_r_tgo}
\end{subfigure}
\begin{subfigure}{0.32\linewidth}
\includegraphics[width=\linewidth]{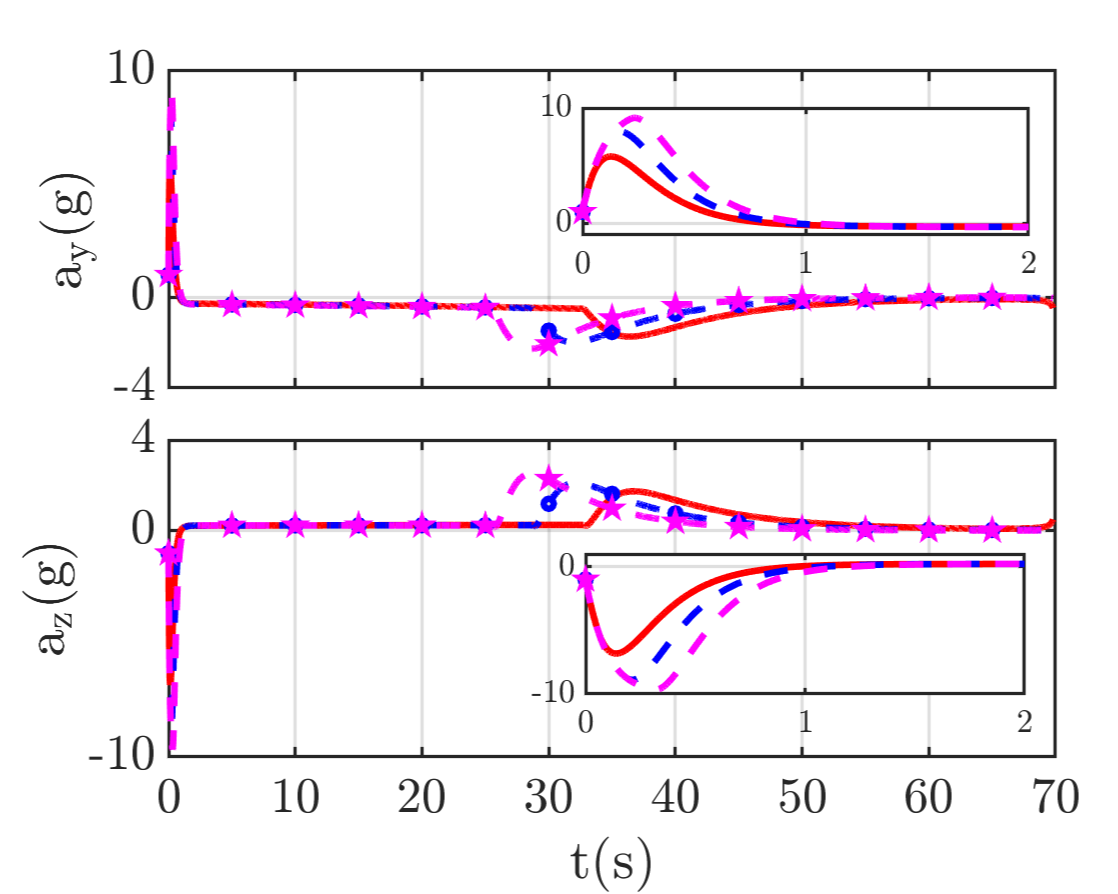}
\caption{Lateral Acceleration}\label{fig:k1_effect_case_e_ay_az}
\end{subfigure}
\begin{subfigure}{0.32\linewidth}
\includegraphics[width=\linewidth]{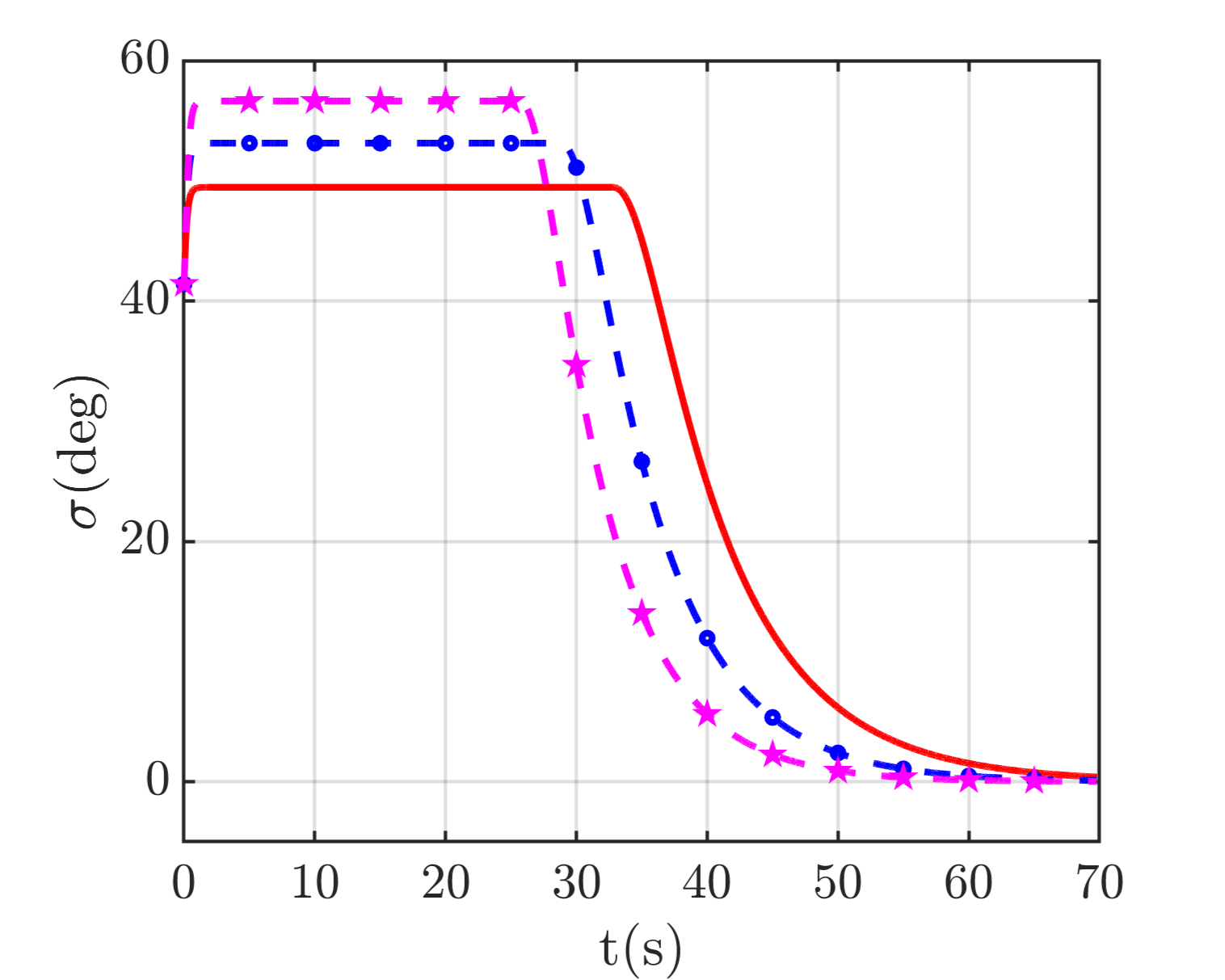}
\caption{Effective Lead Angle}\label{fig:k1_effect_case_e_sigma}
\end{subfigure}
\begin{subfigure}{0.32\linewidth}
\includegraphics[width=\linewidth]{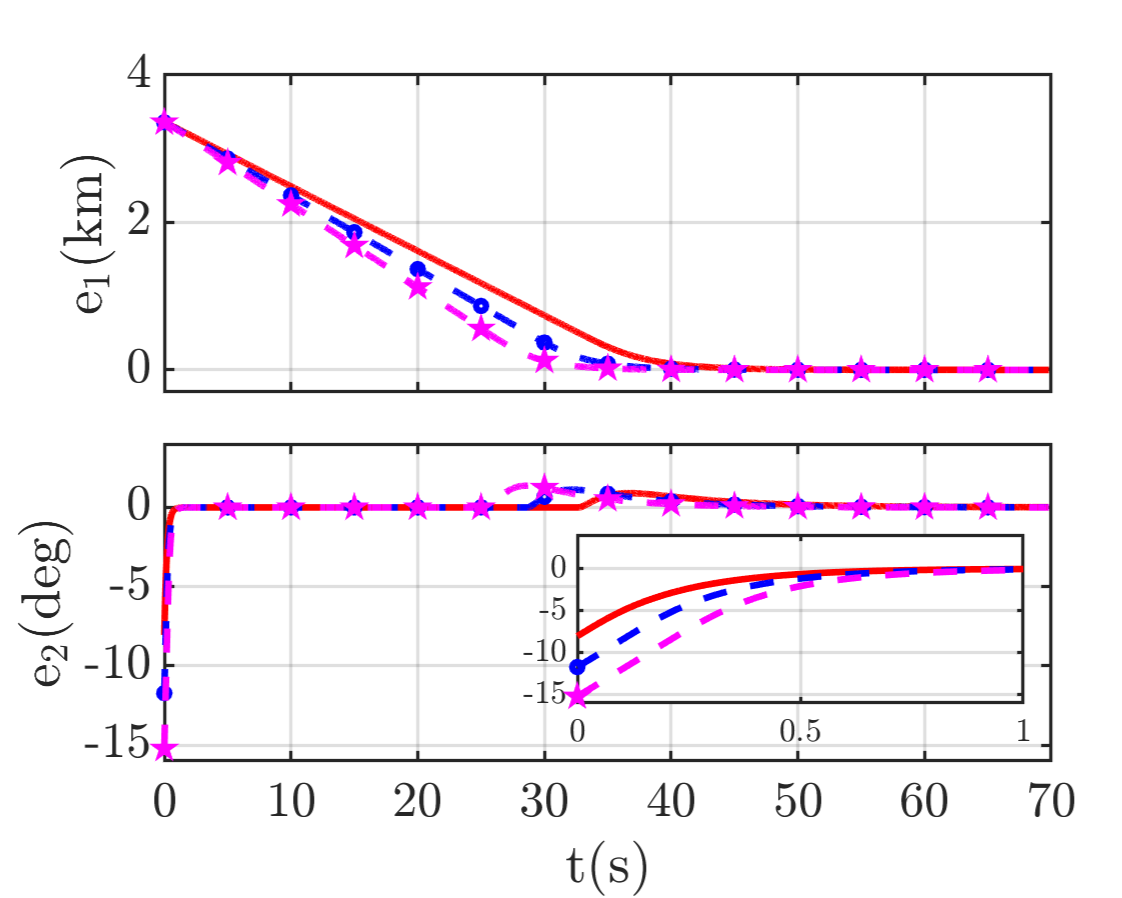}
\caption{Errors $e_1$ and $e_2$}\label{fig:k1_effect_case_e_e1_e2}
\end{subfigure}
\begin{subfigure}{0.32\linewidth}
\includegraphics[width=\linewidth]{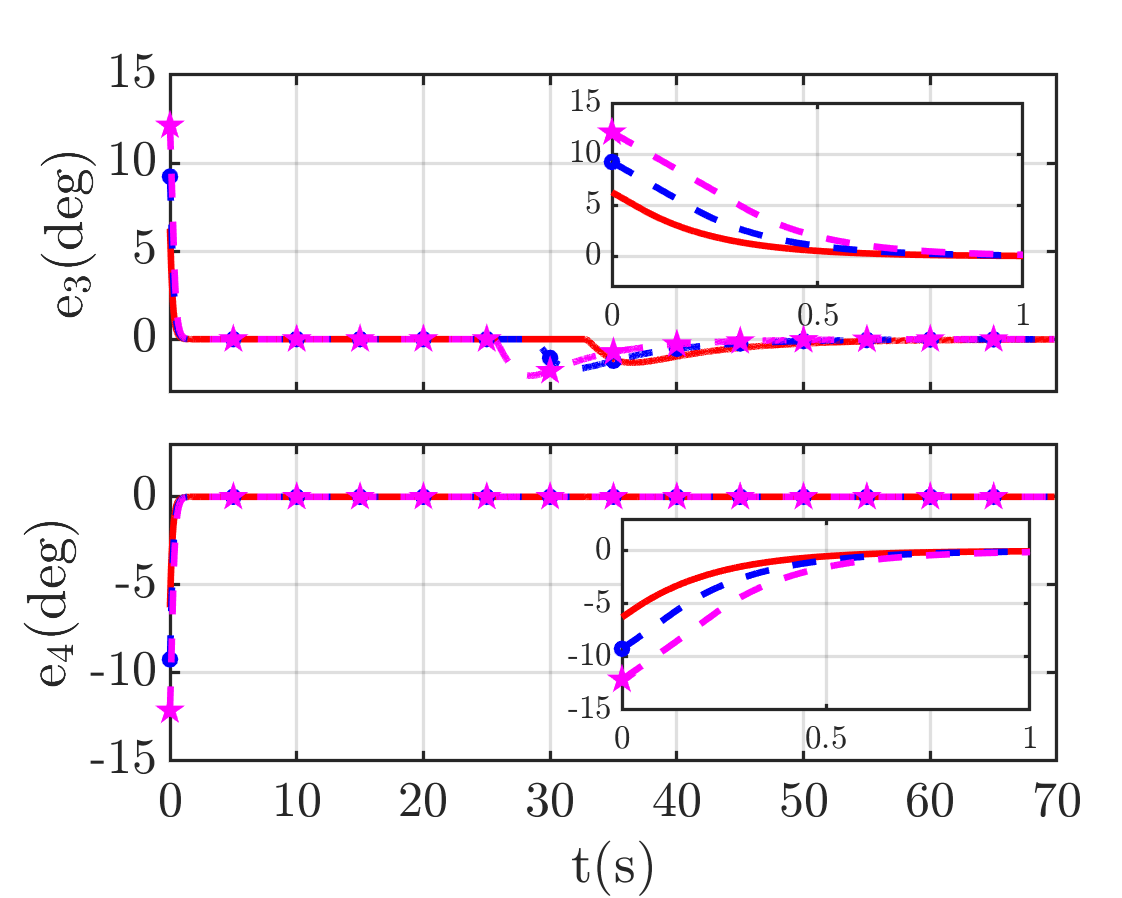}
\caption{Errors $e_3$ and $e_4$}\label{fig:k1_effect_case_e_e3_e4}
\end{subfigure}
\caption{Effect of Variation in Parameter $k_1$.}
\label{fig:k1_effect_case_e}
\end{figure*}

\begin{figure*}[!htpb]
\centering
\begin{subfigure}{0.32\linewidth}
\includegraphics[width=\linewidth]{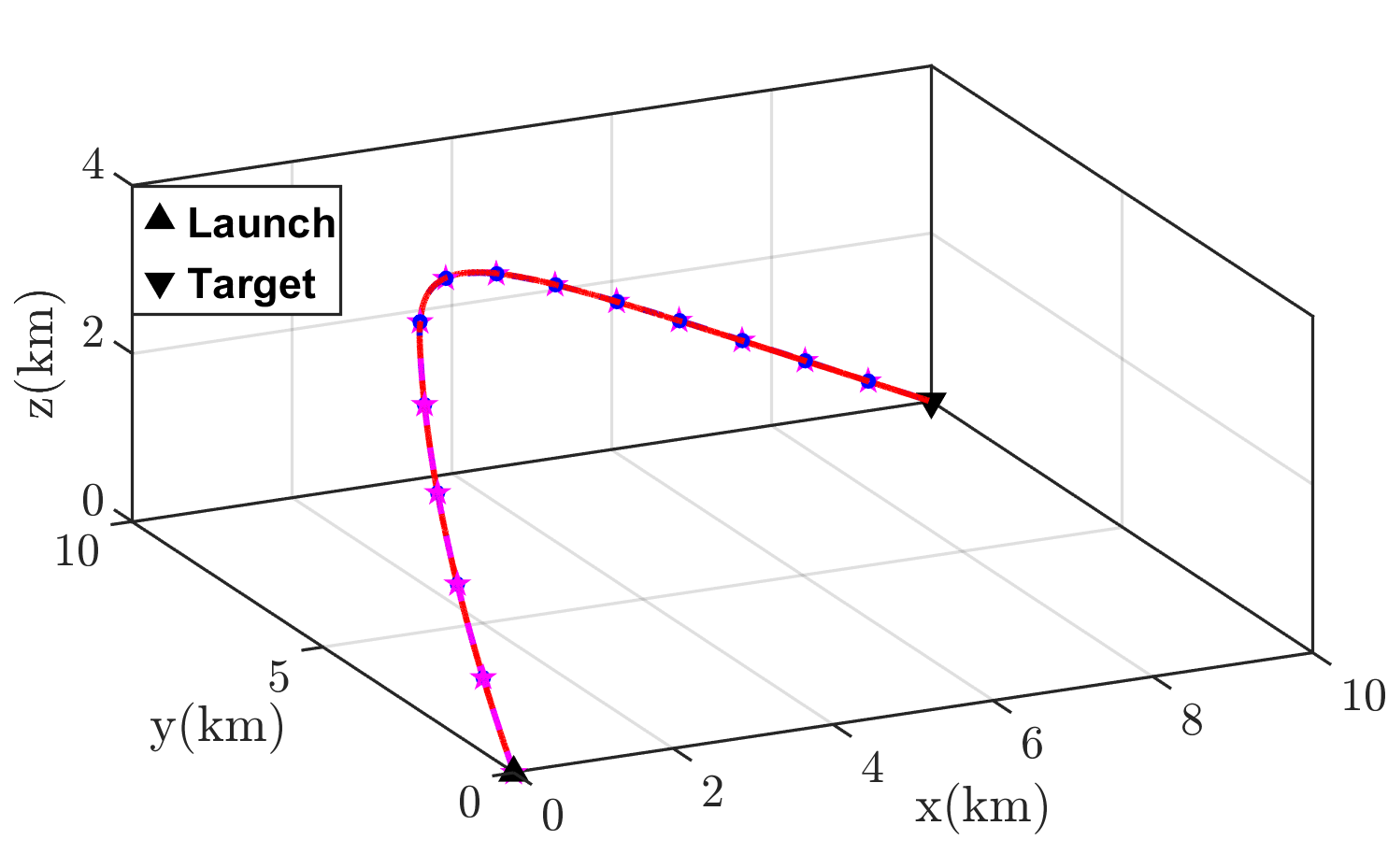}
\caption{Trajectory}\label{fig:k3_effect_case_e_xyz}
\end{subfigure}
\begin{subfigure}{0.32\linewidth}
\includegraphics[width=\linewidth]{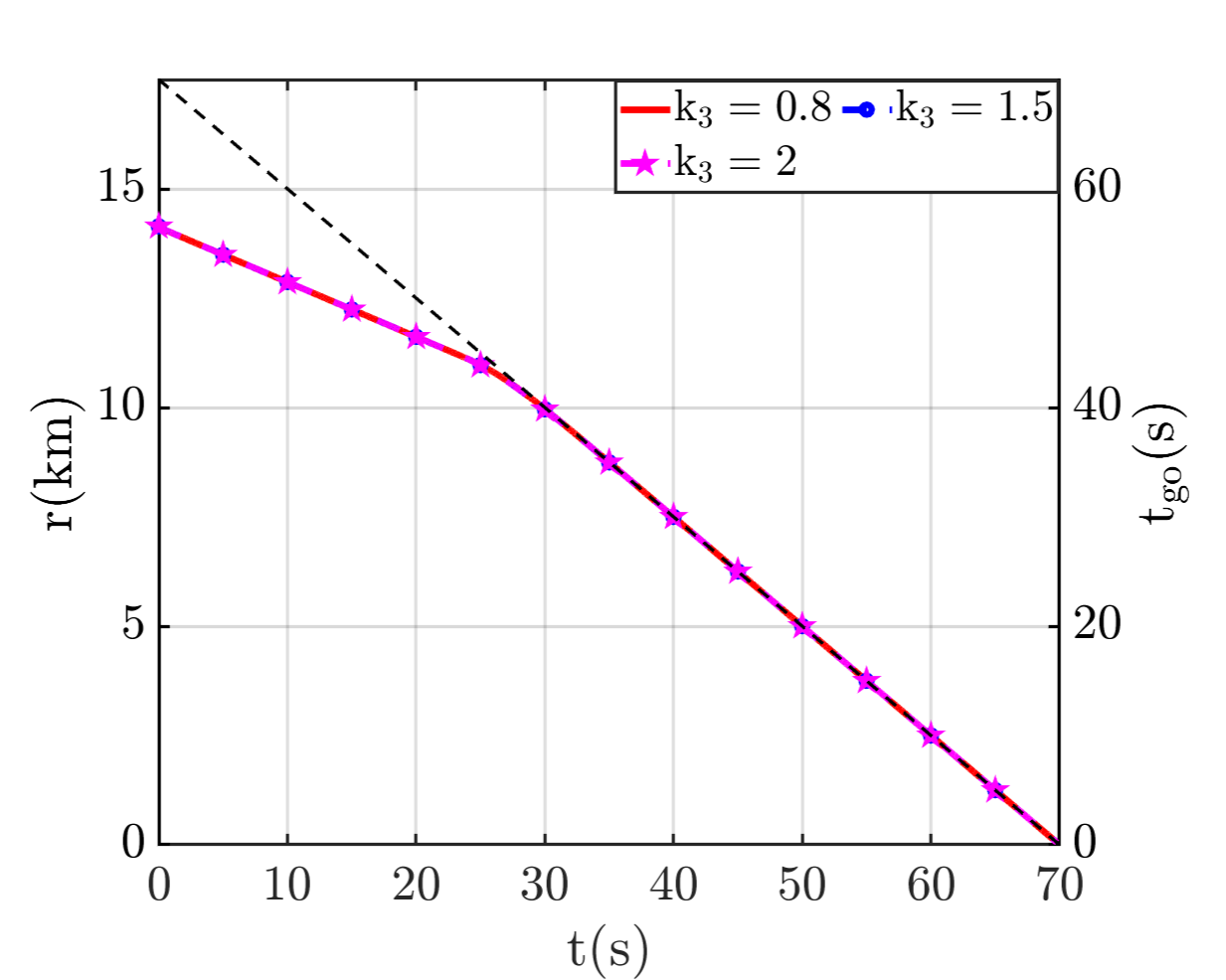}
\caption{Range and Time-to-go}\label{fig:k3_effect_case_e_r_tgo}
\end{subfigure}
\begin{subfigure}{0.32\linewidth}
\includegraphics[width=\linewidth]{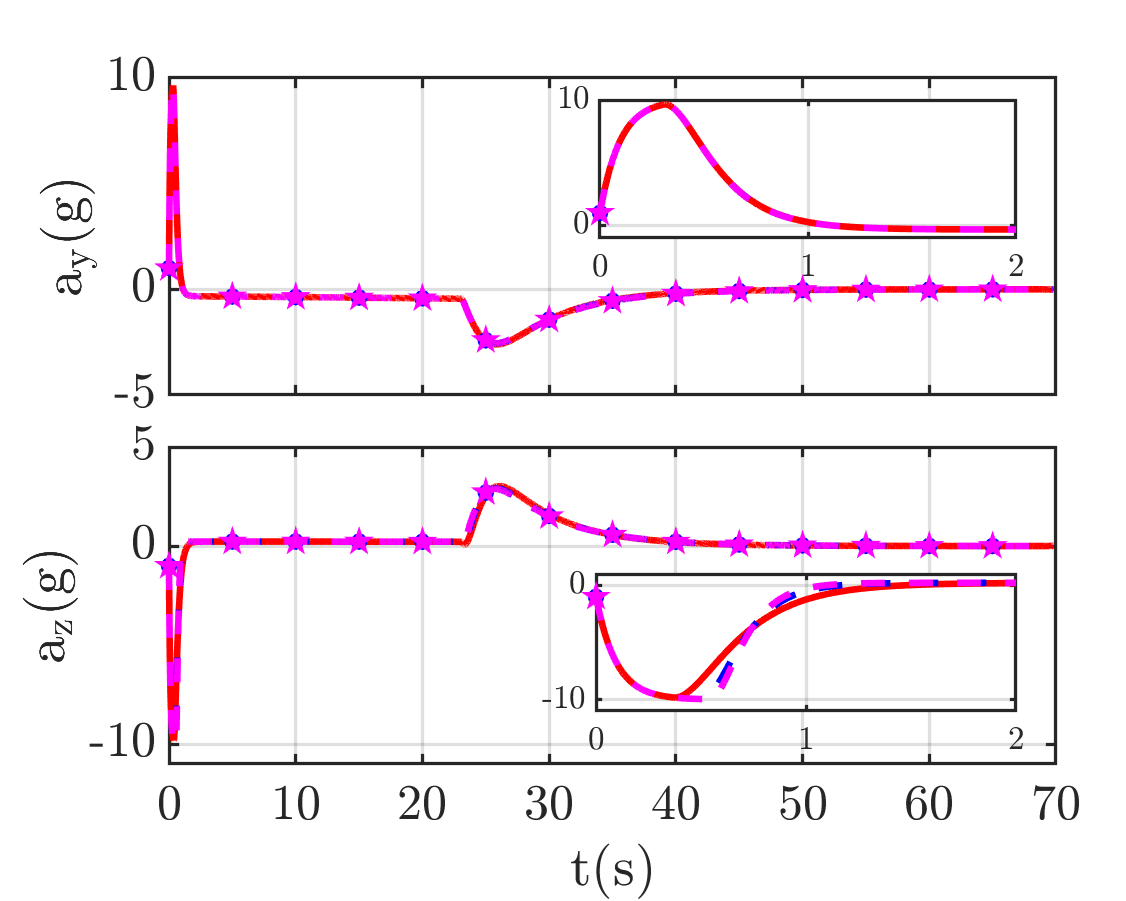}
\caption{Lateral Acceleration}\label{fig:k3_effect_case_e_ay_az}
\end{subfigure}
\begin{subfigure}{0.32\linewidth}
\includegraphics[width=\linewidth]{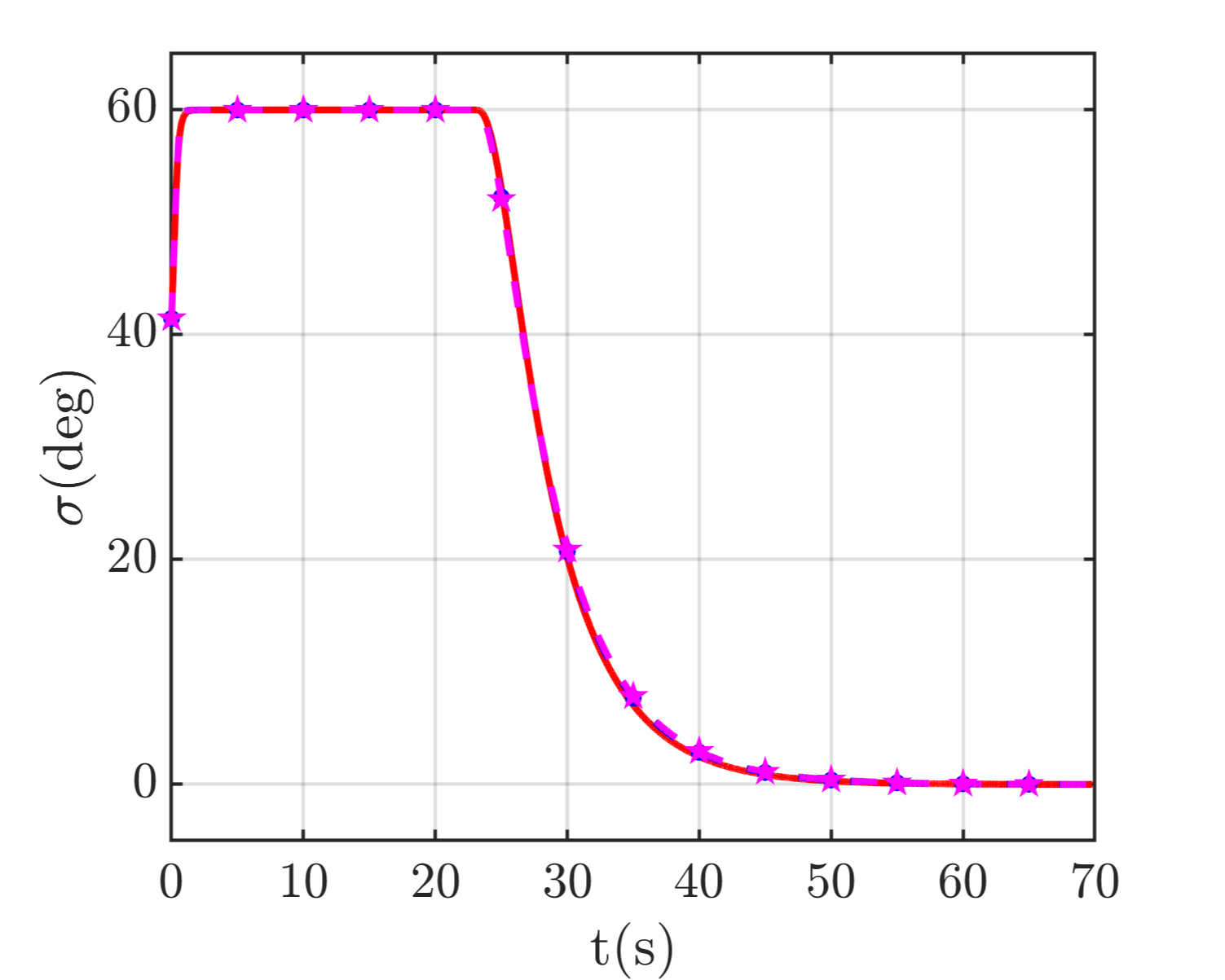}
\caption{Effective Lead Angle}\label{fig:k3_effect_case_e_sigma}
\end{subfigure}
\begin{subfigure}{0.32\linewidth}
\includegraphics[width=\linewidth]{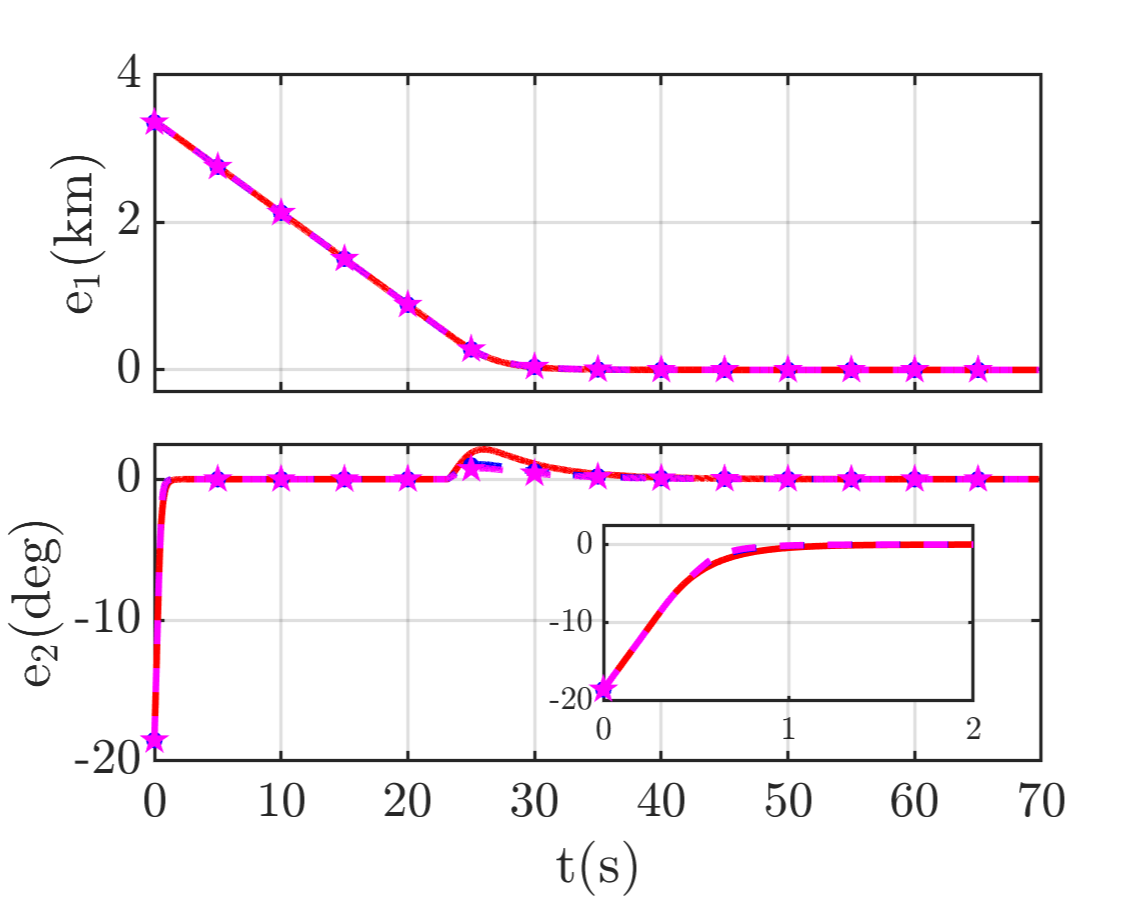}
\caption{Errors $e_1$ and $e_2$}\label{fig:k3_effect_case_e_e1_e2}
\end{subfigure}
\begin{subfigure}{0.32\linewidth}
\includegraphics[width=\linewidth]{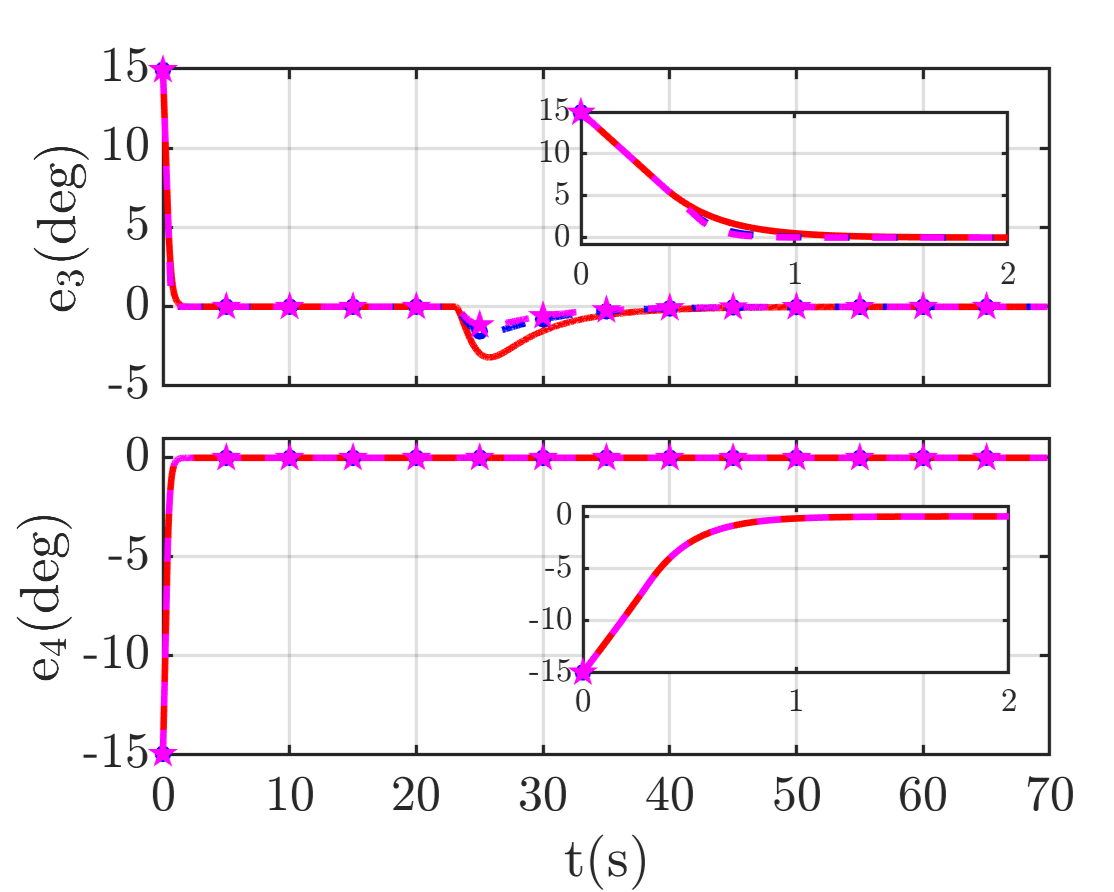}
    \caption{Errors $e_3$ and $e_4$}\label{fig:k3_effect_case_e_e3_e4}
\end{subfigure}
\caption{Effect of Variation in Parameter $k_3$.}
\label{fig:k3_effect_case_e}
\end{figure*}

\begin{figure}[!ht]
\centering
\includegraphics[width=\linewidth]{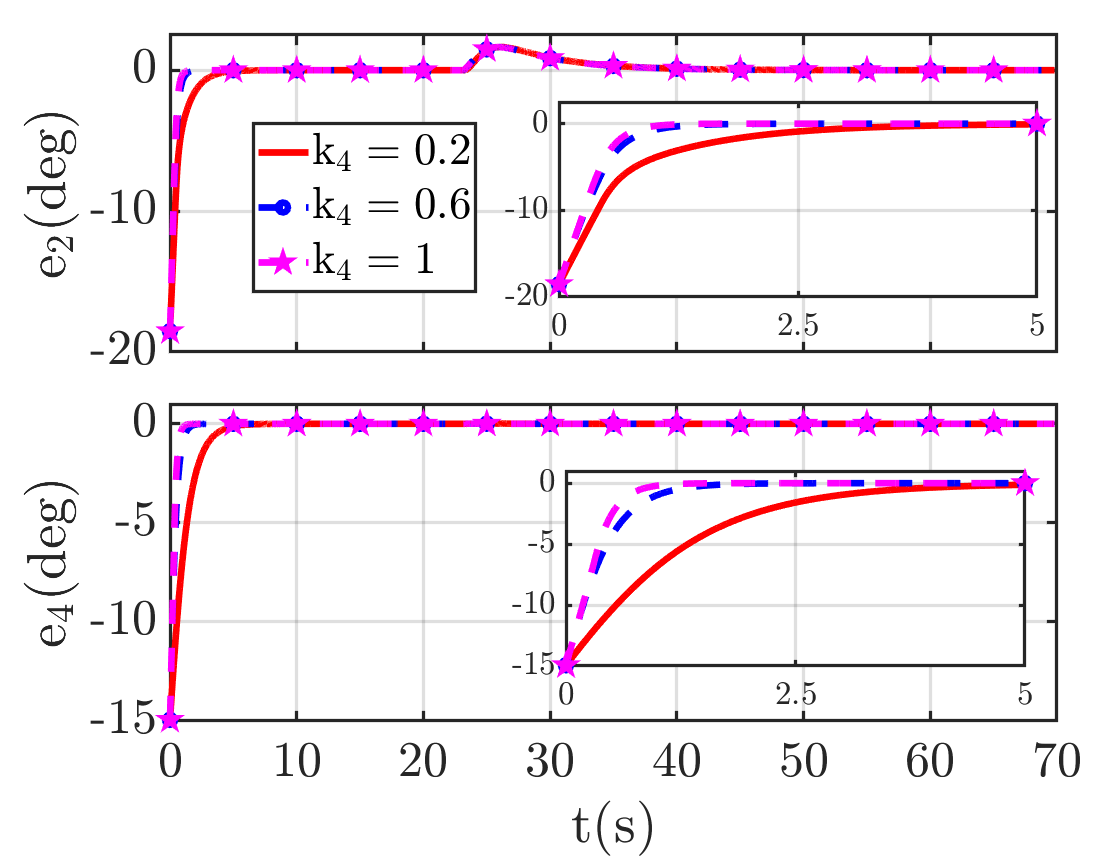}
\label{fig:k4_effect_case_e_e2_e4}
\caption{Effect of  Variation in Parameter $k_4$.}
\label{fig:k4_effect_case_e}
\end{figure}

In this subsection, we discuss the effect of the tunable parameters, namely, $\phi$, $k_1$, $k_2$, $k_3$, and $k_4$ on the performance of the proposed guidance laws. For the simulations, we have chosen an impact time and initial heading angles of $(t_{\rm f}, \theta_{\rm m}(0), \psi_{\rm m}(0)) = (70\,\rm s, -30^\circ, 30^\circ)$, respectively. The effects of varying the parameter $\phi$ are shown in \Cref{fig:Phi_effect_case_e} and consecutively \Cref{fig:k1_effect_case_e,fig:k3_effect_case_e} depicts the effect of variations in $k_1$ and $k_3$, respectively. \Cref{fig:Phi_effect_case_e_sigma} indicates that the value of $\phi$ affects the steepness of the switching in the lead angle, which in turn requires more lateral acceleration as seen from \Cref{fig:Phi_effect_case_e_ay_az}. An increase in the value of $\phi$ decreases the peak in the heading errors, $e_3$, and $e_4$ when the interceptor returns to the collision course, as depicted in \Cref{fig:Phi_effect_case_e_e3_e4}. \Cref{fig:k1_effect_case_e} reveals that the parameter $k_1$ plays a major role in the performance of guidance laws. A higher value of $k_1$ results in a closer proximity of the peak lead angle to the bound, as seen in \Cref{fig:k1_effect_case_e_sigma}. It also controls the slope of decrease of the error $e_1$ and the relative range $r$, as shown in \Cref{fig:k1_effect_case_e_e1_e2} and \Cref{fig:k1_effect_case_e_r_tgo}, respectively. Furthermore, an increase in the value of $k_3$ results in a faster convergence of $e_3$ and subsequently $e_2$, as evidenced by \Cref{fig:k3_effect_case_e_e3_e4}, and \Cref{fig:k3_effect_case_e_e1_e2}. The effect of the parameter $k_4$ is similar to that of $k_3$, with the key distinction being its direct impact on the rate of decrease of $e_4$, as observed in \Cref{fig:k4_effect_case_e}.

\subsection{Performance Comparison with Existing Strategy}
\begin{figure*}[!htpb]
\centering
\begin{subfigure}{0.32\linewidth}
\includegraphics[width=\linewidth]{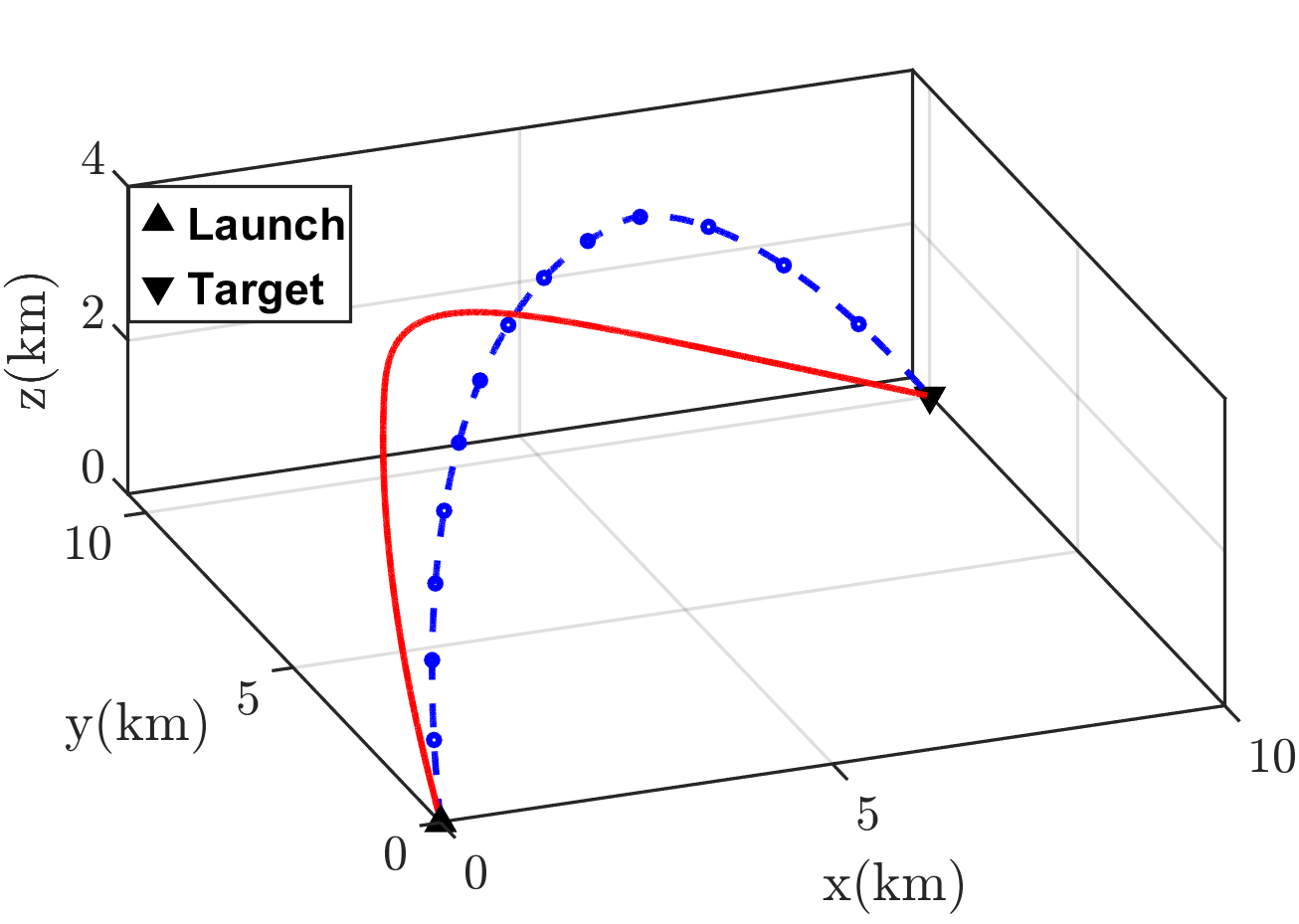}
    \caption{Trajectory}\label{fig:comp_case_e_xyz}
\end{subfigure}
\begin{subfigure}{0.32\linewidth}
\includegraphics[width=\linewidth]{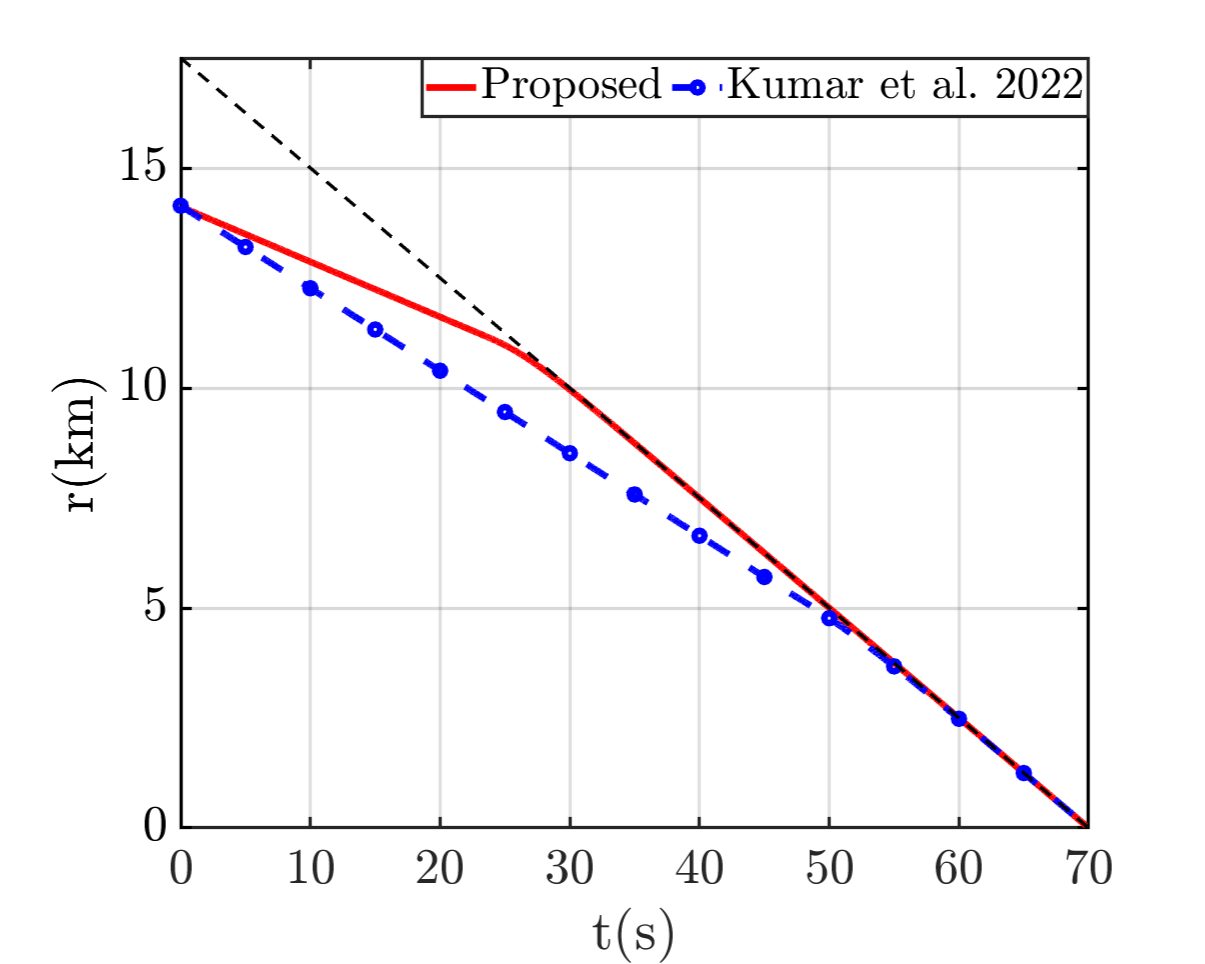}
\caption{Range}\label{fig:comp_case_e_r}
\end{subfigure}
\begin{subfigure}{0.32\linewidth}
\includegraphics[width=\linewidth]{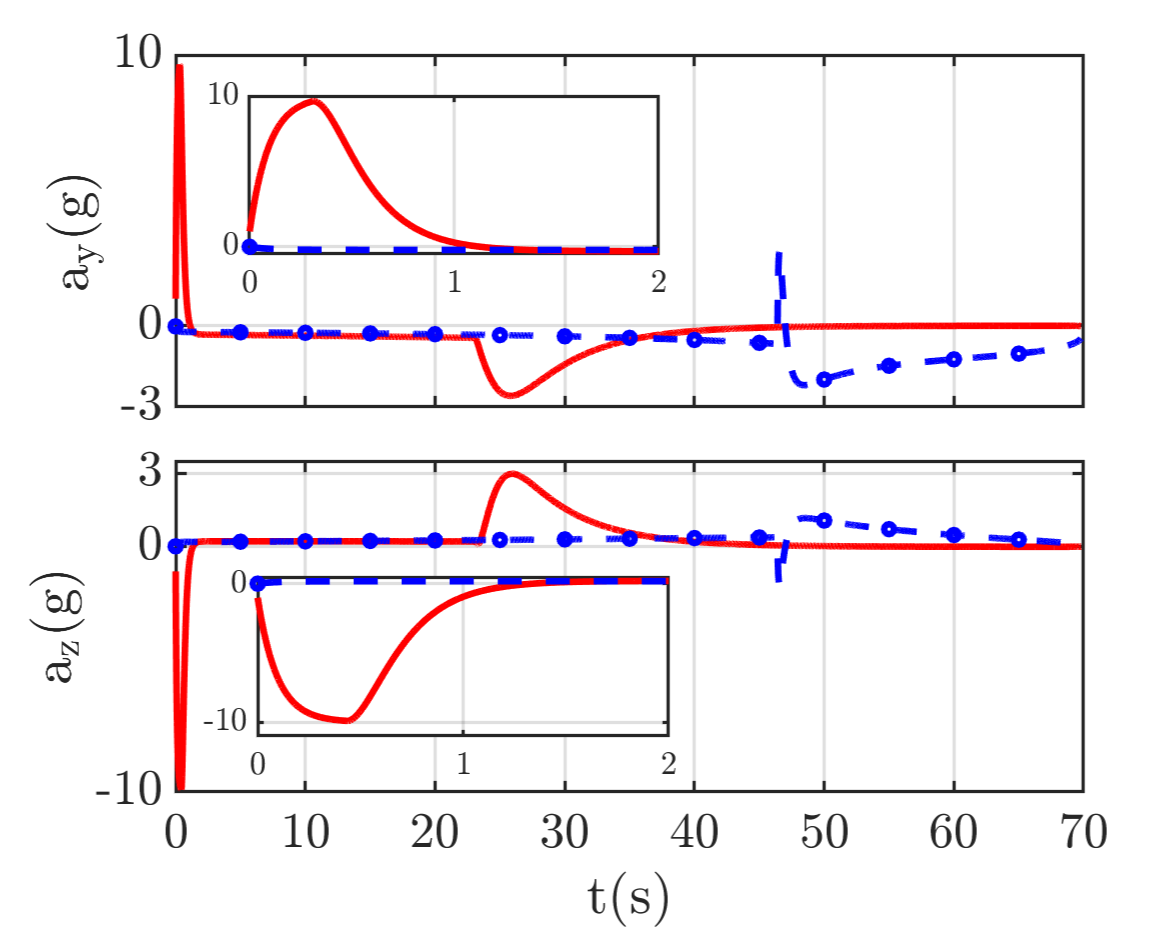}
\caption{Lateral Acceleration}\label{fig:comp_case_e_ay_az}
\end{subfigure}
\begin{subfigure}{0.32\linewidth}
\includegraphics[width=\linewidth]{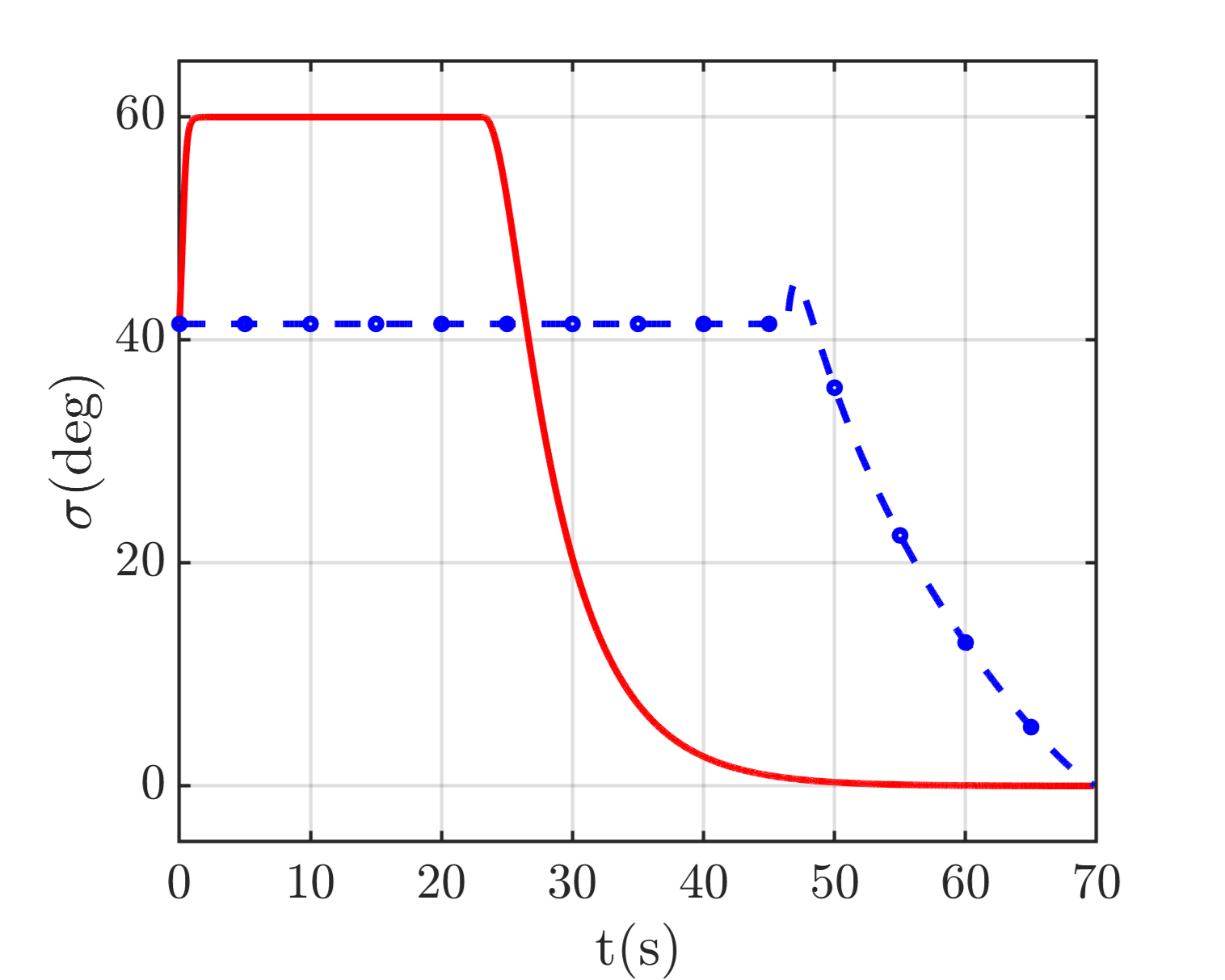}
\caption{Effective Lead Angle}\label{fig:comp_case_e_sigma}
\end{subfigure}
\begin{subfigure}{0.32\linewidth}
\includegraphics[width=\linewidth]{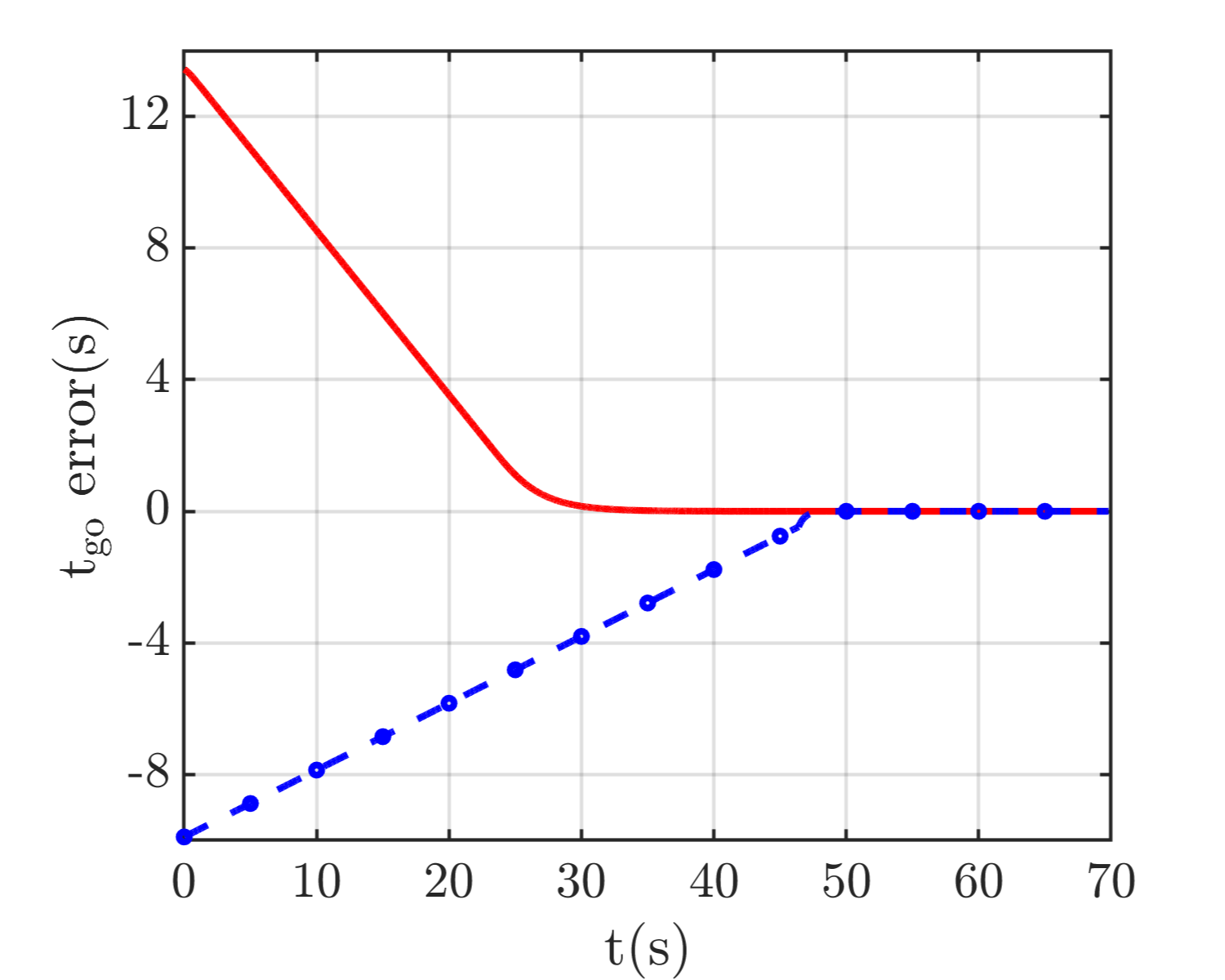}
\caption{Time-to-go Error}\label{fig:comp_case_e_t_go_error}
\end{subfigure}
\caption{Comparison with guidance law in \cite{Kumar2022BarLyap}}
\label{fig:comp_case_e}
\end{figure*}

To demonstrate the advantages of the proposed guidance law, a comparative study is performed with the strategy in \cite{Kumar2022BarLyap}. The simulation parameters for guidance law under comparison are the same as those in \cite{Kumar2022BarLyap}, $\mathcal{K}_1=1$, $\eta=1$, $\mathcal{N}=3$, $\epsilon=0.5$ and $\alpha_{\rm y} = \alpha_{\rm z}$. The desired impact time and the initial heading angles are set to $t_{\rm f}=70\,$s, $\theta_{\rm m}=-30^\circ$ and $\psi_{\rm m}=30^\circ$, respectively. The impact time chosen for this case is greater than the maximum achievable impact time for the strategy in \cite{Kumar2022BarLyap} $(62.6295\,\rm s)$, so the extension for large impact times, proposed in \cite{Kumar2022BarLyap} is used. According to the extended guidance law, the interceptor maintains the initial effective lead angle up to the switching instant, which is calculated to be $46.3948\,$s as observed in \Cref{fig:comp_case_e_sigma}. On the other hand, the proposed guidance law allows the interceptor to maintain a higher lead angle during the initial phase. This causes quicker convergence of the time-to-go error as seen in \Cref{fig:comp_case_e_t_go_error}. It can be observed from \Cref{fig:comp_case_e_xyz} that the target is intercepted successfully under both guidance laws. Note that the time-to-go error for the proposed strategy, plotted in \Cref{fig:comp_case_e_t_go_error}, is just a scaled version of the range error, $t_{\rm go}\ \rm{error} = e_1/V_{\rm m}$. The lead angle reaches zero only at target interception, indicating that the collision course is not attained for the strategy in \cite{Kumar2022BarLyap}. However, in the proposed strategy, lead angles of value close to the FOV bound can be attained by increasing the parameter $k_1$, which enables target interception at larger impact times. The achievable impact time $t_{\rm f}$ range for the proposed strategy is $\in (59.19, 86.27)$ whereas that for the extended strategy in \cite{Kumar2022BarLyap} is $\in [68.17, 70.5]$. Thus, the proposed law enables larger achievable impact time without any degradation in performance of guidance strategy.

\subsection{Performance Comparison for Different Virtual Inputs}
\label{Subsec:all_cases_one_t_f}
In this subsection, the simulation results for all the cases listed in \Cref{tab:achieve_td} with a desired impact time of $70\,$s and initial heading angles of $\theta_{\rm m}=-30^\circ$ and $\psi_{\rm m}=30^\circ$ are presented. The value of $k_1$ was fixed to be $0.001$ lower than the bound given by \eqref{eq:k1_FOV_limit}, and $\phi$ was set as $200$. The remaining parameters are the same as listed in \Cref{tab:Sim_params}. The minimum and maximum achievable impact times are calculated using the relations noted in the previous section, and the same are tabulated in \Cref{tab:achieve_td}. One can verify that the desired impact time of $70\,$s is indeed within the feasible range. To ensure that the trajectory stays within the first octant of the LOS frame, the sign of the desired heading angles is selected in accordance with \Cref{tab:Octant_signs}. 

\begin{figure*}[!htpb]
\centering
\begin{subfigure}{0.32\linewidth}
\includegraphics[width=\linewidth]{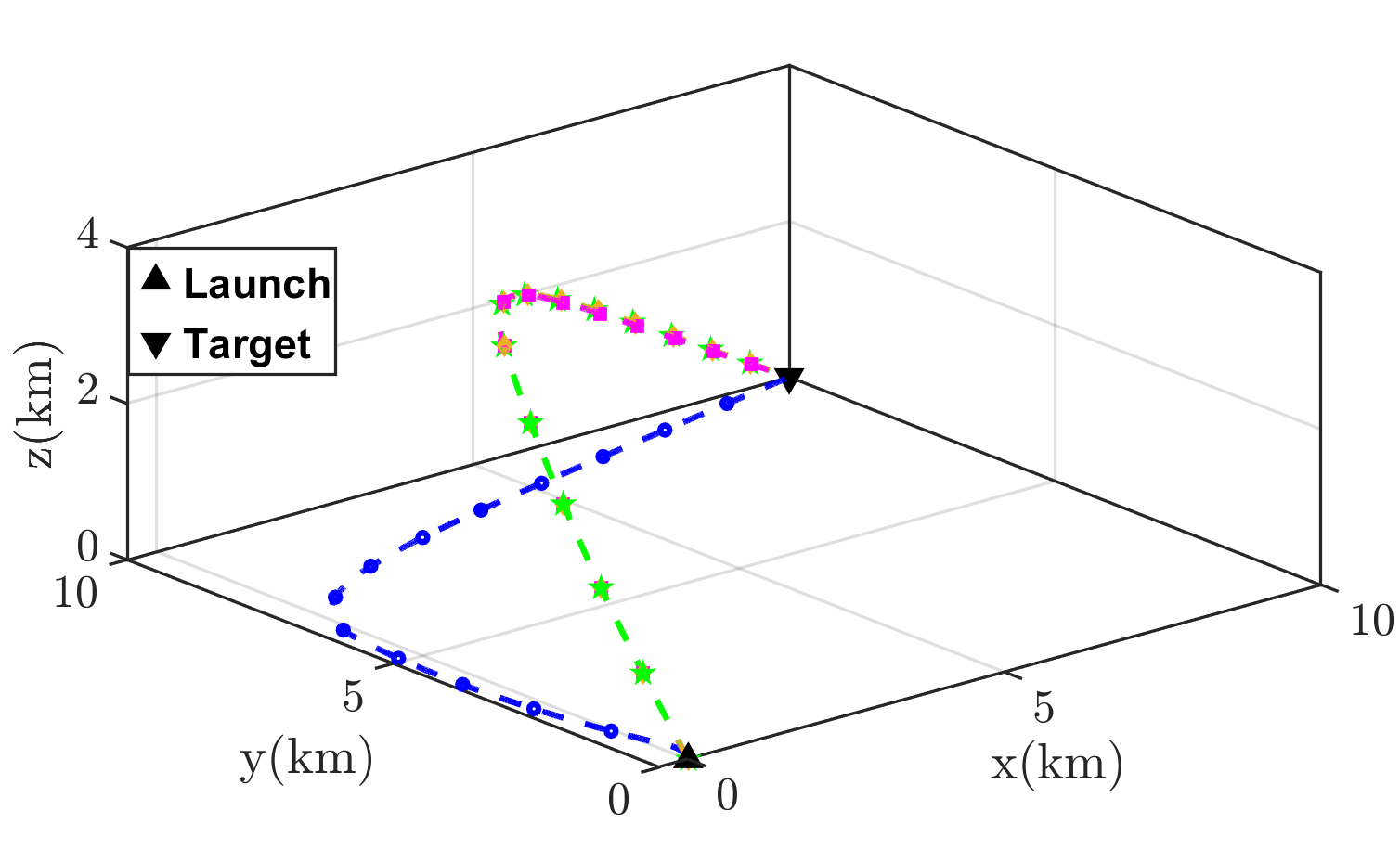}
\caption{Trajectory}\label{fig:one_tf_all_cases_xyz}
\end{subfigure}
\begin{subfigure}{0.32\linewidth}
\includegraphics[width=\linewidth]{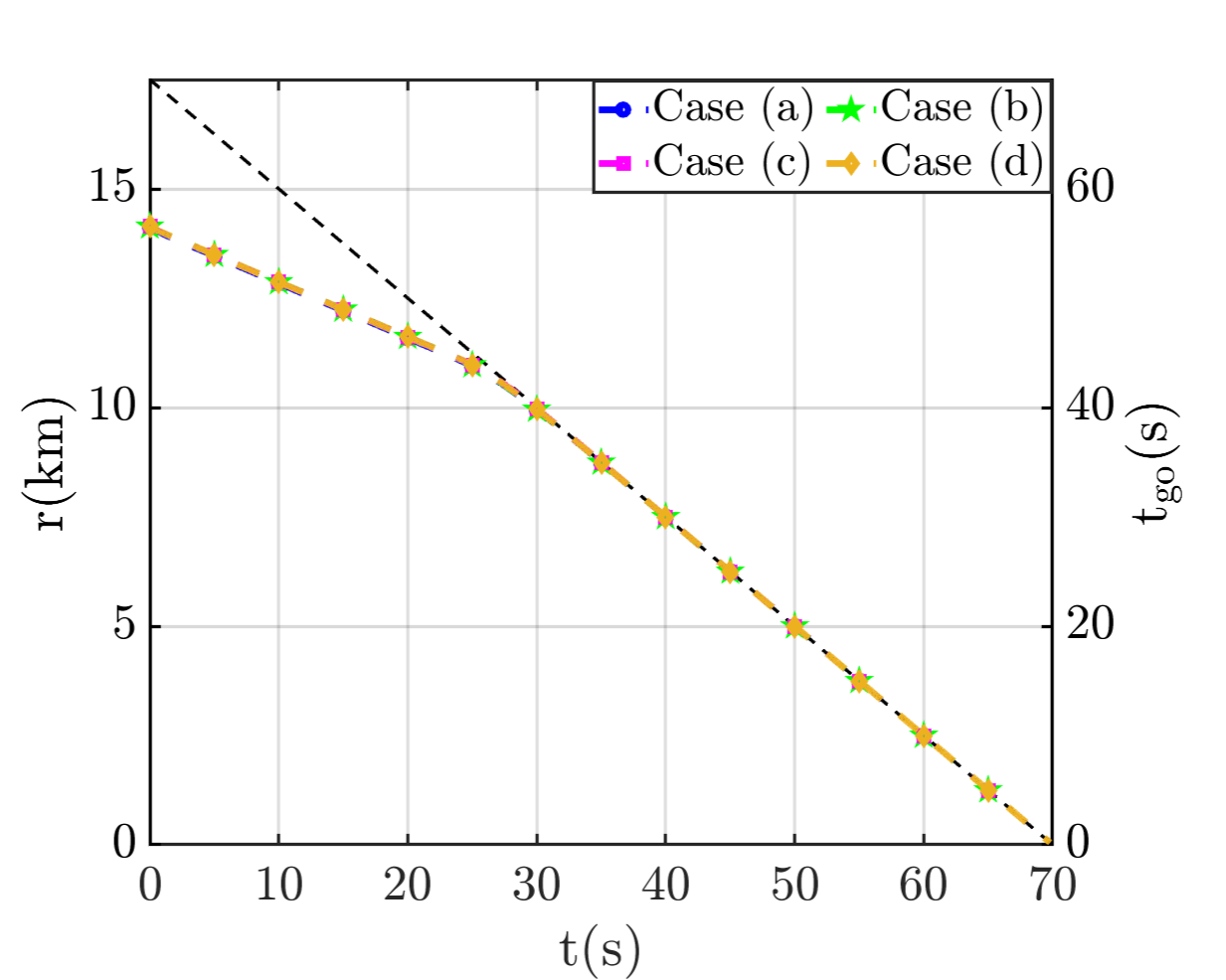}
\caption{Range and Time-to-go}\label{fig:one_tf_all_cases_r_tgo}
\end{subfigure}
\begin{subfigure}{0.32\linewidth}
\includegraphics[width=\linewidth]{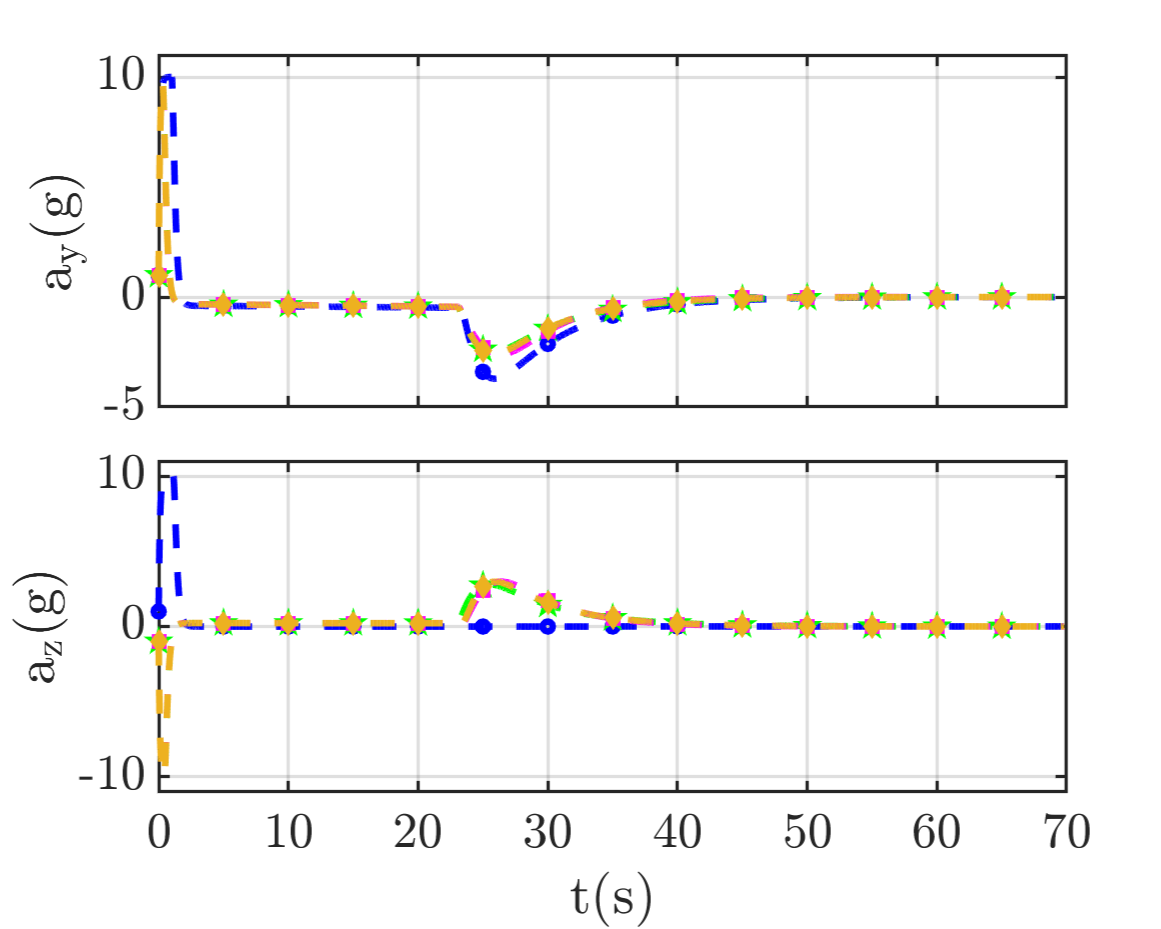}
\caption{Lateral Acceleration}\label{fig:one_tf_all_cases_ay_az}
\end{subfigure}
\begin{subfigure}{0.32\linewidth}
\includegraphics[width=\linewidth]{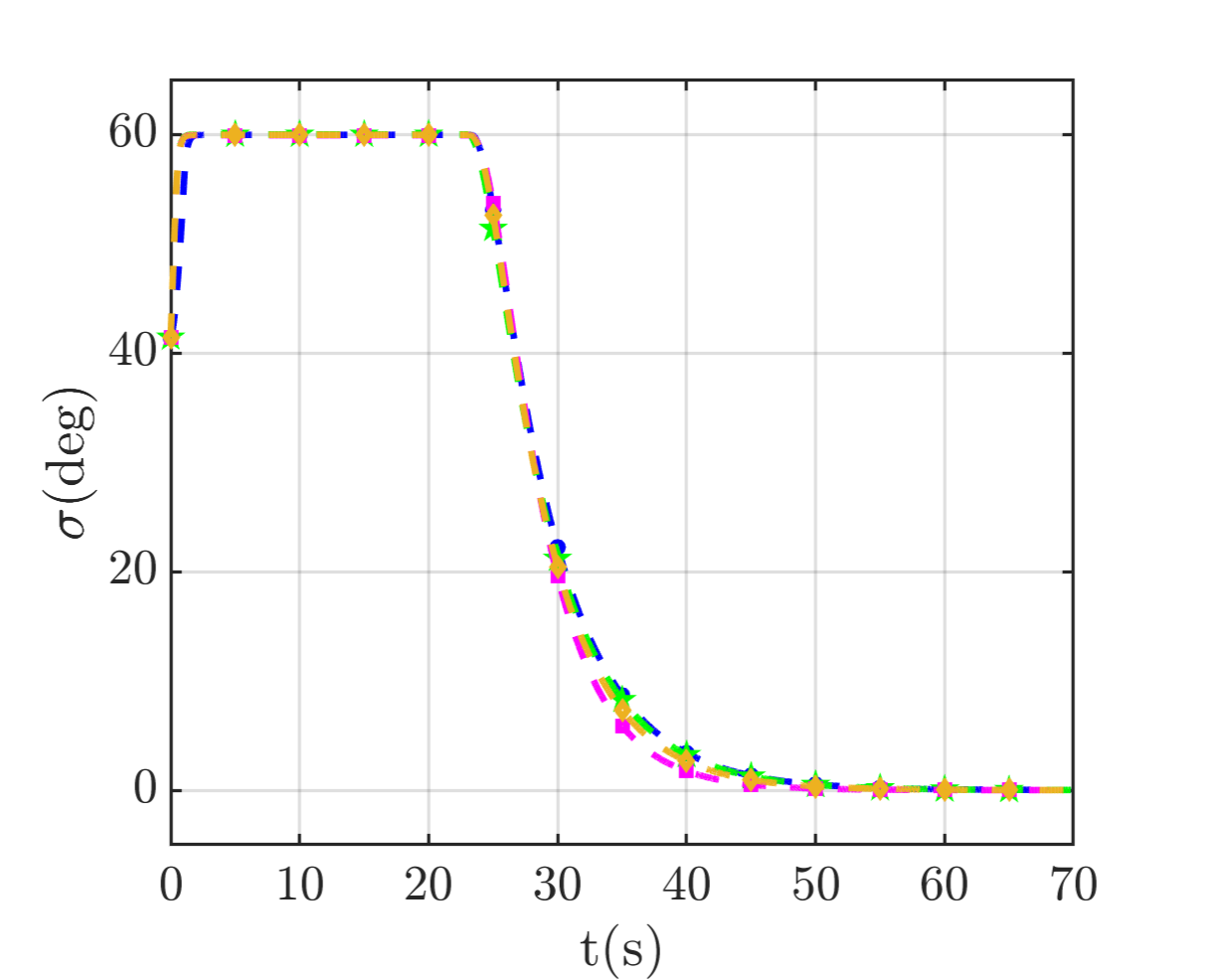}
\caption{Effective Lead Angle}\label{fig:one_tf_all_cases_sigma}
\end{subfigure}
\begin{subfigure}{0.32\linewidth}
\includegraphics[width=\linewidth]{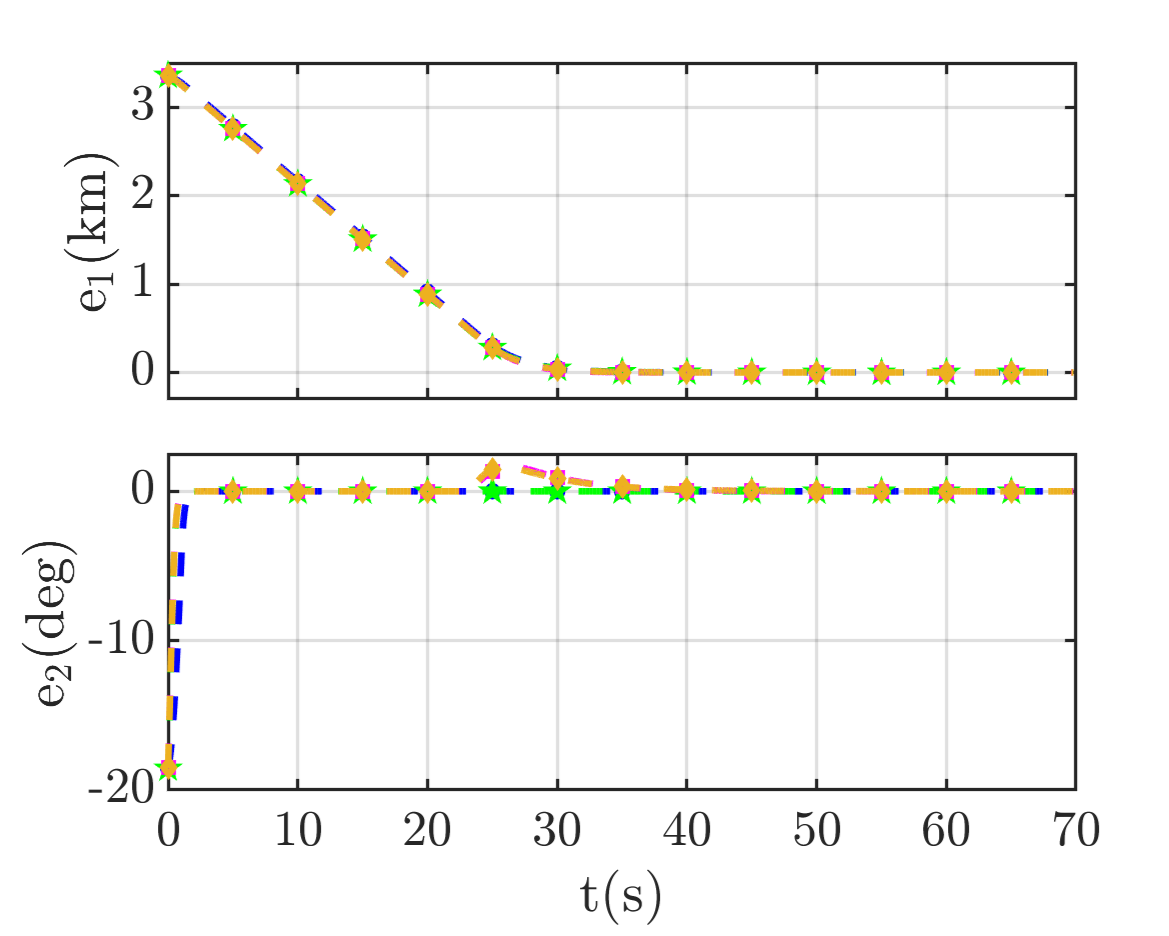}
\caption{Errors $e_1$ and $e_2$}\label{fig:one_tf_all_cases_e1_e2}
\end{subfigure}
\begin{subfigure}{0.32\linewidth}
\includegraphics[width=\linewidth]{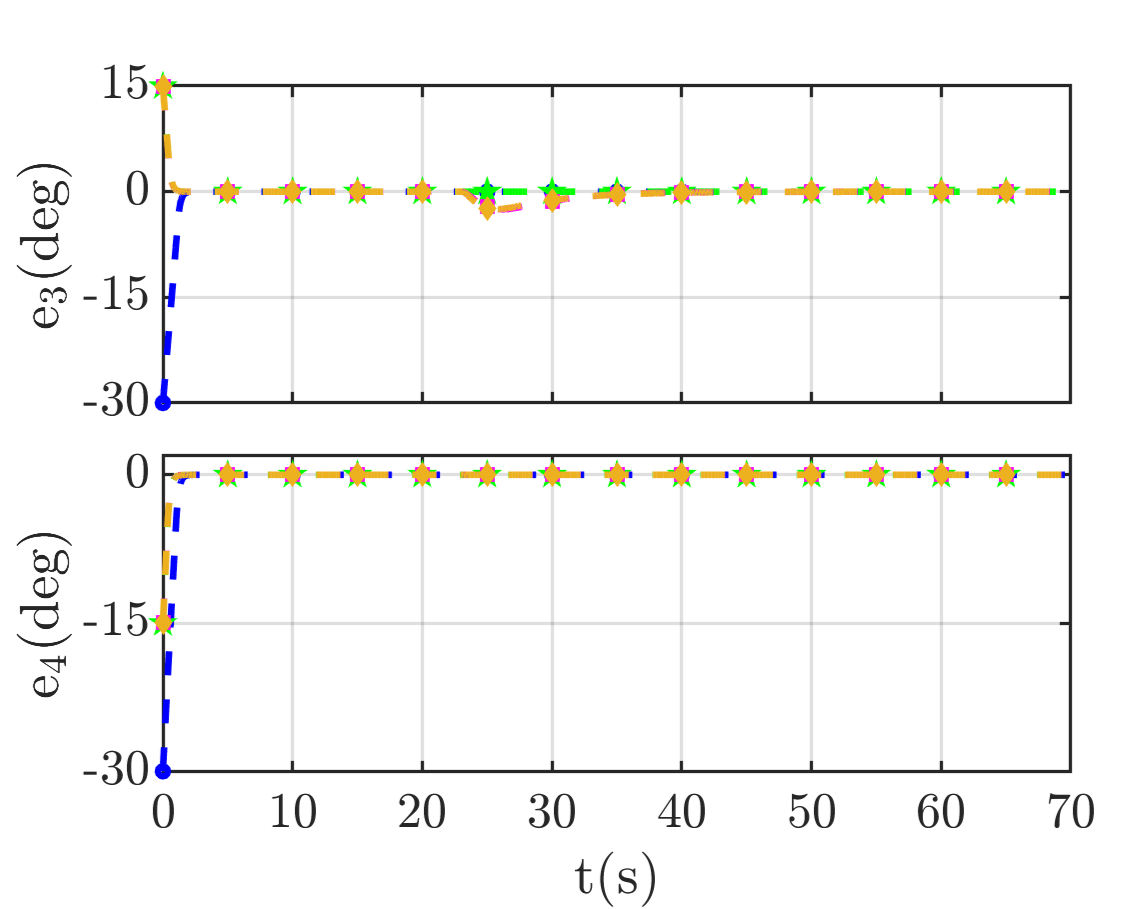}
\caption{Errors $e_3$ and $e_4$}\label{fig:one_tf_all_cases_e3_e4}
\end{subfigure}
\caption{All cases for $t_{\rm f}=70\,\rm s$, initial $\theta_{\rm m}=-30^\circ,\,\psi_{\rm m}=30^\circ$.}
\label{fig:one_tf_all_cases}
\end{figure*}

\Cref{fig:one_tf_all_cases} shows the trajectory, range, time-to-go variation, lateral acceleration commands, lead angles, and variation of errors.
\begin{table}[!ht]
\centering
    \caption{Achievable Impact Time with Different Choice of Virtual Inputs.}
    \label{tab:achieve_td}
    \begin{tabular}{lcc}
        \hline\hline
        Case & Minimum $t_{\rm f}$ (s) & Maximum $t_{\rm f}$ (s)\\ \hline
        (a) & 57.61 & 87.05 \\ 
        (b) & 59.19 & 86.35 \\ 
        (c) & 59.19 & 71.57 \\ 
        (d) & 59.19 & 86.27 \\ \hline\hline
    \end{tabular}
\end{table}
The trajectory for Case (a) in \Cref{tab:Vir_Inp} adheres to the $XY$-plane as intended by the choice of virtual inputs. Cases (b) to (d) exhibit very similar trajectories, suggesting that the heading angle profiles are similar to each other. In \Cref{fig:one_tf_all_cases_sigma}, all the cases show a general trend of two phases: first, maintaining a non-zero effective lead angle to meet the impact time constraint, followed by returning to the collision course to intercept the target, with the switch occurring at around $25\,$s. The interceptor lateral acceleration command $a_{\rm z}$ is zero for Case (a) around $25\,$s, which aligns with the interceptor taking a turn in the $XY$-plane, as observed from \Cref{fig:one_tf_all_cases_ay_az}, and \Cref{fig:one_tf_all_cases_xyz}, respectively. Additionally, from \Cref{fig:one_tf_all_cases_e1_e2,fig:one_tf_all_cases_e3_e4}, it is evident that the errors converge within a finite time indicating a successful interception of the target at the desired impact time. 

\section{Conclusion}\label{Sec:Conclusion} 
A three-dimensional nonlinear guidance law with impact time regulation and FOV constraints was presented in this work. Owing to the nonlinear framework, the proposed guidance schemes are valid even for the large initial heading error scenarios. Simulation studies have confirmed the possibility of noisy behavior in the lateral acceleration command with a backstepping approach using the effective lead angle. Backstepping was performed using the heading angles to prevent such issues from coupling between the heading angles. The control over the octant of the interceptor's trajectory was demonstrated through numerical simulations. The guidance law behaved similar to the deviated pursuit during the initial phase, maintaining a constant lead angle, after which the trajectory smoothly converged to the collision course for successful target interception. Moreover, the guidance law was validated to perform well, exhibiting zero acceleration near target interception, the capability to retain the desired trajectory octant, linear convergence of errors, and consistent performance even at high initial heading angles. These qualities were retained even in the presence of an autopilot with a time constant of 0.1\,s. Studying the effect of the various tunable parameters in the guidance law brought out the full potential of the guidance law. It was also shown that the proposed guidance strategy provided more flexibility and a higher achievable impact time than the existing ones. 

\bibliography{ref_doi} 

\end{document}